\documentclass[lettersize,journal]{IEEEtran}

\usepackage[table]{xcolor}
\usepackage{caption}
\usepackage{soul,framed} 
\usepackage[pdftex]{graphicx}
\graphicspath{{../pdf/}{../jpeg/}}
\DeclareGraphicsExtensions{.pdf,.jpeg,.png}
\usepackage[cmex10]{amsmath}
\usepackage{array}
\usepackage{eqparbox}
\usepackage{url}
\usepackage{graphicx}
\usepackage{amsmath}
\usepackage{amsfonts}
\usepackage{algorithm}
\usepackage{algpseudocode}
\usepackage{multirow}
\usepackage{footnote}
\usepackage{bm}
\usepackage{makecell}
\usepackage{pdflscape}
\usepackage{afterpage}
\usepackage{bbding}
\usepackage{lscape}
\usepackage{booktabs}
\usepackage{tikz}
\usepackage{tabularx}
\usepackage{mathrsfs}
\usepackage[figuresright]{rotating}
\usepackage{amssymb}  
\usepackage{pgfplots}
\pgfplotsset{compat=1.12}
\usepackage{subcaption}
    \pgfplotsset{
        layers/my layer set/.define layer set={
            background,
            main,
            foreground
        }{ },
        set layers=my layer set,
    }

\usepackage[square,sort,comma,numbers]{natbib} 
\usepackage[top=0.4in,bottom=0.4in,left=0.4in,textwidth=7.7in]{geometry}
\usepackage{amsfonts}

\usepackage{pifont}

\begin{document}

\title{FM4SAR-Bench 1.0: Towards A Fair Benchmark for Evaluating Foundation Models for SAR Image Interpretation}

\author{Mingyao Chen, 
Yongxiang Liu,~\IEEEmembership{Member,~IEEE,}
Tianpeng Liu*,
Yuchen Yong,
Jie Zhou,
Zhenhao Yang,
Pinyi Huang,
Weijie Li,
Jiehua Zhang,
Bowen Peng,
Zhimao Peng,
Zhiyuan Zhang, and
Li Liu*,~\IEEEmembership{Senior Member,~IEEE}
\thanks{The authors are with the College of Electronic Science and Technology, National University of Defense Technology (NUDT), Changsha, China.}
\thanks{Corresponding authors: Li Liu (liuli\_nudt$@$nudt.edu.cn) and Tianpeng Liu (liutianpeng2004$@$nudt.edu.cn).}
\thanks{This work was supported by National Natural Science Foundation of China (NSFC) under Grant 62376283 and 62531026, Innovation Research Foundation of National University of Defense Technology (JS2023-03), and the Fundamental and Interdisciplinary Disciplines Breakthrough Plan of the Ministry of Education of China (JYB2025XDXM110).}}

\markboth{Journal of \LaTeX\ Class Files,~Vol.~14, No.~8, August~2021}%
{Shell \MakeLowercase{\textit{et al.}}: A Sample Article Using IEEEtran.cls for IEEE Journals}


\maketitle

\begin{abstract}
Foundation models have rapidly advanced synthetic aperture radar (SAR) image interpretation, while giving rise to a heterogeneous and fragmented model ecosystem. Existing models differ substantially in their input, training data, objectives, architectures, and release policy, yet the field lacks a unified taxonomy and comparative framework. Current evaluations rely primarily on task-specific benchmarks, which are ill-suited to fair comparison among models with fundamentally different input-output formats and task paradigms, and often overlook data transparency, reproducibility, interpretability, and deployment readiness. Moreover, restricted accessibility of SAR data means high reported performance does not necessarily imply access to training data, source code, or weights, limiting reproduction, benchmarking, and deployment. To address these issues, this paper proposes a systematic taxonomy and hierarchical evaluation framework for SAR-related foundation models: five rating levels, six dimensions, and fine-grained criteria, with separate protocols for visual foundation models (VFMs) and vision-language models (VLMs). Beyond overall ratings, it yields interpretable multidimensional capability profiles and supports user-defined weighting, so rankings follow user priorities. Utilizing a fixed auxiliary multimodal large language model in combination with human review, we conduct a systematic meta-analysis and multidimensional assessment of 75 VFMs and 40 VLMs. The results yields the field-wide empirical picture. In terms of SAR benchmark task performance under our framework, SAR-native VLMs do not significantly outperform remote sensing VLMs pretrained on mixed multimodal corpora in which SAR is one of several modalities. This absence of a clear advantage is attributable in large part to data availability: large-scale, high-quality, publicly available annotated SAR data remain scarce, which restricts both the model scale and the training-data scale of SAR-native models. The framework provides a structured basis for requirement-aware model selection and remains extensible as new models emerge.
\end{abstract}


\section{Introduction}
\IEEEPARstart{S}{ynthetic} aperture radar (SAR) is a key active microwave remote sensing modality with the advantages of all-time, all-weather, and active imaging capabilities \cite{survey-fifty-zhou,survey-A-tutorial-Alberto}. It can provide reliable ground observations under adverse weather conditions, during nighttime, and over large-scale areas. As a result, SAR imagery has become increasingly important in a wide range of applications, including disaster monitoring \cite{survey-Landslide-Nicola,survey-Urban-flood-Jie,survey-Earthquake-Eleanor}, maritime surveillance \cite{survey-oil-spill-Brekke,survey-SAR4Ocan-Asiyabi}, urban change detection \cite{survey-SARbuilding-Ge}, and target recognition \cite{survey-recent-advance-Ping,dataset-ATRNet-STAR-Liu,atr-PGD-Li,detect-Madinet-Zhou}. Despite these advantages, SAR image understanding remains challenging due to several unique factors, such as strong speckle noise \cite{survey-Deep-Learning4sar-Liao}, complex imaging geometry, ambiguous scattering mechanisms, expensive annotation \cite{survey-A-Review-Yin,survey-UDA-Shi}, and limited generalization across sensors, regions, and acquisition conditions \cite{survey-Transfer-Adaption-Yang,survey-Toward-generalizing-Liao}. These characteristics make SAR both a highly valuable and technically demanding domain for visual representation learning.

Recently, the rapid development of large-scale models and foundation models (FMs) has significantly reshaped the research paradigm of remote sensing and SAR image understanding \cite{benchmark-SARChat-Bench-2M-Ma,model-RingMoE-Bi,Gene-HuiYanEarth-SAR-Liu}. Inspired by the success of visual foundation models (VFMs) and vision-language models (VLMs) in natural image domains, a growing number of models have emerged in remote sensing \cite{model-Ringmo-Xian} and SAR communities \cite{model-SAR-ATR-X-Li}. These include remote-sensing visual foundation models, remote-sensing vision-language models, SAR-specific visual foundation models, and SAR-specific vision-language models. Such models have shown promising potential in a variety of tasks, including scene classification \cite{model-VHM-Pang}, object detection \cite{model-SAR-ATR-X-Li}, semantic segmentation \cite{model-FUSAR-KLIP-Yi}, change detection \cite{model-falcon-Yao}, image-text understanding \cite{model-FUSAR-GPT-Zhang,model-SARCLIP-Jiang}, and visual question answering \cite{model-SAR-TEXT-He}. They also provide new opportunities for general-purpose SAR representation learning and multi-task transfer.

However, the rapid increase in model numbers has also led to a highly heterogeneous and fragmented model ecosystem. The gap between general-purpose foundation model, remote-sensing foundation models, and SAR-specific foundation models is substantial \cite{benchmark-Choice-AN}. Existing models differ significantly in input modality, training data, pretraining objectives, network architecture, parameter scale, task formulation, and release policy. Due to the lack of a unified taxonomy and comparison coordinate system, researchers struggle to systematically understand the relationships, capability boundaries, and development trajectories of different models \cite{survey-fifty-zhou}. Meanwhile, engineers face difficulties in selecting suitable models for practical deployment \cite{survey-Foundation-Models-Xiao,survey-Generative-Artificial-Intelligence-Huang}.

Existing evaluation practices are mainly based on task-specific benchmark leaderboards, where metrics such as classification accuracy, detection precision, segmentation quality, or change detection performance are used as the primary criteria \cite{benchmark-SARChat-Bench-2M-Ma,benchmark-SAREval-Wang,benchmark-SARLANG-1M-wei}. However, such evaluation protocols often overlook broader evaluation dimensions, including model influence, data transparency, openness, reproducibility, architectural interpretability, domain adaptability, and deployment readiness. This issue is particularly critical in the SAR domain. Since SAR data are often sensitive or subject to access restrictions, some models with strong reported performance may not provide sufficient access to training data, source code, or pretrained weights. As a result, they may be difficult to reproduce, evaluate, or deploy in real-world applications. Moreover, due to heterogeneous training architectures and divergent task paradigms, many models cannot be fairly compared using the same evaluation benchmark. For instance, CLIP-based models output a scalar similarity score rather than a natural language sequence, which makes it infeasible for mainstream SAR multi-modal large model evaluation frameworks (e.g., SARLANG-1M \cite{benchmark-SARLANG-1M-wei}, SARChat-Bench-2M \cite{benchmark-SARChat-Bench-2M-Ma}) to directly and fairly benchmark them.


Furthermore, different user groups have substantially different expectations for SAR foundation models. Academic researchers usually emphasize architectural innovation, methodological novelty, task performance, and scientific impact \cite{survey-Foundation-Models-Xiao,survey-VFM-in-RS-Lu}. In contrast, engineering practitioners prioritize model stability, deployability, maintainability, inference efficiency, data compliance, security, and integration with operational systems \cite{survey-Open-Sourcing-Seger}. Therefore, a single fixed evaluation scheme cannot satisfy the diverse requirements of both scientific research and engineering deployment. A systematic framework that jointly considers model capability, openness, feasibility, and user-specific preferences is urgently needed for the sustainable development of SAR foundation models.

To address these issues, this paper presents a systematic taxonomy and a hierarchical evaluation framework for SAR-related foundation models, implemented as an extensible rating tool. The taxonomy organizes heterogeneous models into a unified structure, while the framework evaluates them along multiple dimensions beyond task performance and adapts to different model types and user requirements. The main contributions are summarized as follows:

\begin{itemize}
    \item We establish a six-category taxonomy of SAR-related foundation models and systematically compile and analyze existing SAR-related benchmarks, summarizing their task coverage and evaluation characteristics. This provides a structured overview of SAR-related foundation models and the benchmarks utilized to evaluate them.
    
    \item We propose a hierarchical rating framework that evaluates models across five rating levels, six dimensions, and multiple fine-grained criteria. The framework extends beyond task performance to assess influence, technical architecture, SAR domain adaptability, training data, and open-sourcing and reproducibility.
    
    \item We develop dual-track evaluation protocols for VFMs and VLMs, retaining shared high-level dimensions while adapting criteria to each model type. We also design a dynamic weight allocation scheme based on user profiling, comprising seven predefined weighting strategies and support for user-customized configurations, enabling one-click switching of model ratings and rankings under different requirement priorities.
    
    \item We apply the framework to 115 SAR-related foundation models, including 75 VFMs and 40 VLMs. For the SAR benchmark task performance dimension, we conduct a meta-analysis of the experimental evidence reported in the original publications, enabling scalable evaluation without requiring all models to be re-evaluated on a common benchmark.
\end{itemize}

\begin{table*}[t]
\centering
\caption{Taxonomy of SAR-related foundation models.}
\label{tab:sar_model_taxonomy}

\renewcommand{\arraystretch}{1.18}
\setlength{\tabcolsep}{3.5pt}
\scriptsize

\begin{tabularx}{\textwidth}{
    >{\raggedright\arraybackslash}p{2.0cm}
    >{\raggedright\arraybackslash}X
    >{\raggedright\arraybackslash}X
    >{\raggedright\arraybackslash}p{2.9cm}
    >{\raggedright\arraybackslash}X
    >{\raggedright\arraybackslash}X
}
\toprule
\textbf{Category}
&
\textbf{Operational definition}
&
\textbf{Core inclusion criteria}
&
\textbf{Representative examples}
&
\textbf{Expected strengths}
&
\textbf{Potential improvements}
\\
\midrule

\makecell[l]{\textbf{RS-VFM}\\
\textbf{without SAR}\\
\textit{Transfer-relevant}}
&
A visual foundation model pretrained on optical or multispectral
remote-sensing imagery without documented use of SAR data.
&
Provides transferable representations for multiple remote-sensing
tasks; uses overhead imagery during pretraining; does not explicitly include SAR in the disclosed pretraining or adaptation data.
&
Prithvi \cite{model-Prithvi-Johannes}, RingMo \cite{model-Ringmo-Xian}, RingMo-Aerial \cite{model-RingMo-Aerial-Diao}, RingMo-Galaxy \cite{model-RingMo-Galaxy-Wang}, RingMo-Sense \cite{model-RingMo-sense-Yao}, CrossEarth \cite{model-CrossEarth-Gong}
&
Improved perception of nadir or near-nadir views, arbitrary object
orientations, multiscale geographic structures, and spatially organized remote-sensing scenes. These properties may support zero-shot or few-shot transfer to SAR.
&
Does not explicitly learn radar-specific radiometry or scattering
mechanisms; the degree of transfer to SAR remains task-, sensor-, and preprocessing-dependent.
\\
\midrule

\makecell[l]{\textbf{RS-VFM}\\
\textbf{with SAR}\\
\textit{Directly SAR-aware}}
&
A remote-sensing visual foundation model pretrained using SAR together
with optical, multispectral, or other Earth-observation modalities.
&
SAR is explicitly included in representation learning, multimodal
alignment, continued pretraining, or systematic transfer evaluation.
&
CROMA \cite{model-CROMA-Aoh}, MAPEX \cite{model-MAPEX-Hanna}, SkySense \cite{model-SkySense-Guo}, RingMoE \cite{model-RingMoE-Bi}, TerraFM \cite{model-TerraFM-Muhammad}, MaRS \cite{model-MaRS-Yang}, FlexiMo \cite{model-FlexiMo-Li}, RingMo-SAM \cite{model-RingMo-SAM-Yan}
&
Supports multisensor representation learning, optical-SAR alignment,
cloud-independent observation, and improved initialization for
downstream SAR tasks.
&
SAR may constitute only a limited proportion of the pretraining corpus;
performance can remain sensitive to sensor type, frequency band,
polarization, resolution, and preprocessing.
\\
\midrule

\makecell[l]{\textbf{RS-VLM}\\
\textbf{without SAR}\\
\textit{Transfer-relevant}}
&
A language-enabled remote-sensing model trained mainly on optical
overhead imagery, without documented SAR exposure.
&
Supports remote-sensing captioning, visual question answering, retrieval,
grounding, or dialogue; uses remote-sensing imagery and text but does not
explicitly include SAR data.
&
RemoteCLIP \cite{model-RemoteCLIP-Liu}, GeoChat \cite{model-GeoChat-Kuckreja}, LHRS-Bot \cite{model-LHRS-Bot-Muhtar}, SkyEyeGPT \cite{model-SkyEyeGPT-Zhan}, SkySense-O \cite{model-SkySense-O-Zhu}, GeoLLaVA \cite{model-GeoLLaVA-Wang}, GeoPixel \cite{model-GeoPixel-Shabbir}
&
Stronger geospatial vocabulary, overhead-scene understanding, spatial
reasoning, and orientation robustness than models trained predominantly
on ground-level natural images.
&
Potential SAR capability is indirect and originates mainly from
overhead-view and geospatial priors rather than radar-specific
knowledge.
\\
\midrule

\makecell[l]{\textbf{RS-VLM}\\
\textbf{with SAR}\\
\textit{Directly SAR-aware}}
&
A remote-sensing vision-language model explicitly exposed to SAR
together with optical or other Earth-observation modalities.
&
SAR data are included in pretraining, cross-modal alignment, supervised fine-tuning, or instruction tuning.
&
EarthGPT \cite{model-EarthGPT-Zhang}, EarthDial \cite{model-EarthDial-Soni}, Falcon \cite{model-falcon-Yao}, VHM \cite{model-VHM-Pang}, EarthGPT-X \cite{model-EarthGPT-X-Zhang}, TerraScope \cite{model-TerraScope-Shu}
&
Combines geospatial language capability with multi-sensor awareness and may support cross-modal reasoning over optical and SAR observations.
&
SAR training data are often limited, and evaluation may not cover
classical SAR problems such as scattering interpretation, aspect-angle variation, or coherent change detection.
\\
\midrule

\makecell[l]{\textbf{SAR-VLM}\\
\textit{Directly SAR-aware}}
&
A vision-language model specifically designed or adapted for
natural-language interaction with SAR imagery.
&
SAR-centered pretraining or instruction data are used, and a language
interface is an essential component of the model; evaluation primarily
targets SAR understanding.
&
FUSAR-KLIP \cite{model-FUSAR-KLIP-Yi}, SARCLIP \cite{model-SARCLIP-Jiang}, FUSAR-GPT \cite{model-FUSAR-GPT-Zhang}, SARVLM \cite{model-SARVLM-Ma}, FUSAR-R1 \cite{model-FUSAR-R1-Yi}, SAR-GPT \cite{model-SAR-TEXT-He}
&
Supports SAR image description, visual question answering, semantic recognition, instruction following, and natural-language interaction.
&
Reliability on precise localization, counting, fine-grained target
recognition, physical interpretation, and out-of-domain transfer remains insufficiently evaluated.
\\
\midrule

\makecell[l]{\textbf{SAR-VFM}\\
\textit{Directly SAR-aware}}
&
A transferable visual foundation model whose pretraining data,
objectives, or intended downstream applications are centered on SAR.
&
Uses large-scale SAR data for representation learning and demonstrates transfer across multiple SAR datasets, sensors, or downstream tasks.
&
SARATR-X \cite{model-SAR-ATR-X-Li}, SUMMIT \cite{model-SUMMIT-Du}, SARViT \cite{model-SARViT-Ma}, SAR-MAE \cite{model-SARMAE-Liu}, PICV \cite{model-PICV-Wang}, SAR-W-MixMAE \cite{model-SAR-W-MixMAE-Ali}, SAMBA \cite{model-SAMBA-Wang}
&
High deployment value for target recognition, object detection,
segmentation, change detection, and general-purpose SAR feature
extraction.
&
Typically lacks a natural-language interface and is often omitted from VLM-centered benchmarks; transfer evaluation may also be affected by overlap between pretraining and evaluation data.
\\

\bottomrule
\end{tabularx}

\end{table*}

\section{Taxonomy of SAR-Related Benchmarks and Models}
\subsection{Scope and Operational Definitions}

Following the scale-and-capability perspective adopted in recent reviews of foundation models \cite{survey-Foundation-Models-Xiao,survey-VFM-in-RS-Lu}, we do not define a foundation model using a fixed parameter threshold. Instead, a model is considered a foundation model when it is pretrained on large and diverse datasets, provides transferable representations or interfaces, and can be adapted to multiple downstream tasks. Model size is therefore an important but insufficient condition. Pretraining breadth, transferability, and task generality are equally important.

Within this scope, two model families should be distinguished. A vision-language model (VLM) establishes an explicit connection between visual observations and natural language. It supports at least one language-conditioned function, such as image-text retrieval, visual question answering (VQA), image captioning, referring-expression grounding, or instruction following. A visual foundation model (VFM), by contrast, primarily learns transferable visual representations for classification, detection, segmentation, change detection, or other perception tasks. The two categories are not mutually exclusive at the architectural level: a model may provide both general-purpose visual features and a language interface. In this work, models are assigned according to their primary training objective and intended usage.

In this study, SAR-related models are defined from the perspective of their relevance to SAR understanding and transfer, rather than solely by whether SAR data are included during training. The resulting landscape therefore covers both models with direct SAR exposure and models with plausible transferability to SAR applications. We distinguish two levels of relevance. Directly SAR-aware models explicitly incorporate SAR data during pretraining, representation alignment, supervised fine-tuning, or instruction tuning. By contrast, SAR-transfer-relevant models may not have encountered SAR data but are pretrained on large-scale optical or multi-spectral remote-sensing imagery and are therefore potentially transferable to SAR tasks. Compared with foundation models trained predominantly on ground-level natural images, remote-sensing models are more extensively exposed to nadir or near-nadir viewpoints, arbitrary object orientations, large variations in spatial scale, and geographically structured scenes \cite{benchmark-OVRSISBenchV2-Li}. Natural-image models, in contrast, often learn orientation priors from gravity-constrained, canonical-view imagery and may consequently be less robust to the rotational variability of overhead observations. Remote-sensing pretraining may thus provide useful geometric and scene-level priors for zero-shot or few-shot transfer to SAR, even without direct radar supervision. This potential advantage remains an empirical hypothesis and should be assessed through controlled comparisons among general models, optical-only remote-sensing models, and SAR-aware models.

Based on the above definition, We organize the SAR-related foundation model landscape using six labels, as shown in Table \ref{tab:sar_model_taxonomy}. It should be noticed that this taxonomy is based on training evidence rather than model names. 

\subsection{SAR-Related Foundation Models}
\label{sec-2.2}
\begin{figure}[htbp]
\centering
\includegraphics[width=1.0\columnwidth]{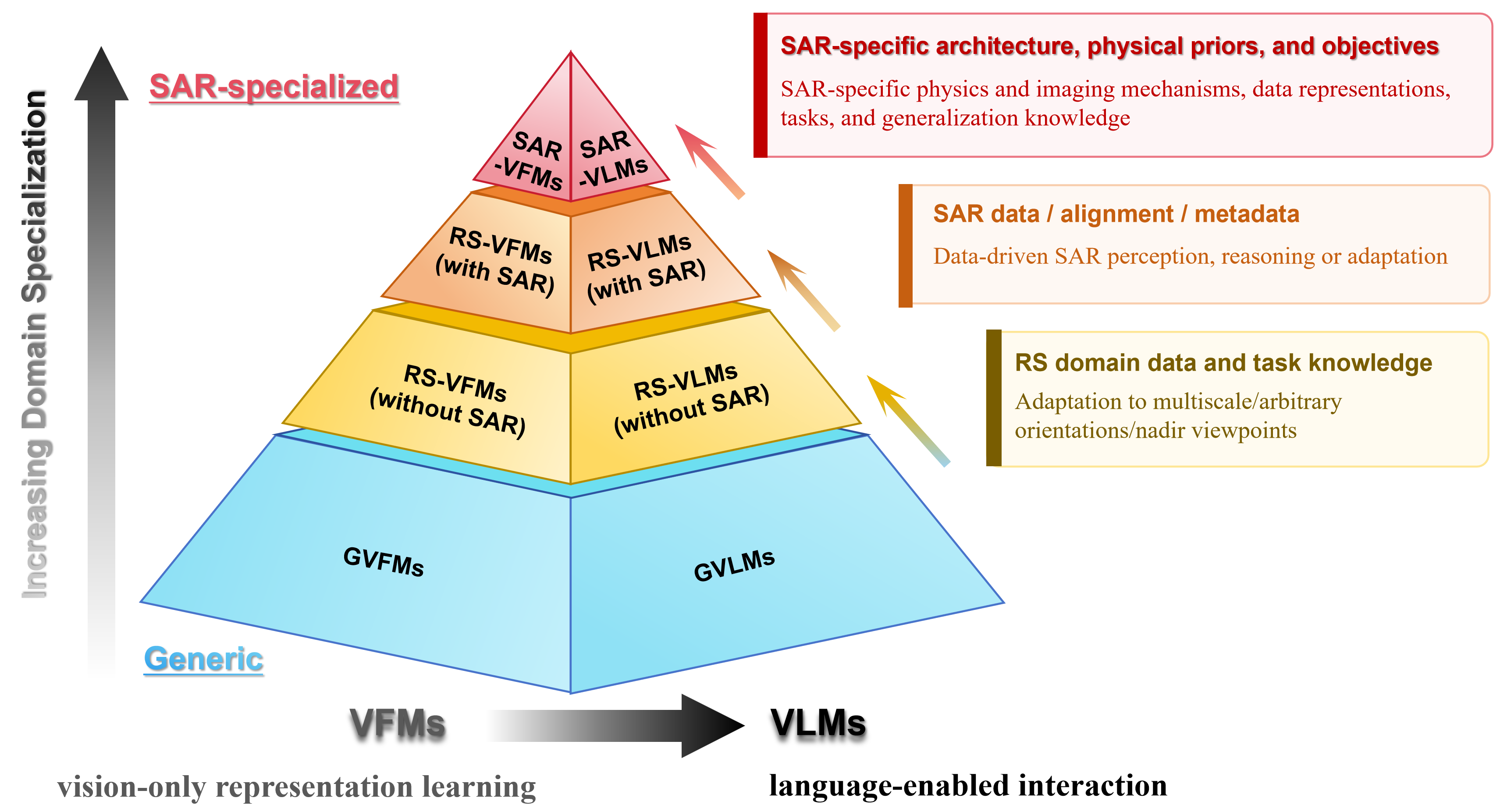}
\caption{Relationship pyramid of SAR-related foundation models.}
\label{fig2}
\end{figure}

The relationships among various SAR-related foundation models can be represented as a pyramid, as illustrated in Fig. \ref{fig2}. The horizontal dimension represents the transition from vision-only representation learning to language-enabled interaction, whereas the vertical dimension represents increasing domain specialization. At the bottom of the pyramid, generic vision-language models (GVLMs) acquire broad semantic knowledge from natural images and web documents but have limited exposure to overhead geometry and radar phenomenology. Remote-sensing models introduce nadir-view imagery, multi-scale geographic structures, multi-spectral observations, and geospatial terminology. Models that further incorporate SAR learn representations that are more tolerant of speckle, imaging geometry, scattering variation, and modality-specific intensity distributions. At the top of the pyramid, SAR-specialized VFMs and VLMs are optimized for radar-native interpretation and task transfer.

\subsubsection{Remote-sensing visual foundation models (RS-VFM)}
RS-VFMs aim to learn transferable visual representations from large-scale earth observation imagery. Unlike conventional RS models designed for a single task, RS-VFMs are generally pretrained using self-supervised, supervised, or multi-modal objectives and can subsequently be adapted to scene classification, semantic segmentation, object detection, change detection, and other downstream applications.

In this survey, RS-VFMs are divided into two groups according to whether SAR imagery is used during pretraining. This is a data-centric taxonomy. A model that can technically accept SAR images at inference time is not considered an RS-VFM with SAR unless SAR data participate in its training process. Moreover, an RS-VFM with SAR is not necessarily a SAR foundation model. The former may treat SAR as one of several auxiliary modalities, whereas the latter should utilize SAR as the primary modality and explicitly model its physical or statistical characteristics.

\textbf{RS-VFMs without SAR: }
At present, a substantial portion of RS-VFMs are trained primarily on optical, RGB, multi-spectral, or hyperspectral imagery. By exploiting large-scale image collections together with masked image modeling (MIM), temporal-consistency learning, multi-scale learning, and spectral-reconstruction objectives, these models substantially improve the transferability of RS visual representations. The visual representations learned by these models are particularly effective for optical RS phenomena, including reflectance patterns, color and texture distributions, cloud and shadow characteristics, vegetation appearance, and man-made structures. As a result, they generally perform well on optical scene classification, land-cover mapping, object detection, and segmentation. These optical successes, however, are not indicative of SAR-specific understanding. Although a RS-VFM trained solely on optical data can transfer some spatial priors to SAR tasks after fine-tuning, its zero-shot SAR capability is typically limited. In particular, such models may fail to distinguish an object's appearance from its radar-scattering behavior, and may struggle to interpret bright or dark regions in an image in terms of the SAR-specific imaging mechanism.

\textbf{RS-VFMs with SAR: }
Another class of foundation models incorporates SAR together with optical imagery or other earth-observation modalities during pre-training, employing techniques such as cross-modal contrastive learning and masked reconstruction. Introducing SAR into pre-training offers several potential advantages. First, SAR can provide complementary information under cloud cover or at night. Second, pre-training on multi-sensor data may improve robustness to missing modalities, varying illumination conditions, and diverse imaging environments. These properties are valuable for applications such as land-cover mapping, disaster monitoring, flood detection, infrastructure monitoring, and change detection.

Nevertheless, the mere existence of a SAR branch or a small number of SAR images does not guarantee strong SAR representations. The actual contribution of SAR during training depends on the quantity, diversity, and proportion of SAR data, as well as the loss weighting across modalities. If the training samples and task annotations are drawn predominantly from optical imagery, the optimization signal may still be dominated by optical data. In this case, the model may exploit SAR merely as an auxiliary cross-modal alignment cue rather than modeling it as an independent physical information source. Consequently, to clarify a model's capacity to process SAR data, the proportion of SAR data used during training should be reported. It is worth noting that simply increasing the SAR proportion without also increasing data diversity may lead to overfitting to specific sensors. In summary, RS-VFMs with SAR can be regarded as an intermediate class between optical-centric RS-VFMs and SAR-VFMs.

\subsubsection{Remote-sensing vision-language models (RS-VLM)}
RS-VLMs are designed to bridge RS imagery and natural language. Their tasks include image captioning, VQA, image-text retrieval, referring-expression grounding, spatial reasoning, and interactive dialogue. Compared with RS-VFMs, RS-VLMs offer users a more intuitive natural-language interface. However, it should be noted that fluent language generation does not necessarily imply reliable visual or physical understanding. Following the same data-centric principle, we divide RS-VLMs into RS-VLMs without SAR and RS-VLMs with SAR, based on whether SAR imagery is used for pre-training, supervised fine-tuning, instruction tuning, or cross-modal alignment.

\textbf{RS-VLMs without SAR: }
These models are built primarily around optical-dominated or RGB-dominated RS data. By mapping remote-sensing imagery and textual descriptions into a shared semantic space, they have advanced RS vision-language understanding. They can describe land-cover categories, identify objects and their spatial relationships, answer image-based questions, generate relatively detailed captions, and perform referring-expression segmentation or pixel-level grounding. Nevertheless, when SAR is not included in the training corpus, these capabilities do not directly generalize to SAR imagery. A language model may have encountered the term "SAR" and related knowledge during general text pre-training, but such text-level knowledge does not establish a visual mapping between SAR image patterns and their physical causes. In practice, an RS-VLM trained solely on optical imagery may mistake speckle noise for ordinary image noise, interpret radar shadow as optical shadow, describe strongly scattering SAR targets using optical color concepts, or even fail to determine whether an image belongs to the SAR modality. This also explains why a large RS-VLM can sometimes recognize the modality name "SAR" yet fail to understand an input SAR image.

\textbf{RS-VLMs with SAR: }
Multi-sensor RS-VLMs seek to reduce reliance on optical data by combining SAR with other earth-observation data such as optical, multi-spectral and infrared. Incorporating SAR into RS-VLM training serves two purposes. On the visual side, the model can learn that SAR and optical images represent the same geographic scene through different sensing mechanisms. On the language side, SAR-related captions, questions, and instructions can introduce concepts such as speckle noise, radar shadow, target structure, and the complementarity of sensor parameters. This may improve modality recognition and basic cross-sensor semantic alignment.

However, the strength of SAR understanding still depends heavily on the effective proportion of SAR data and the quality of SAR-related language supervision. If SAR imagery accounts for only a small fraction of the visual training data while the vast majority of instructions and descriptions are derived from optical imagery, the model may still learn primarily optical concepts, with its cross-modal understanding of SAR remaining at the level of mere modality recognition. If SAR imagery is paired only with generic labels such as "ship," "building," or "urban area," the model may acquire object-level semantics but fail to understand why these targets appear bright, dark, or fragmented in SAR imagery. Therefore, to assess and compare a model's level of SAR understanding, each model should report the quantity, proportion, and sensor parameters of its SAR data, as well as its performance on SAR-only subsets. In addition, SAR results should be reported separately from optical results. In particular, the ability to interpret SAR-specific characteristics such as speckle noise, layover, and radar shadow warrants dedicated analysis. 

\subsubsection{SAR vision foundation models (SAR-VFM)}
SAR-VFMs are stricter and more specialized than RS-VFMs. A model can be classified as a SAR-VFM only when SAR occupies a central role in its data, model architecture, pre-training objectives, and evaluation protocols. Developing a SAR-VFM is more challenging than developing an optical-centric RS-VFM. SAR data are affected by factors such as sensor configuration, frequency band, polarization, incidence angle, spatial resolution, radiometric-calibration status, and processing level, so the same target may exhibit markedly different appearances under different observation conditions. Moreover, complex-valued SAR data contain both amplitude and phase information, whereas many public datasets retain only amplitude or log-amplitude images. Consequently, if a model is trained only on processed intensity images, it may fail to learn phase-related or polarization-related information.

SAR-specific pre-training objectives may include masked modeling of amplitude and phase, polarization-aware feature learning, contrastive learning across different imaging conditions and predictive learning over multi-temporal SAR sequences. These objectives help the model become more robust to speckle noise and imaging variations while preserving information critical to scattering mechanisms. Nevertheless, many current SAR representation-learning models remain task-specific rather than foundation models in the strict sense. They may use only a single sensor, a limited geographic region, or a few polarization modes, or may be trained solely for specific tasks such as automatic target recognition (ATR), ship detection, or land-cover classification. Such models may achieve strong performance on their specific tasks, yet they do not necessarily meet the requirements of foundation models in terms of data scale, scene diversity, transferability, and multi-task adaptability.

The scarcity of strict SAR-VFMs is closely linked to the lack of large-scale, diverse, and openly available SAR datasets. Future SAR-VFM training sets should span multiple SAR tasks and sensors, covering different frequency bands, polarization modes, incidence angles, spatial resolutions, geographic regions, seasons, and imaging conditions. Otherwise, a model may overfit to a few sensors rather than learn broadly transferable SAR representations.
\subsubsection{SAR vision-language models (SAR-VLM)}
A SAR-VLM should take SAR images as its primary visual input and be pre-trained or instruction-tuned on a large volume of SAR-related captions, question-answering data, or instruction data. Beyond recognizing the SAR modality, it should be able to describe and reason about targets, scenes, spatial relationships, temporal changes, and SAR-specific phenomena. Notably, the capabilities demonstrated by a SAR-VLM must go beyond modality recognition and cross-modal alignment: if a model can determine that an image is SAR but cannot reliably interpret speckle noise, radar shadow, layover, target scattering, or polarization-related characteristics, it cannot yet be considered a mature SAR-VLM. At the semantic level, a mature SAR-VLM should be able to recognize targets such as ships, vehicles, and buildings, as well as scenes such as harbors, roads, farmland, and urban areas. At the reasoning level, the model should further be able to explain the relationship between image patterns and radar imaging mechanisms, including scattering, shadow formation, layover, polarization, speckle noise, and incidence-angle variation. For multi-temporal SAR data, the model must distinguish genuine physical change from changes in imaging conditions. The depth of a model's SAR image understanding is a key focus in the construction of experimental SAR-VLM benchmarks \cite{benchmark-SARChat-Bench-2M-Ma,benchmark-SAREval-Wang,benchmark-SARLANG-1M-wei}.

Language supervision is especially important in training models toward stronger SAR perception and reasoning. Generic captions can teach a model that a scene contains ships or buildings, but not necessarily how these targets appear in SAR imagery or why the associated scattering patterns arise. SAR-specific instruction data should therefore include questions on target recognition, scattering behavior, imaging geometry, sensor conditions, and uncertainty. The proportion of SAR data is one of the core factors in SAR-VLM development: a certain scale of SAR image-text pairs can improve cross-modal alignment and scene-level semantic understanding. However, strong SAR reasoning typically requires a sufficiently large and diverse SAR corpus, together with SAR-specific training objectives and language annotations. Accordingly, a model's SAR understanding should not be measured by image scale alone, but should also account for SAR-related training mechanisms, the amount of instruction data, data diversity, and the handling of SAR-specific characteristics. This further motivates us to incorporate these factors into the rating tool we propose.

Overall, progressing from the "RS-VLM with SAR" to a genuine SAR-VLM cannot be achieved merely by adding SAR images to a multi-modal corpus. It also requires SAR-centric representation learning, physically meaningful language supervision, SAR-specific evaluation, and transparent reporting of SAR data usage. This distinction also explains why some early GVLMs or RS-VLMs could fluently answer SAR-related questions yet still fail to recognize or interpret specific visual evidence in SAR images.

\subsection{Remote Sensing and SAR-Specific Benchmarks}

\subsubsection{Remote-sensing benchmarks}

\begin{table*}[htbp]
    \centering
    \caption{Overview of Remote-Sensing Benchmarks (Abbreviations adopted: OP=Optical, BT=Bi-Temporal, MT=Multi-Temporal, FC=False Color, MS=Multi-Spectral, HS=Hyperspectral, DEM=Digital Elevation Models, IF=Infrared, A=Automatic, M=Manual, V=Verify, IC=Image Caption, BBox=Bounding Box, MCQ=Multiple Choice Questions, Seg=Segmentation Mask, CoT=Chain-of-Thought, IoU=Intersection over Union, MCC=Matthews Correlation Coefficient, LoRA=Low-Rank Adaptation, SFT=Supervised Fine-Tuning).}
    \label{tab:RS-benchmark}
    
    \resizebox{0.99\linewidth}{!}{%
        \begin{tabular}{|>{\centering\arraybackslash}m{2.2cm}|>{\centering\arraybackslash}m{0.6cm}|>{\centering\arraybackslash}m{1.4cm}|>{\centering\arraybackslash}m{2.0cm}|>{\centering\arraybackslash}m{1.5cm}|>{\centering\arraybackslash}m{1.2cm}|c|c|c|>{\centering\arraybackslash}m{3.0cm}|>  
        {\centering\arraybackslash}m{1.3cm}|>{\centering\arraybackslash}m{1.3cm}|}
            \hline
            \multirow{2}{*}{\shortstack{\\Name}}  & \multirow{2}{*}{\shortstack{\\Year}} & \multirow{2}{*}{\shortstack{\\Modalities}} & \multirow{2}{*}{\shortstack{Total Annotated\\ Pairs}} &  \multirow{2}{*}{\shortstack{Annotation\\Method}} & \multirow{2}{*}{\shortstack{ Answer\\Type}}&\multirow{2}{*}{\shortstack{Open\\Tasks}} & \multirow{2}{*}{\shortstack{Task\\Dimensions}}&\multirow{2}{*}{\shortstack{Evaluation\\Settings}}&\multirow{2}{*}{\shortstack{Core Metrics}}&  \multicolumn{2}{c|}{Evaluated Models} \\
            \cline{11-12}
            &  &  &  &  &  &  &  &  &   & GVLMs & RS-VLMs \\
            \hline
            \shortstack{VRSBench\\  \cite{benchmark-VRSBench-Li}}  & 2024 & OP & 205,307 & A+M+V &IC, BBox& \large \ding{51} & 12 &Zero-shot/LoRA&  BLEU, ROUGE\_L, METEOR, CIDEr, Acc@0.5, Acc@0.7, Accuracy&4& 1\\
            \hline
            \shortstack{GeoChat-Bench\\  \cite{model-GeoChat-Kuckreja}}  & 2024 & OP & 8,966 & A+M+V & IC, BBox & \large \ding{51} & 6 &Zero-shot & Accuracy, Acc@0.5, Acc@0.25, METEOR, ROUGE-1, ROUGE\_L & 3  & 6\\
            \hline
            \shortstack{VLEO-Bench\\  \cite{benchmark-VLEO-Bench-Zhang}}  & 2024 & OP, BT  & 3,000+ & A+M+V & IC, BBox, MCQ & \large \ding{51} & 10 &Zero-shot & Accuracy, MAE, BLEU, METEOR, ROUGE, CIDEr, SPICE, mIoU, CLIPScore, $R^2$, F1, RefCLIPScore, Pr@0.5, Precision, Recall, MAPE, Refusal Rate & 5  & 0\\
            \hline
            \shortstack{XLRS-Bench\\  \cite{benchmark-XLRS-Bench-Wang}}  & 2025 & OP, BT & 45,942 & A+M+V  & IC, BBox, MCQ & \large \ding{51} & 16 &Zero-shot & Accuracy, Acc@0.7, Acc@0.5, BLEU, METEOR, ROUGE\_L & 9 & 1\\
            \hline
            \shortstack{CHOICE\\  \cite{benchmark-Choice-AN}}  & 2025 & OP, SAR, MT, BT, FC & 10,507 & A+M+V &  BBox, MCQ, Seg & \large \ding{55} & 23 &Zero-shot & Accuracy, mIoU  & 18  &  6\\
            \hline
            \shortstack{RSVLM-QA\\  \cite{benchmark-RSVLM-QA-Zi}}  &2025  &  OP& 400,830 & A+V &  IC& \large \ding{51} & 6 & Zero-shot &Accuracy, BLEU, ROUGE\_L, METEOR, Caption Score&  6 & 0\\
            \hline
            \shortstack{UrBench\\  \cite{benchmark-Urbench-Zhou}}  &2025  & OP & 11,600+ & A+M+V &  IC& \large \ding{51} & 14 &Zero-shot & Accuracy & 21  &0 \\
            \hline
            \shortstack{OmniEarth-Bench\\  \cite{benchmark-OmniEarth-Bench-Wang}}  & 2025 & OP, MS, SAR, MT & 29,855 & A+M+V & IC, BBox, MCQ, CoT & \large \ding{55} & 109 &Zero-shot & Accuracy, Acc@0.5, Acc@0.7, Precision, Recall, F1, BLEU, METEOR, ROUGE\_L, CIDEr & 9  &  0\\
            \hline
            \shortstack{RSMMVP\\  \cite{benchmark-RSMMVP-Adejumo}}  & 2025 & OP & 300 & M & MCQ& \large \ding{55} & 6 &Zero-shot & Accuracy & 3  &  2\\
            \hline
            \shortstack{GEOBench-VLM\\  \cite{benchmark-GEOBENCH-VLM-Danish}}  & 2025 & OP, MS, SAR, MT, BT  & 10,000+ & A+M+V & IC, BBox, MCQ, Seg & \large \ding{51} & 31 &Zero-shot & Accuracy, Prec@0.5, Prec@0.25, mIoU, BERTScore & 8  &  5\\
            \hline
            \shortstack{GeoMMBench\\  \cite{benchmark-GeoMMBench-Xiao}}  & 2026 & OP, MS, HS, SAR, LiDAR, Thermal, DEM  & 1,053 & M+V & MCQ & \large \ding{55} & 6 &Zero-shot & Micro-averaged accuracy & 31  &  5\\
            \hline
            \shortstack{TerraScope-Bench\\  \cite{model-TerraScope-Shu}}  & 2026 & OP, SAR, MT & 3,837 & A+M+V & MCQ, Seg & \large \ding{55} & 6 &Zero-shot/LoRA & Accuracy, IoU & 6  & 6\\
            \hline
            \shortstack{OVEarth-Bench\\  \cite{benchmark-OVEarth-Bench-Li}}  & 2026 & OP & 5,023 & A+M+V & BBox, Seg & \large \ding{51} & 3 &Zero-shot & macro-P/R/IoU, micro-IoU, MCC, micro-P/R/F1@0.5, micro-F1@0.5:0.95 & 36  & 13\\
            \hline
            \shortstack{RSVideo-Bench\\  \cite{benchmark-RSVideo-Zhou}}  &  2026&OP  &  10,773 & A+M+V & MCQ, Evidence & \large \ding{55} & 17 &Zero-shot & Accuracy, T-Hit, R-Hit, Format Valid, Evidence Length & 26  & 3\\
            \hline
            \shortstack{LongEarth-Bench\\  \cite{benchmark-LongEarth-R1-Ding}} &2026 & OP & 120,367 & A+M+V & IC, CoT & \large \ding{51} &  12& Zero-shot/SFT & Accuracy, Temporal F1, Spatial F1& 3 & 4 \\
            \hline
            \shortstack{GeoReason-Bench\\  \cite{benchmark-GeoReason-Li}}  &  2026 &  OP& 4,000 & A+V & MCQ, CoT & \large \ding{55}  & 5 & Zero-shot& Accuracy & 3  &4 \\
            \hline
            \shortstack{ChatEarthBench\\  \cite{benchmark-ChatEarthBench}}  &2026  &  OP, MS, SAR&112,000+  & A+M+V & IC & \large \ding{51} & 4 & Zero-shot& Accuracy, BLEU, CIDEr, METEOR, ROUGE\_L, SPICE, MAE, Precision, Recall, F1-score &  4 & 7\\
            \hline
            \shortstack{SenseBench\\  \cite{benchmark-SenseBench-Zhong}}  &2026  &  OP, SAR, MT&10,815  & A+M+V & IC, MCQ & \large \ding{51} & 22 & Zero-shot& Accuracy, Sum Score, P0, P1, P2,  &  17 & 5\\
            \hline
        \end{tabular}%
    }
    
    \vspace{4pt}  
            
\end{table*}

With the development of RS models, comparing them fairly has become a pressing need for the field. The emergence of numerous evaluation benchmarks has provided a platform for such comparison, yet it exhibits a clear paradigm divergence. The overwhelming majority of benchmarks target the question-answering, reasoning, and generation capabilities of VLMs, whereas evaluations aimed at VFM tasks such as segmentation, detection, and few-shot learning remain comparatively scarce. Although efforts such as PANGAEA \cite{benchmark-PANGAEA-Valerio}, GEO-Bench \cite{benchmark-GeoBench-Lacoste}, PhilEO-Bench \cite{benchmark-PhilEO-Bench-Casper}, and Copernicus \cite{model-Copernicus-FM-Wang} have attempted to build VFM evaluation systems, the models they assess are largely early architectures such as Prithvi \cite{model-Prithvi-Johannes} and SatMAE \cite{model-SatMAE-Cong}, lagging noticeably behind the iteration pace of mainstream VLMs. This asymmetry in benchmark development reflects that the academic community's attention to interactive model understanding has outpaced its concern for pixel-level precision, even though the latter remains an indispensable requirement in real-world remote sensing applications. Accordingly, this paper focuses on existing VLM-oriented remote sensing benchmarks, as summarized in Table \ref{tab:RS-benchmark}.

Overall, the development of RS-specific VLM benchmarks has followed a trajectory from absence to presence, from presence to breadth, and from breadth to depth. Early GVLMs (e.g., GPT-4V) were not yet capable of interpreting non-optical modalities such as multi-spectral and SAR data, nor multi-temporal data \cite{benchmark-VLEO-Bench-Zhang}. Consequently, early benchmarks relied almost entirely on optical RGB imagery. As the technology matured, GVLMs evolved from merely possessing prior knowledge of non-optical RS modalities to incorporating such data into their pretraining scope, steadily improving their ability to process multi-modal RS data. The benchmarks proposed thereafter began to exhibit greater modal breadth: CHOICE \cite{benchmark-Choice-AN} explicitly incorporates SAR into its VQA question design, GEOBench-VLM \cite{benchmark-GEOBENCH-VLM-Danish} systematically evaluates multi-spectral and SAR inputs, and OmniEarth-Bench \cite{benchmark-OmniEarth-Bench-Wang} systematically covers multi-source data across the six earth spheres.

Alongside this expansion in breadth, the focus of benchmark construction has gradually shifted toward greater depth, that is, toward evaluating VLMs within a constrained range of functions or scenarios. In terms of data types, for example, XLRS-Bench \cite{benchmark-XLRS-Bench-Wang} focuses on assessing model capabilities on ultra-high-resolution RS imagery. UrBench \cite{benchmark-Urbench-Zhou} considers only the cross-view alignment problem for multi-view urban data. And RSVideo-Bench \cite{benchmark-RSVideo-Zhou} evaluates VLM performance across multiple tasks on RS videos. In terms of problem scenarios, RSMMVP \cite{benchmark-RSMMVP-Adejumo} targets the failure of models to distinguish RS images that are visually distinct yet semantically similar. LongEarth-R1 \cite{benchmark-LongEarth-R1-Ding} specializes in long-horizon spatiotemporal reasoning, and OVEarth-Bench \cite{benchmark-OVEarth-Bench-Li} evaluates only open-vocabulary localization. This shift from breadth to depth is, in essence, a refinement of evaluation granularity, extending assessment from the outcome level down to the process level.

Although current RS-specific benchmarks now possess both breadth and depth, several difficulties and limitations remain:

\textbf{Task heterogeneity and incomparability of results: } 
The best-performing model is not consistent across different benchmarks, and the relative ranking of models likewise varies. This occurs because differences in task design amplify a model's respective strengths and weaknesses. For instance, the cross-sphere interaction reasoning in OmniEarth-Bench \cite{benchmark-OmniEarth-Bench-Wang} demands broad geoscientific domain knowledge, whereas GeoReason-Bench \cite{benchmark-GeoReason-Li} focuses on whether a model's output reasoning chain is self-consistent. A model that performs poorly in geoscientific breadth does not necessarily lag on benchmarks that focus on reasoning quality. In addition, differences in the choice of core metrics are an important cause of inconsistent evaluation results.

\textbf{Low adoption leading to poor extensibility: }
Despite the rapid growth of benchmarks, newly published RS-VLM models rarely use them for comparison, instead favoring evaluation on representative classic datasets or constructing new benchmarks of their own. For example, Falcon \cite{model-falcon-Yao}, TerraScope \cite{model-TerraScope-Shu}, SkyMoE \cite{model-skymoe-Liu}, GeoPixel \cite{model-GeoPixel-Shabbir}, and SkySense-O \cite{model-SkySense-O-Zhu} all report results on private evaluation sets rather than on existing benchmarks. Such fragmented usage makes it difficult to extend older benchmarks to newly released models and weakens the standardizing role that benchmarks are meant to play.

\textbf{Insufficient focus on non-optical modalities: }
Although existing benchmarks incorporate a variety of non-optical modalities into their evaluation data, they do not construct questions around these modalities systematically, often employing them only sporadically as auxiliary inputs. In OmniEarth-Bench \cite{benchmark-OmniEarth-Bench-Wang} and CHOICE \cite{benchmark-Choice-AN}, for example, SAR is present but constitutes an extremely small proportion, and thus cannot comprehensively reflect model performance under this modality. Across existing RS-specific benchmarks, in terms of both breadth and depth, SAR has yet to be adequately foregrounded. It remains merely one modal option in the construction of multi-modal benchmarks, and even lacks a broad-coverage starting point.

These limitations are the core motivation for constructing a SAR-oriented evaluation tool in this paper. A rating benchmark with strong comparability, extensibility, and a dedicated focus on SAR is urgently needed to advance governance in this field.

\subsubsection{SAR-specific benchmarks}
As an important vertical domain of remote sensing, SAR has seen a gradual emergence of benchmarks dedicated to evaluating model performance on SAR tasks, following the broader development of remote sensing evaluation benchmarks. Unlike other RS benchmarks, SAR benchmarks contain only SAR imagery and focus exclusively on typical SAR tasks and applications. Building a SAR benchmark relies on annotating large volumes of SAR images. Because such annotation is difficult and requires expert verification and screening, relatively few SAR benchmarks have been publicly released. Existing SAR benchmarks and their related information are summarized in Table \ref{tab:SAR-benchmark}.

Among them, SARChat-Bench-2M \cite{benchmark-SARChat-Bench-2M-Ma} and SARLANG-1M \cite{benchmark-SARLANG-1M-wei} each derive a complete VQA dataset by annotating public datasets, supporting both training and testing under a LoRA fine-tuning evaluation setting. SAREval \cite{benchmark-SAREval-Wang}, by contrast, provides only a test benchmark. SARChat-Bench-2M argues that existing GVLMs lack basic SAR comprehension without fine-tuning, so it primarily evaluates multiple GVLMs after fine-tuning and includes only a single model to assess zero-shot capability for comparison. SAREval evaluates only the zero-shot capability of GVLMs on SAR data, while SARLANG-1M considers both LoRA fine-tuning and zero-shot settings.

These benchmarks provide an evaluation platform and fair comparison for the development of SAR-related models, and offer a basis for model selection in engineering deployment. However, several limitations remain:

\textbf{Incomplete coverage of evaluated models: } 
Existing benchmarks focus on popular GVLMs and do not evaluate VLMs specialized for remote sensing or SAR. Because GVLMs are predominantly trained on natural optical images, the objects in these images generally follow gravity-constrained spatial relationships, such as a top-to-bottom orientation, and lack rotation invariance \cite{benchmark-OVRSISBenchV2-Li}. Remote sensing and SAR VLMs are pre-trained on large volumes of remote sensing or SAR images and should have improved perception of the overhead viewpoint. Such models may therefore hold greater potential than GVLMs for zero-shot SAR tasks and merit evaluation. In addition, model selection has not considered CLIP-based models or VFMs. These model types, particularly VFMs, have considerable value for deployment and practical applications, and their evaluation should not be overlooked.

\textbf{Incomplete task coverage and ambiguous task definitions: } 
Existing benchmarks generally lack change-detection tasks, despite the fact that related applications, including flood detection \cite{dataset-DAM-Net-Saleh}, agriculture survey \cite{CD-Description-Jose}, urban planning \cite{CD-Multitemporal-Spaceborne-Ban}, and military monitoring \cite{CD-Patch-Similarity-Sun}, are important applications of SAR. Moreover, task definitions are not standardized across benchmarks. The definitions of specific task dimensions differ substantially, causing cross-benchmark comparisons of the same task to carry different meanings.

\begin{figure}[htbp]
\centering
\includegraphics[width=1.0\columnwidth]{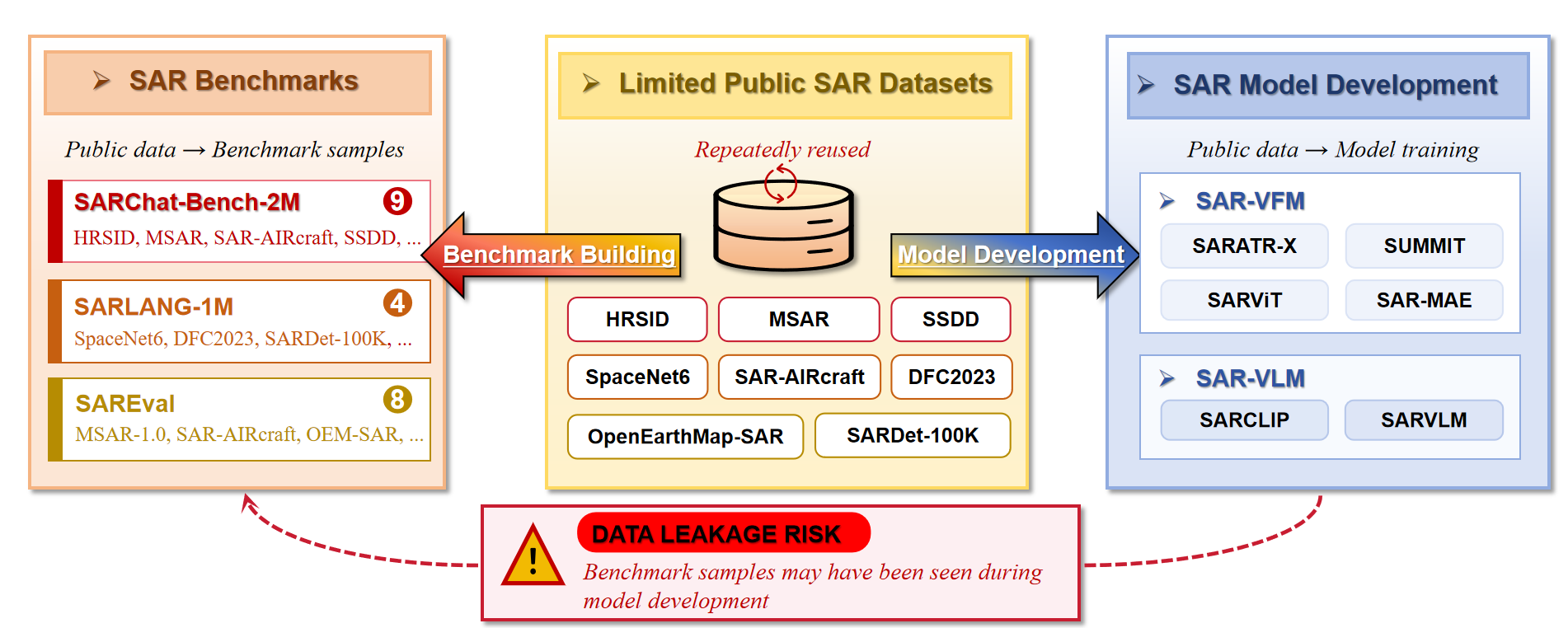}
\caption{Data leakage risk example. The construction of SAR benchmarks and model pre-training involve substantial dataset overlap, posing data leakage risks when using these benchmarks to evaluate SAR-related models.}
\label{fig3}
\end{figure}

\textbf{Risk of data leakage: }
The risk of data leakage has a relatively limited impact on GVLM and RS-VLM benchmarks but is particularly serious in the SAR domain. Since only a small number of public datasets are available in the community, existing benchmarks are often constructed based on high-impact datasets. Consequently, when these benchmarks are used to evaluate SAR-related large models, overlap between the datasets used for model pretraining and the benchmark datasets is inevitable, as illustrated in Fig. \ref{fig3}.

\textbf{Excessive focus on task performance and neglect of other evaluation dimensions: }
Existing benchmarks consider performance alone and lack evaluations of factors such as the difficulty of cross-domain adaptation, openness, and reproducibility. Since the SAR community has a relatively low level of openness, a model that performs well on a benchmark is not necessarily open-source or easy to deploy.

In summary, constructing a comprehensive evaluation benchmark with broad model coverage, representative tasks, and complete evaluation dimensions remains an urgent need for the development of SAR-related large models.

\begin{table*}[htbp]
    \centering
    \caption{Overview of SAR-specific Benchmarks (Abbreviations adopted: CD=Change Detection, IC=Image Captioning, VG=Visual Grounding, FGD=Fine-Grained Description, CMR=Cross-Modal Reasoning, MCQ=Multiple Choice Questions, Q=Questions, VQA=Visual Question Answering, IoU=Intersection over Union, LoRA=Low-Rank Adaptation).}
    \label{tab:SAR-benchmark}
    
    \resizebox{0.99\linewidth}{!}{%
        \begin{tabular}{|>{\centering\arraybackslash}m{2.55cm}|>{\centering\arraybackslash}m{0.6cm}|>{\centering\arraybackslash}m{1.8cm}|>{\centering\arraybackslash}m{2.8cm}|>{\centering\arraybackslash}m{1.3cm}|c|c|c|c|>{\centering\arraybackslash}m{0.8cm}|>  
        {\centering\arraybackslash}m{3.0cm}|>{\centering\arraybackslash}m{2.4cm}|>{\centering\arraybackslash}m{1.8cm}|>{\centering\arraybackslash}m{1.7cm}|>{\centering\arraybackslash}m{0.7cm}|}
            \hline
            \multicolumn{2}{|c|}{\textbf{Basic information}} & 
            \multicolumn{13}{c|}{\textbf{Dataset Details}} \\
            \hline
            \multirow{2}{*}{\shortstack{\\ \\ \\ \\ \\Name}}  & \multirow{2}{*}{\shortstack{\\ \\ \\ \\ \\Year}} & \multirow{2}{*}{\shortstack{\\ \\ \\ \\ \\Data Modality}} & \multirow{2}{*}{\shortstack{\\ \\ \\ \\ \\Data Source}} &  \multirow{2}{*}{\shortstack{\\ \\ \\ \\ \\Total Data}} & \multicolumn{2}{c|}{Split Details} & \multicolumn{4}{c|}{SAR Data Specifics}  &  \multicolumn{4}{c|}{Image-Text Data Quality} \\
            \cline{6-6}
            \cline{7-15}
            &  &  &  &  & Train & Test & Format & Resolution & Band  & Platform & Text Source & Image-Text Pairing Method & SAR-Specific Property Description & Open Tasks \\
            \hline
            \shortstack{SARChat-Bench-2M\\  \cite{benchmark-SARChat-Bench-2M-Ma}}  & 2025 & SAR image + text & AIR-SARShip; HRSID; MSAR; SADD; SAR-AIRcraft; ShipDataset; SSDD; OGSOD; SIVED & 2.06 M & 89\% & 11\% & L2 &0.1-3.0 m & \parbox{0.8cm}{\centering C/X/\\Ku/Ka} &  GF-3; Sentinel-1; TerraSAR-X; HISEA-1; Sentinel-1B; RadarSat-2; TanDEMX; Airborne-SAR-Synthetic-Slice; & Detection dataset annotation + fixed text template  & \multicolumn{1}{>{\raggedright\arraybackslash}m{1.8cm}|}{1. VG + Single image;\newline 2. IC + Single image} & \large \ding{55} & \large \ding{55}  \\
            \hline
            \shortstack{SARLANG-1M\\  \cite{benchmark-SARLANG-1M-wei}}  & 2026 & SAR image + text & SpaceNet6; DFC2023; OpenEarth-Map-SAR; SARDet-100k & 1.13 M  & 87\% & 13\% & L2 &0.1-25.0 m & \parbox{0.8cm}{\centering C/X/\\Ku/Ka} &  Capella Space’s Satellite; SuperView-1; Gaofen-2; Gaofen-3; Umbra-Satellites; Sentinel-1; TerraSAR-X; RadarSat-2; TanDEMX; HISEA-1; Airborne-SAR-Synthetic-Slice; &\shortstack{1. GPT-4o cross-\\modal annotation +\\ manual filtering;\\2. Bbox-driven text \\generation +\\ manual verification}     & \multicolumn{1}{>{\raggedright\arraybackslash}m{1.8cm}|}{IC + Single image} & \large \ding{55} & \large \ding{51} \\
            \hline
            \shortstack{SAREval\\  \cite{benchmark-SAREval-Wang}}  & 2025 & SAR image + text + mask & MSAR-1.0; SAR-AIRcraft; OpenEarth-Map-SAR; Deep-SAR-Oil-Spill; FAIR-CSAR; BHdataset; OpenSARWake; ATRNet-STAR & 14,950 & 0\% & 100\% & L2 &0.1-12.5 m & \parbox{0.8cm}{\centering C/X/\\Ku/Ka\\/L} &  Haisi-1; Gaofen-3; Sentinel-1B; Airborne SAR; ALOS-PALSAR; Sentinel-1A; TerraSAR-X; & LLM + manual verification  & \multicolumn{1}{>{\raggedright\arraybackslash}m{1.8cm}|}{1. MCQ + Single image;\newline 2. Mask + Q + image; \newline 3. VG + Single image; \newline4. IC + Single image} & \large \ding{51} & \large \ding{51}  \\
            \hline
        \end{tabular}%
    }
    
    \vspace{4pt}  
    
    \resizebox{0.99\linewidth}{!}{%
        \begin{tabular}{|>{\centering\arraybackslash}m{2.55cm}|>{\centering\arraybackslash}m{0.6cm}|>{\centering\arraybackslash}m{1.4cm}|>{\centering\arraybackslash}m{3.0cm}|>{\centering\arraybackslash}m{1.5cm}|>{\centering\arraybackslash}m{1.3cm}|>{\centering\arraybackslash}m{0.8cm}|>{\centering\arraybackslash}m{0.8cm}|>{\centering\arraybackslash}m{0.8cm}|>{\centering\arraybackslash}m{0.8cm}|>{\centering\arraybackslash}m{0.8cm}|>  
        {\centering\arraybackslash}m{0.8cm}|>{\centering\arraybackslash}m{0.8cm}|>{\centering\arraybackslash}m{0.8cm}|>{\centering\arraybackslash}m{0.8cm}|>{\centering\arraybackslash}m{0.8cm}|>{\centering\arraybackslash}m{0.8cm}|>{\centering\arraybackslash}m{0.8cm}|>{\centering\arraybackslash}m{1.6cm}|}
            \hline
            \multicolumn{2}{|c|}{\textbf{Basic information}} & \multicolumn{3}{c|}{\textbf{Evaluated Model Selection}}&
            \multicolumn{14}{c|}{\textbf{Typical SAR Task Coverage}} \\
            \hline
            \multirow{2}{*}{\shortstack{\\ \\ \\Name}}  & \multirow{2}{*}{\shortstack{\\ \\ \\Year}} & \multirow{2}{*}{\shortstack{\\ \\Traditional\\ Models}} & \multirow{2}{*}{\shortstack{\\ \\ \\GVLMs}} &\multirow{2}{*}{\shortstack{\\ \\ SAR-related \\FMs}}  &\multirow{2}{*}{\shortstack{\\ \\ Dimension\\ Count}} & \multicolumn{2}{c|}{Classification} & \multicolumn{2}{c|}{Segmentation}  & \multicolumn{3}{c|}{Detection}&\multirow{2}{*}{\shortstack{\\ \\ \\CD}}  &\multirow{2}{*}{\shortstack{\\ \\ \\IC}}&\multirow{2}{*}{\shortstack{\\ \\ \\VG}}&\multirow{2}{*}{\shortstack{\\ \\ \\FGD}}& \multirow{2}{*}{\shortstack{\\ \\ \\CMR}}& \multirow{2}{*}{\shortstack{\\ \\ \\ Other Tasks}}\\
            \cline{7-13}
            &  &  &  &  &  & Target & Scene & Land Cover & Ship & Ship  & Aircraft & Multi-class &  &  &  &  & & \\
            \hline
            \shortstack{SARChat-Bench-2M\\  \cite{benchmark-SARChat-Bench-2M-Ma}}  & 2025 & None & Qwen2-VL; InternVL2.5; DeepSeekVL; Phi-3.5-vision; GLM-Edge-V; mPLUG-Owl3; YiVL; LLaVA-1.5; & None & 6 & \large \ding{51} & \large \ding{55} &\large \ding{55} & \large \ding{55} &  \large \ding{51} & \large \ding{51} & \large \ding{51} & \large \ding{55} & \large \ding{51} & \large \ding{51} & \large \ding{51} &\large \ding{51} &Counting\\
            \hline
            \shortstack{SARLANG-1M\\  \cite{benchmark-SARLANG-1M-wei}}  & 2026 & MLAT; HCNet; & BLIP; LLaVA1.5-7B; LLaVA1.5-13B; QWEN2-VL-7B; QWEN2.5-VL-7B; & None & 10 & \large \ding{51} & \large \ding{51} & \large \ding{55} &\large \ding{55} & \large \ding{55} &  \large \ding{55} & \large \ding{55} & \large \ding{55} & \large \ding{51} & \large \ding{51} & \large \ding{51} & \large \ding{51} &Counting; Shape Description; Road Orientation Description\\
            \hline
            \shortstack{SAREval\\  \cite{benchmark-SAREval-Wang}}  & 2025 & None & Phi-3.5-vision; Qwen2.5-VL-7B; Deepseek-VL2-tiny; Mono-InternVL-2B; InternVL2-4B; InternVL2.5-4B; InternVideo2.5-8B; InternVL3-9B; LLaVA-1.5-7B; LLaVA-Next; LLaVA-onevision-qwen2-7b; & None & 20 & \large \ding{51} & \large \ding{51} & \large \ding{51} &\large \ding{55} & \large \ding{55} &  \large \ding{55} & \large \ding{55} & \large \ding{55} & \large \ding{51} & \large \ding{51} & \large \ding{51} & \large \ding{51} &Counting; Ship Heading; Tank Volume; Imaging Parameter Inversion; Building Height Estimation\\
            \hline
        \end{tabular}%
    }

    \vspace{4pt}  
    
    \resizebox{0.99\linewidth}{!}{%
        \begin{tabular}{|>{\centering\arraybackslash}m{2.55cm}|>{\centering\arraybackslash}m{0.6cm}|>{\centering\arraybackslash}m{1.4cm}|>{\centering\arraybackslash}m{1.6cm}|>{\centering\arraybackslash}m{0.9cm}|>{\centering\arraybackslash}m{0.9cm}|>{\centering\arraybackslash}m{0.9cm}|>{\centering\arraybackslash}m{3.2cm}|>{\centering\arraybackslash}m{8.2cm}|>{\centering\arraybackslash}m{5.8cm}|}
            \hline
            \multicolumn{2}{|c|}{\textbf{Basic information}} & \multicolumn{2}{c}{\textbf{Evaluation Protocol}}&
            \multicolumn{6}{|c|}{\textbf{Results and Analysis}} \\
            \hline
            \multirow{2}{*}{\shortstack{Name}}  & \multirow{2}{*}{\shortstack{Year}} & \multirow{2}{*}{\shortstack{Model \\ Adaptation}} & \multirow{2}{*}{\shortstack{Core Metrics}} &\multicolumn{3}{c|}{Reproducibility}   &  \multirow{2}{*}{\shortstack{Main Contribution}}& \multirow{2}{*}{\shortstack{Main Conclusion}}& \multirow{2}{*}{\shortstack{Potential Extensions}} \\
            \cline{5-7}
            &  &  &  & Code & Weights & Data &  &  &  \\
            \hline
            \shortstack{SARChat-Bench-2M\\  \cite{benchmark-SARChat-Bench-2M-Ma}}  & 2025 & LoRA & Accuracy, IoU & \large \ding{51} & \large \ding{51} & \large \ding{55} & \multicolumn{1}{>{\raggedright\arraybackslash}m{3.2cm}|}{Large-scale standardized image-text dataset.} & \multicolumn{1}{>{\raggedright\arraybackslash}m{8.2cm}|}{1. Poor performance in both single-target and multi-target text-based grounding tasks highlights the limitations in establishing precise text–object correspondences within the SAR domain.\newline 2. High recognition accuracy demonstrates the strong capability of VLMs in cross-modal recognition tasks.\newline  3. In fine-grained description tasks, models with larger parameter scales perform better, indicating sensitivity to model size. \newline 4. Without SAR-domain fine-tuning, general-purpose models exhibit near-zero performance on SAR tasks, except for counting. This indicates that general foundation models are unable to comprehend SAR targets.} 
            & \multicolumn{1}{>{\raggedright\arraybackslash}m{5.8cm}|}{1. A dedicated zero-shot evaluation protocol could be introduced to assess generalization in SAR imagery.\newline 2. The task coverage could be broadened to include scene classification and other commonly studied SAR tasks.\newline 3. Incorporating multi-temporal tasks, such as change detection.}

            \\
            \hline
            \shortstack{SARLANG-1M\\  \cite{benchmark-SARLANG-1M-wei}}  & 2026 & Zero-shot/LoRA & Accuracy, BLEU, ROUGE\_L, CIDEr & \large \ding{51} & \large \ding{55} & \large \ding{55} &\multicolumn{1}{>{\raggedright\arraybackslash}m{3.2cm}|}{Large-scale SAR image-text dataset for GVLM fine-tuning.} &\multicolumn{1}{>{\raggedright\arraybackslash}m{8.2cm}|}{1. Fine-tuning on SARLANG-1M significantly boosts static SAR interpretation, rivaling human experts. \newline 2. Proper SAR image preprocessing improves zero-shot GVLMs performance.} &\multicolumn{1}{>{\raggedright\arraybackslash}m{5.8cm}|}{1. VQA and captioning could include coordinate-level grounding. \newline 2. Task-wise results could complement overall VQA accuracy.\newline 3. Class balance could be further improved to ensure that no single category (e.g., ships) is disproportionately represented.} 
            \\
            \hline
            \shortstack{SAREval\\  \cite{benchmark-SAREval-Wang}}  & 2025 & Zero-shot & Acc$@$0.5, Acc$@$0.25, BLEU, ROUGE\_L, BERTScore, $\triangle$Accuracy & \large \ding{51} & \large \ding{55} & \large \ding{51} & \multicolumn{1}{>{\raggedright\arraybackslash}m{3.2cm}|}{1. Added SAR-specific tasks: polarization inference, speckle robustness, geometric distortion correction; \newline 2. Novel distractor generation strategy for more reasonable evaluation.} & \multicolumn{1}{>{\raggedright\arraybackslash}m{8.2cm}|}{Current VLMs exhibit inherent optical prior bias due to modality imbalance in pre-training, demonstrating inadequate adaptation to geometric distortions and scattering characteristics in SAR imagery, thereby limiting their effectiveness in fine-grained object recognition and complex reasoning tasks.} & \multicolumn{1}{>{\raggedright\arraybackslash}m{5.8cm}|}{1. The definition of ``detection'' could be further clarified, as it refers to chip-level recognition in this work, differing from the coordinate-level localization used in conventional detection. \newline 2. Given the distinct requirements of basic SAR understanding and advanced quantitative inference, complex tasks such as depression angle, building height, oil tank storage volume, and resolution estimation could be evaluated separately in zero-shot settings.}  
            
            \\
            \hline
        \end{tabular}%
    }
\end{table*}

\section{The Model Rating Tool Framework}
\begin{figure*}[htbp]
\centering
\includegraphics[width=1.0\textwidth]{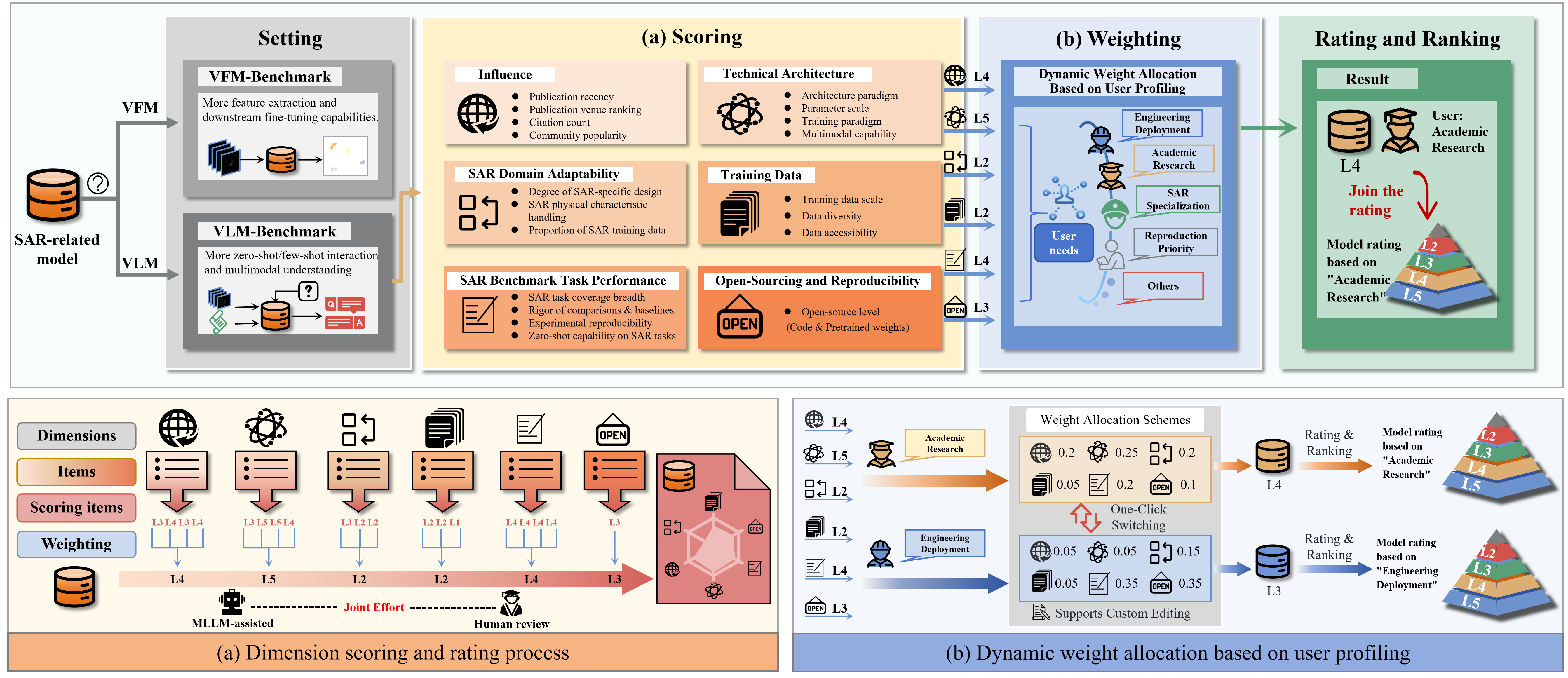}
\caption{Framework of the SAR-related model rating tool.}
\label{fig4}
\end{figure*}

In this section, we establish a standardized rating tool for SAR-related models, addressing the limitation of existing task-only SAR benchmarks, which fail to reflect a model's capability levels, domain adaptability, and openness and reproducibility. The proposed tool evaluates models across six dimensions: influence, technological and architectural advancement, adaptability to the SAR domain, training data, SAR task performance, and openness and reproducibility. Each dimension is assigned a hierarchical score, and the scores are aggregated using weighted evaluation. To accommodate the personalized needs of different user types, we further propose a dynamic weight allocation scheme, enabling model rating and ranking under varying requirement priorities. The framework of the evaluation toolkit is shown in Fig. \ref{fig4}.

\subsection{Dual-Track Evaluation Benchmark}
Within the proposed rating framework, both SAR-related VLMs and VFMs are included in the evaluation. However, the two model categories differ substantially in their functional roles, training and inference paradigms, and deployment scenarios. Therefore, applying an identical set of rating criteria to both would be inappropriate. For example, when evaluating performance on SAR benchmark tasks, VFMs generally do not regard language-guided zero-shot capabilities as a primary evaluation criterion, whereas such capabilities are central to assessing the generalization ability and practical value of VLMs. More specifically, VFMs emphasize effective visual feature extraction and transfer, and their performance is typically evaluated through few-shot learning, linear probing, or fine-tuning on downstream tasks. By contrast, VLMs place greater emphasis on vision-language alignment and semantic understanding, as well as cross-modal interaction, instruction following, and task generalization under zero-shot or few-shot settings.

Despite these differences in capability emphasis, the high-level dimensions of interest to researchers and practitioners are largely consistent across the two model categories. These dimensions include model influence, technological architecture advancement, SAR domain adaptability, training data, SAR benchmark task performance, and open-sourcing and reproducibility. Accordingly, the proposed framework adopts a unified set of primary evaluation dimensions for VFMs and VLMs, thereby ensuring framework consistency and broad comparability across model categories. Meanwhile, differentiated secondary indicators, scoring items, and evaluation requirements are designed under each dimension to account for the distinct technical characteristics and application objectives of the two model categories. The proposed rating criteria for VLMs are presented in Table \ref{tab:VLM-bench}. 

For VFMs, the rating criteria follow the same dimensions as those used for VLMs in terms of influence, SAR domain adaptability, SAR benchmark task performance, and open-sourcing and reproducibility. However, differences arise in the dimensions of technological architecture advancement and training data, as well as in the specific treatment of SAR benchmark task performance, where the evaluation emphasis and indicator definitions are adapted to the characteristics of foundation models. The rating criteria for VFMs in these three dimensions are presented in Table \ref{tab:VFM-bench}.

\begin{table*}[htbp]
\centering
\caption{Rating Criteria for SAR-related VLMs (Abbreviations adopted: SCI=Science Citation Index, JCR=Journal Citation Reports, RLHF=Reinforcement Learning from Human Feedback, CLIP=Contrastive Language-Image Pre-Training, SAM=Segment Anything Model, SOTA=State‑of‑the‑Art, CNN=Convolutional Neural Network) (Part 1 of 2).}
\label{tab:VLM-bench}
\footnotesize
\renewcommand{\arraystretch}{1.3}

\begin{tabular}{p{1.9cm}  p{3.4cm} c p{7.5cm}}
\toprule
\textbf{Dimensions} & \textbf{Items} &\textbf{Weight} & \textbf{Criteria} \\
\midrule
\multirow{16}{=}{Influence}
 & Publication recency &0.25 & \multicolumn{1}{>{\raggedright\arraybackslash}m{11.5cm}}{\textbf{Time Since Publication} \newline
L5: $\leq 1$ year \newline
L4: $>1$ and $\leq 2$ years \newline
L3: $>2$ and $\leq 3$ years \newline
L2: $>3$ and $\leq 5$ years \newline
L1: $>5$ years} \\

\cline{2-4}
 & Publication venue ranking & 0.25& \multicolumn{1}{>{\raggedright\arraybackslash}m{11.5cm}}{\textbf{Publication venue influence} \newline
L5: Top conferences (e.g., CVPR, ICCV, ECCV, NeurIPS, ICLR, ICML); Nature/Science sub-journals; top journals in the field (e.g., TPAMI, GRSM). \newline
L4: JCR Q1 SCI journals / leading conferences (e.g., TGRS, TIP, IJCV, AAAI, ISPRS) \newline
L3: JCR Q2 SCI journals; general international and regional conferences\newline
L2: JCR Q3 / Q4 SCI journals; general national conferences \newline
L1: arXiv preprint only / not formally published; other journals or conferences} \\
\cline{2-4}

  & Citation count & 0.25& \multicolumn{1}{>{\raggedright\arraybackslash}m{11.5cm}}{\textbf{Total citations (Google Scholar)} \newline
L5: $>$1000 \newline
L4: 300--1000 \newline
L3: 100--300 \newline
L2: 20--100 \newline
L1: $<$20} \\
\cline{2-4}

  & Community popularity & 0.25& \multicolumn{1}{>{\raggedright\arraybackslash}m{11.5cm}}{\textbf{Community popularity (GitHub Stars)} \newline
L5: $>$1000 \newline
L4: 300--1000 \newline
L3: 100--300 \newline
L2: 50--100 \newline
L1: $<$50} \\
\midrule

\multirow{16}{=}{Technical Architecture}
  & Architecture paradigm & 0.25 & \multicolumn{1}{>{\raggedright\arraybackslash}m{11.5cm}}{\textbf{Architectural Innovation} \newline
L5: Original and novel architecture (e.g., a first-of-its-kind multimodal fusion architecture or SAR-specific encoder) \newline
L4: Substantially redesigns an established architecture, including major changes to core modules or information flow (e.g., extensive modification of ViT-, SAM-, or CLIP-based architectures) \newline
L3: Domain adaptation of a mature architecture (e.g., fine-tuning or adapter integration) \newline
L2: Applies minor architectural modifications or a standard assembly of existing components \newline
L1: Directly adopts an existing architecture without meaningful structural innovation} \\
\cline{2-4}

  & Parameter scale & 0.25 & \multicolumn{1}{>{\raggedright\arraybackslash}m{11.5cm}}{\textbf{Parameter Scale} \newline
L5: $>$10 B \newline
L4: 1--10 B \newline
L3: 300 M--1 B \newline
L2: 50--300 M \newline
L1: $<$50 M} \\
\cline{2-4}

  & Training paradigm & 0.25 & \multicolumn{1}{>{\raggedright\arraybackslash}m{11.5cm}}{\textbf{Training Paradigm (Complexity and Innovation)} \newline
L5: Uses a comprehensive multi-stage pipeline, including large-scale pretraining, multimodal / instruction tuning, and preference or reinforcement alignment; constituting a complete modern training pipeline \newline
L4: Combines domain-specific self-supervised, contrastive, or multimodal pretraining with multi-task or instruction tuning \newline
L3: Adapts pretrained models through full or parameter-efficient fine-tuning for domain-specific tasks \newline
L2: Uses single-stage supervised fine-tuning or conventional task-specific training \newline
L1: Uses no dedicated model training or only minimal task adaptation} \\
\cline{2-4}

  & Multimodal capability & 0.25 & \multicolumn{1}{>{\raggedright\arraybackslash}m{11.5cm}}{\textbf{Degree of Supported Multimodal Interaction} \newline
L5: Supports multi-turn dialogue and unified multitask capabilities across multiple fine-grained levels (e.g., image-, region-, or pixel-level interaction) \newline
L4: Image–text alignment + text generation (e.g., captioning / VQA), or dialogue-based image understanding without fine-grained interaction \newline
L3: Image–text retrieval or matching or classification-level alignment, but not substantive free-form text generation \newline
L2: Weak text guidance only, without substantial image–text alignment or language generation \newline
L1: Purely visual model with no language modality} \\

\midrule
\multirow{1}{=}{Open-Sourcing and Reproducibility}
  & Open-source level & 1.0 & \multicolumn{1}{>{\raggedright\arraybackslash}m{11.5cm}}{\textbf{Extent of Open-Source Availability} \newline
L5: Source code, pretrained weights, and training scripts are all publicly available \newline
L4: Pretrained weights only \newline
L3: Pretrained weights available upon request \newline
L2: Source code only \newline
L1: Closed source} \\

\bottomrule
\end{tabular}

\end{table*}
\begin{table*}[htbp]
\centering
\setcounter{table}{\value{table}}     
\addtocounter{table}{-1}              
\caption{Rating Criteria for SAR-related VLMs (Part 2 of 2).}
\label{tab:VLM-bench2}
\footnotesize
\renewcommand{\arraystretch}{1.3}

\begin{tabular}{p{1.9cm}  p{3.4cm} c p{7.5cm}}
\toprule
\textbf{Dimensions} & \textbf{Items} &\textbf{Weight} & \textbf{Criteria} \\
\midrule
\multirow{12}{=}{SAR Domain Adaptability}
 & Degree of SAR-specific design &0.4 & \multicolumn{1}{>{\raggedright\arraybackslash}m{11.5cm}}{\textbf{Degree of SAR-Oriented Design} \newline
L5: SAR-exclusive design with SAR-specific components or objectives (e.g., tokenizer, encoder, loss function) \newline
L4: Joint optical--SAR modeling with dedicated SAR-specific modules \newline
L3: General-purpose RS model trained with a certain proportion of SAR data, or a model trained primarily on SAR data but without SAR-specific architectural improvements \newline
L2: General-purpose model fine-tuned for SAR tasks \newline
L1: Purely optical or natural-image model directly applied to SAR in a zero-shot manner} \\

\cline{2-4}
 & SAR physical characteristic handling & 0.3& \multicolumn{1}{>{\raggedright\arraybackslash}m{11.5cm}}{\textbf{Consideration of SAR-Specific Characteristics (e.g., speckle, geometric distortions, complex-valued data/phase information, polarization, and incidence angle)} \newline
L5: More than three types of SAR-specific characteristics are considered \newline
L4: Two types of SAR-specific characteristics are considered \newline
L3: One type of SAR-specific characteristic is considered \newline
L2: SAR images are treated merely as grayscale images \newline
L1: SAR-specific characteristics are completely ignored} \\
\cline{2-4}

  & Proportion of SAR training data & 0.3& \multicolumn{1}{>{\raggedright\arraybackslash}m{11.5cm}}{\textbf{Proportion and Scale of SAR Training Samples} \newline
L5: SAR samples constitute the majority of the training data or $>$ 1 M; L4: 100 K--1 M SAR samples; L3: 10 K--100 K SAR samples; L2: SAR samples account for only a small proportion of the training data or are at the thousand scale or below; L1: No SAR data are used for training} \\

\midrule

\multirow{12}{=}{Training Data}
  & Training data scale & 0.4 & \multicolumn{1}{>{\raggedright\arraybackslash}m{11.5cm}}{\textbf{Total Training Data Volume (Image--Text Pairs)} \newline
L5: $>$10 M; L4: 1 M--10 M; L3: 100 K--1 M; L2: 10 K--100 K; L1: $<$10 K} \\
\cline{2-4}

  & Data diversity & 0.3 & \multicolumn{1}{>{\raggedright\arraybackslash}m{11.5cm}}{\textbf{Training Data Diversity (Sensors, Frequency Bands, Spatial Resolutions, Scenes, and Geographic Coverage)} \newline
L5: Multiple sensors + multiple frequency bands + multiple spatial resolutions + diverse scenes with global geographic coverage \newline
L4: Three diversity dimensions are covered \newline
L3: Two diversity dimensions are covered \newline
L2: Multiple scenes acquired by a single sensor \newline
L1: A single data source covering a single scene} \\
\cline{2-4}

  & Data accessibility & 0.3 & \multicolumn{1}{>{\raggedright\arraybackslash}m{11.5cm}}{\textbf{Public Availability of Training Data} \newline
L5: All training data are publicly available \newline
L4: Most training data are publicly available \newline
L3: Some training data are publicly available, while the remainder are documented only through descriptions \newline
L2: Training data are described but not publicly available \newline
L1: Training data are neither described nor publicly available} \\

\midrule
\multirow{16}{=}{SAR Benchmark Task Performance}
  & SAR task coverage breadth & 0.2 & \multicolumn{1}{>{\raggedright\arraybackslash}m{11.5cm}}{\textbf{Number of Supported SAR Task Types Reported in the Paper} \newline
L5: More than five task types; L4: Three to four task types; L3: Two task types; L2: A single task type with outstanding performance; L1: A single task type, or none at all} \\
\cline{2-4}

  & Rigor of comparisons \& baselines & 0.3 & \multicolumn{1}{>{\raggedright\arraybackslash}m{11.5cm}}{\textbf{Rigor and Representativeness of Comparative Baselines Reported in the Paper} \newline
L5: All tasks are evaluated against recent and competitive baselines, with comprehensive evaluation dimensions and sufficiently detailed dataset descriptions \newline
L4: The main experiments include strong baselines, ablation studies, discussions of alternative explanations, and evaluations on multiple datasets \newline
L3: The comparisons are relatively comprehensive, cover multiple types of methods, and include evaluations on publicly available datasets \newline
L2: Mainstream baselines are included, and the comparisons are generally reasonable \newline
L1: Very few baselines are included, or the comparisons are evidently selective} \\
\cline{2-4}

  & Experimental reproducibility & 0.2 & \multicolumn{1}{>{\raggedright\arraybackslash}m{11.5cm}}{\textbf{Reproducibility of Experiments Reported in the Paper} \newline
L5: Highly reproducible, with complete training and inference pipelines publicly available and detailed reproduction instructions provided \newline
L4: Code, model weights, and experimental protocols are comprehensively organized in open-source repositories, enabling the reproduction of all experiments \newline
L3: Code, model weights, and experimental protocols are relatively complete, but some resources must be collected independently based on the paper's descriptions (e.g., datasets or materials for ablation studies) \newline
L2: Only partially open, with certain resources unavailable, such as weights for ablation models or portions of the test data \newline
L1: No code or experimental protocols are available, and weights for ablation models are not released} \\

\cline{2-4}

  & Zero-shot capability on SAR tasks & 0.3 & \multicolumn{1}{>{\raggedright\arraybackslash}m{11.5cm}}{\textbf{Reported Zero-Shot Capability on SAR Data} \newline
L5: Zero-shot capabilities are explicitly reported and evaluated across multiple SAR tasks, achieving state-of-the-art performance in certain capability dimensions \newline
L4: Demonstrates zero-shot capabilities on several tasks and can process and interpret SAR imagery based on SAR-specific characteristics \newline
L3: Can correctly recognize SAR imagery as a distinct modality but lacks interpretation capabilities or awareness of SAR-specific characteristics \newline
L2: Possesses general SAR-related knowledge but cannot recognize or interpret SAR imagery \newline
L1: Possesses no prior SAR-related knowledge} \\

\bottomrule
\end{tabular}

\end{table*}
\begin{table*}[htbp]
\centering
\caption{Rating Criteria for SAR-related VFMs in Three Dimensions Distinct from VLMs.}
\label{tab:VFM-bench}
\footnotesize
\renewcommand{\arraystretch}{1.3}

\begin{tabular}{p{1.9cm}  p{3.4cm} c p{7.5cm}}
\toprule
\textbf{Dimensions} & \textbf{Items} &\textbf{Weight} & \textbf{Criteria} \\
\midrule
\multirow{8}{=}{Technical Architecture}
  & Architecture paradigm & 0.4 & \multicolumn{1}{>{\raggedright\arraybackslash}m{11.5cm}}{\textbf{Architectural innovation} \newline
L5: Original novel architecture or new pre-training paradigm (e.g., pioneering visual self-supervised framework, SAR / RS-specific encoder, entirely new backbone or representation-learning mechanism) \newline
L4: Significant improvement over mainstream architectures / pre-training paradigms (deep modification of ViT / SAM / DINO / MAE families, e.g., attention redesign, multi-scale backbone, substantive masked-modeling innovation) \newline
L3: Domain adaptation of mature architectures (fine-tuning, adapter / LoRA, lightweight prompt tuning, or input-layer adaptation for modalities such as SAR / multispectral) \newline
L2: Applies minor architectural modifications or a standard assembly of existing components\newline
L1: Directly adopts an existing architecture without meaningful structural innovation} \\
\cline{2-4}

  & Parameter scale & 0.3 & \multicolumn{1}{>{\raggedright\arraybackslash}m{11.5cm}}{\textbf{Parameter Scale} \newline
L5: $>$10 B; L4: 1--10 B; L3: 300 M--1 B; L2: 50--300 M; L1: $<$50 M} \\
\cline{2-4}

  & Training paradigm & 0.3 & \multicolumn{1}{>{\raggedright\arraybackslash}m{11.5cm}}{\textbf{Training Paradigm (Complexity and Innovation)} \newline
L5: Ultra-large-scale self-supervised pre-training + complete modern pipeline with multi-task / multi-stage adaptation \newline
L4: Self-supervised pre-training + downstream fine-tuning \newline
L3: Supervised pretraining + transfer learning \newline
L2: Purely supervised training \newline
L1: Training from scratch for a single task} \\

\midrule

\multirow{12}{=}{Training Data}
  & Training data scale & 0.4 & \multicolumn{1}{>{\raggedright\arraybackslash}m{11.5cm}}{\textbf{Total Training Data Volume (Images)} \newline
L5: $>$10 M; L4: 1 M--10 M; L3: 100 K--1 M; L2: 10 K--100 K; L1: $<$10 K} \\
\cline{2-4}

  & Data diversity & 0.3 & \multicolumn{1}{>{\raggedright\arraybackslash}m{11.5cm}}{\textbf{Training Data Diversity (Sensors, Frequency Bands, Spatial Resolutions, Scenes, and Geographic Coverage)} \newline
L5: Multiple sensors + multiple frequency bands + multiple spatial resolutions + diverse scenes with global geographic coverage \newline
L4: Three diversity dimensions are covered \newline
L3: Two diversity dimensions are covered \newline
L2: Multiple scenes acquired by a single sensor \newline
L1: A single data source covering a single scene} \\
\cline{2-4}

  & Data accessibility & 0.3 & \multicolumn{1}{>{\raggedright\arraybackslash}m{11.5cm}}{\textbf{Public Availability of Training Data} \newline
L5: All training data are publicly available \newline
L4: Most training data are publicly available \newline
L3: Some training data are publicly available, while the remainder are documented only through descriptions \newline
L2: Training data are described but not publicly available \newline
L1: Training data are neither described nor publicly available} \\

\midrule
\multirow{24}{=}{SAR Benchmark Task Performance}
  & SAR task coverage breadth & 0.2 & \multicolumn{1}{>{\raggedright\arraybackslash}m{11.5cm}}{\textbf{Number of Supported SAR Task Types Reported in the Paper} \newline
L5: More than five task types; L4: Three to four task types; L3: Two task types; L2: A single task type with outstanding performance; L1: A single task type, or none at all} \\
\cline{2-4}

  & Rigor of comparisons \& baselines & 0.3 & \multicolumn{1}{>{\raggedright\arraybackslash}m{11.5cm}}{\textbf{Rigor and Representativeness of Comparative Baselines Reported in the Paper} \newline
L5: All tasks are evaluated against recent and competitive baselines, with comprehensive evaluation dimensions and sufficiently detailed dataset descriptions \newline
L4: The main experiments include strong baselines, ablation studies, discussions of alternative explanations, and evaluations on multiple datasets \newline
L3: The comparisons are relatively comprehensive, cover multiple types of methods, and include evaluations on publicly available datasets \newline
L2: Mainstream baselines are included, and the comparisons are generally reasonable \newline
L1: Very few baselines are included, or the comparisons are evidently selective} \\
\cline{2-4}

  & Experimental reproducibility & 0.2 & \multicolumn{1}{>{\raggedright\arraybackslash}m{11.5cm}}{\textbf{Reproducibility of Experiments Reported in the Paper} \newline
L5: Highly reproducible, with complete training and inference pipelines publicly available and detailed reproduction instructions provided \newline
L4: Code, model weights, and experimental protocols are comprehensively organized in open-source repositories, enabling the reproduction of all experiments \newline
L3: Code, model weights, and experimental protocols are relatively complete, but some resources must be collected independently based on the paper's descriptions (e.g., datasets or materials for ablation studies) \newline
L2: Only partially open, with certain resources unavailable, such as weights for ablation models or portions of the test data \newline
L1: No code or experimental protocols are available, and weights for ablation models are not released} \\

\cline{2-4}

  & Few-shot/transfer capability on SAR tasks & 0.3 & \multicolumn{1}{>{\raggedright\arraybackslash}m{11.5cm}}{\textbf{Few-shot / transfer capability on SAR} \newline
L5: Reaches SOTA on multiple SAR downstream tasks with a frozen backbone (frozen backbone + linear probe) or few-shot transfer (explicitly reports multi-task, multi-dimensional evaluation, leading feature-transfer performance) \newline
L4: Frozen features perform well on some SAR tasks, with a degree of cross-domain transfer capability \newline
L3: Can extract effective general features from SAR images, but not optimized for SAR characteristics; frozen-feature transfer limited \newline
L2: Pre-trained features adapt poorly to the SAR domain; frozen features nearly unusable, require extensive fine-tuning \newline
L1: Features do not transfer to SAR at all} \\

\bottomrule
\end{tabular}

\end{table*}

\subsection{Multidimensional System of Evaluation Indicators}
The proposed framework rates both VLMs and VFMs across six dimensions. The rationale for selecting these dimensions and the necessity of the corresponding subcriteria are discussed in detail below.

\subsubsection{Influence}
Model rating requires a fair comparison between newly developed and earlier models. Although some older models may no longer match the performance of more recent models, they may have once dominated benchmarks or played a leading role in the field before a certain point in time. As conventional baselines, such models continue to attract considerable attention from academic researchers and practitioners pursuing cutting-edge innovation. Therefore, model impact is included as one of the evaluation dimensions.

This dimension comprises four criteria: publication recency, publication venue ranking, citation count, and community popularity. Equal weights are assigned to these four criteria when calculating the rating for this dimension. We recognize that newly published, high-quality models may have greater potential for widespread dissemination and practical deployment than older models, while their citation counts and community popularity have not yet had sufficient time to accumulate. Conversely, although classic older models may be outperformed by newer models, they may have accumulated substantial citations and community recognition through their contributions to the development of the field. These indicators should not be disregarded when assessing their impact. Accordingly, publication recency is designed to balance citation count and community popularity, enabling a fairer comparison of the impact of new and established models.

It is worth noting that unpublished preprints are assigned the lowest level under the publication-venue criterion. This is because papers that remain unpublished for an extended period may be at greater risk of being abandoned by their authors. In such cases, associated community links may not be regularly updated or maintained, and a larger number of issues may remain unresolved.

\subsubsection{Technical architecture}
Whether the technical architecture is advanced is an important basis for model selection in both engineering deployment and academic research. In this dimension, VLMs and VFMs are evaluated according to different rating criteria:

\textbf{Criteria for VLMs: }
We rate VLMs on four items: architecture paradigm, parameter scale, training paradigm, and multimodal capability. Whether an architecture embodies original innovation reflects its advancement, so we assign qualitative ratings from high to low along the progression of original design, improvement, adaptation, and direct application. Regarding parameter count, although a larger parameter count does not necessarily mean better performance, the development trajectory of GVLM shows that trading increased parameters for higher performance and broader domain adaptability is the current trend, and models with more parameters require higher training costs and more difficult pre-training. In light of this, we assign higher rating scores to models with more parameters. For the training paradigm, we formulate the relevant criteria based on the evolution of GVLM pre-training paradigms. Finally, we evaluate the degree of image-text modality interaction supported by the model architecture, with architectures capable of more interaction considered more advanced.

\textbf{Criteria for VFMs: }
VFMs are still rated on architecture paradigm, parameter scale, and training paradigm, following the same construction logic as for VLMs, but with the specific content adjusted to the characteristics of VFMs. For example, in the training paradigm item, we consider multi-task, multi-stage domain-adaptation training to be advanced. In addition, since VFMs do not involve the text modality, we remove the multimodal capability item from the rating.

\subsubsection{SAR domain adaptability}
As our evaluation tool is designed for the SAR domain, it is necessary to assess the extent to which a model is adapted to SAR applications. We assume that models with a higher degree of SAR-specific design are more likely to be better suited to SAR-domain applications. Accordingly, the degree of SAR-specific design criterion is rated by examining whether key components of the model, including its loss functions, embedding modules, and encoders, are specifically designed for SAR data.

Moreover, whether the training process explicitly considers SAR-specific characteristics, such as speckle, geometric distortions, complex-valued data and phase information, polarization, and incidence angle, is also regarded as an indicator of SAR-domain adaptability \cite{survey-SAR-CHINA-Hou}. Compared with directly transferring methods developed for natural or optical imagery, dedicated modeling and fine-grained treatment of these SAR characteristics can better accommodate the properties of SAR data.

We further include the proportion of SAR data in the training set as a rating criterion. In general, a higher proportion of SAR data during pretraining is associated with a lower difficulty in adapting the model to downstream SAR tasks. RS models that have seen SAR images and those that have not differ significantly in the difficulty of fine-tuning and adapting to SAR downstream tasks. In particular, SAR-VFMs trained primarily on SAR data can learn more effective SAR representations than RS-VFMs in which SAR data constitute only a small fraction of the training corpus.
\subsubsection{Training data}
Foundation models rely heavily on large-scale, high-quality data during training. Therefore, training data constitute an important dimension of the rating process. The scale and diversity of training data determine the breadth and depth of the knowledge acquired by a model. Accordingly, larger and more diverse training datasets are assigned higher rating scores. When evaluating data scale, we consider the number of images used to train VFMs and the number of image-text pairs used to train VLMs.

Notably, SAR data acquired from different platforms and under different imaging conditions can vary substantially. To ensure that trained models generalize effectively to downstream tasks, the training data should cover as many platforms and imaging conditions as possible. We therefore use the number of diversity dimensions covered by the training data, including sensors, frequency bands, spatial resolutions, scene types, and geographic regions, as a rating criterion. We also assess the public availability of the training data, as data accessibility directly affects the reproducibility and extensibility of model training.
\subsubsection{SAR benchmark task performance}
Model performance across various SAR benchmark tasks is a key concern for different user groups. When models are released, their performance on selected representative benchmarks is typically reported. However, since models differ in their research objectives and intended capabilities, the benchmarks adopted for evaluation are often inconsistent. Although the introduction of multitask benchmarks such as SARLANG-1M and SARChat-Bench-2M has provided platforms for model comparison, assessments produced by different benchmarks may still be inconsistent because of differences in task composition, test data, and evaluation metrics. For example, in the zero-shot image captioning task, Qwen2.5-VL-7B scores substantially higher than LLaVA1.5-7B on SARLANG-1M, whereas LLaVA1.5-7B outperforms Qwen2.5-VL-7B on multiple metrics in SAREval.

We also consider the cost of model rating. Evaluating multiple existing models on the same benchmark entails substantial time costs associated with environment setup and experimental reproduction. Moreover, the limited openness of the SAR community prevents several representative closed-source models from being evaluated and compared using these benchmarks. For these reasons, we do not adopt an evaluation approach that requires experimental reproduction. Instead, we assess the strength of the publicly available experimental evidence for each model, thereby enabling efficient, broadly applicable, and highly extensible model rating. The specific rating and scoring methods are described in detail in Section \uppercase\expandafter{\romannumeral3}. C.

\subsubsection{Open-sourcing and reproducibility}
SAR-specific models generally exhibit a lower level of open-source availability than remote sensing models. Models with greater openness can more effectively advance the field and facilitate practical deployment. Therefore, open-sourcing and reproducibility constitute an important dimension of our rating framework. We assess the availability of source code, pretrained weights, and training scripts. Models that release a broader range of resources and require less effort to reproduce receive higher ratings. In addition, the availability of training data is somewhat related to this dimension, but as we have already assigned it to the training data dimension, we do not include it in the consideration of open-sourcing and reproducibility.

\subsubsection{Overall rating method}
As illustrated in Fig. \ref{fig4}, once the model under evaluation is identified as either a VFM or a VLM, it is scored across six dimensions according to the rating criteria specific to that model type. Within each dimension, the scores of individual items are aggregated using predefined weights, and the weighted score is rounded to obtain the corresponding dimension-level rating. These dimension-level ratings provide an intuitive representation of the model’s strengths and limitations. To enable direct comparisons across models, the ratings of all dimensions are further weighted and rounded to produce an overall model rating.

However, the relative importance of these dimensions varies across user groups with different needs. For example, an engineer selecting a model for practical deployment and a beginner selecting a model for learning may assign markedly different levels of importance to the influence and open-sourcing and reproducibility dimensions. Engineers generally place greater emphasis on open-sourcing and reproducibility because deploying a closed-source model can be substantially more difficult. They may therefore assign a higher weight to open-sourcing and reproducibility when calculating the overall model rating. In contrast, beginners may prefer models with greater academic and community influence, as such models can provide a clearer understanding of the development and evolution of the field. A rating scheme that assigns a higher weight to influence would thus better support their model selection. To accommodate the diverse requirements of different user groups, we propose Dynamic Weight Allocation Based on User Profiling, which is described in detail in Section \uppercase\expandafter{\romannumeral3}.D.

\subsection{Meta-Analysis-Based Performance Evaluation of SAR Benchmark Tasks}
As discussed in Section \uppercase\expandafter{\romannumeral3}. B. 5), considering the level of openness within the SAR community and the efficiency of model rating, the proposed framework adopts a meta-analysis-based approach for the SAR benchmark task performance dimension. Specifically, it evaluates the strength and quality of the experimental evidence reported in the original model publications. This approach derives rating scores directly from the comparative experimental results presented in each publication, thereby enabling new models to be incorporated into the rating system more efficiently and enhancing the scalability of the framework. Given the distinct characteristics of VFMs and VLMs, different rating criteria are applied to the two model types, as described below:

\textbf{Criteria for VLMs: }
First, we assess the breadth of SAR task coverage reported in the original model publication, with higher scores assigned to models evaluated on a wider range of task types. Second, we evaluate the rigor of comparisons and baselines, focusing primarily on the competitiveness of the selected baselines, the comprehensiveness of the evaluation dimensions, the appropriateness of the experimental settings, and the diversity of the test data. When a model is released, its reported experimental results often achieve the best performance in some respects, but the baselines it compares against may not be the previous state-of-the-art methods, or selective comparison may be present. Through this item, we therefore provide a comprehensive assessment of the rigor of the comparative experiments.

In addition, the reproducibility of the reported experiments directly affects the credibility of the results and is thus included as a scoring item. We assign a qualitative tiered score based on the availability of the experimental code, model weights, and data, as well as the clarity of the experimental protocol. Finally, because zero-shot performance provides an important foundation for the practical deployment of VLMs, we include zero-shot capability on SAR tasks as a scoring item. This item is qualitatively rated according to the model’s understanding of the SAR modality and its demonstrated capabilities on zero-shot tasks. All four criteria can be assessed by analyzing the original model publication without conducting additional experiments, making the proposed evaluation procedure straightforward and efficient.

\textbf{Criteria for VFMs: }
VFMs share the same evaluation criteria as VLMs on the SAR task coverage breadth, rigor of comparisons \& baselines, and experimental reproducibility items of this dimension, but differ in the focus of their SAR task capability. Since VFMs are typically applied by selecting an extension head for a downstream task and performing targeted fine-tuning, researchers are more concerned with the effectiveness and transferability of the features they extract than with zero-shot performance. Accordingly, we qualitatively evaluate the transfer performance of a model's extracted features in the SAR domain based on the results reported in its original paper. Such performance can be reflected through experiments such as few-shot classification presented in the model's original paper.

\subsection{Dynamic Weight Allocation Based on User Profiling}
To enable the proposed evaluation tool to support the model-selection needs of diverse user groups, we establish multiple weight-allocation schemes. After the existing models have been evaluated across six dimensions, users can select a weighting scheme to calculate the overall ratings and rankings of the models included in the tool. The tool allows users to switch between weighting schemes with a single click and also supports customized weight configurations, demonstrating strong flexibility and extensibility in evaluation-scheme management, as illustrated in Fig. \ref{fig4}(b). Seven predefined weight-allocation schemes are incorporated into the proposed tool to address common model-selection requirements. Although this finite set of schemes cannot cover every possible evaluation scenario, users can tailor the weight configurations to their specific needs. The detailed weight assignments of the predefined schemes are presented in Table \ref{tab:weight-configurations}.

\begin{table*}[htbp]
    \centering
    \caption{Predefined Weight Allocation Schemes.}
    \label{tab:weight-configurations}
    \resizebox{\textwidth}{!}{%
        \begin{tabular}{cclccccccc}
            \toprule
            No. & Scheme & Positioning & Influence & \makecell{Technical Architecture} & \makecell{SAR Domain\\Adaptation}
                & Training Data & \makecell{SAR Benchmark\\Task Performance} & \makecell{Open-Sourcing and\\Reproducibility} & Total \\
            \midrule

            1 & Academic Research
              & Literature Survey/Review
              & 0.20 & 0.25 & 0.20 & 0.05 & 0.20 & 0.10 & 1.00 \\

            2 & Engineering Deployment
              & Selection for Project Deployment
              & 0.05 & 0.05 & 0.15 & 0.05 & 0.35 & 0.35 & 1.00 \\

            3 & SAR Specialization
              & SAR-Adapted Application
              & 0.05 & 0.10 & 0.40 & 0.05 & 0.30 & 0.10 & 1.00 \\

            4 & Reproduction First
              & Quick Start/Teaching
              & 0.05 & 0.05 & 0.15 & 0.10 & 0.25 & 0.40 & 1.00 \\

            5 & Data Resources
              & Dataset/Foundation Building
              & 0.10 & 0.05 & 0.20 & 0.30 & 0.20 & 0.15 & 1.00 \\

            6 & Frontier Innovation
              & Tracking Emerging Paradigms
              & 0.15 & 0.35 & 0.20 & 0.05 & 0.15 & 0.10 & 1.00 \\

            7 & Performance Priority
              & Pursuing Optimal Performance
              & 0.05 & 0.05 & 0.15 & 0.05 & 0.50 & 0.20 & 1.00 \\
            \bottomrule
        \end{tabular}%
    }
\end{table*}

\textbf{Academic Research: }
This scheme primarily targets model-selection needs in academic research. While maintaining relatively balanced weights across all dimensions, it assigns greater importance to influence and technical architecture.

\textbf{Engineering Deployment: }
This scheme is designed for model selection in practical deployment scenarios. In such scenarios, users typically place less emphasis on model influence and focus more on actual performance. Moreover, the degree of open-source availability directly affects the difficulty of model deployment. Therefore, open-sourcing and reproducibility is assigned a relatively high weight.

\textbf{SAR Specialization: }
This scheme targets model-selection scenarios involving transfer to the SAR domain. It emphasizes both SAR adaptability and performance to identify models that perform effectively on SAR tasks.

\textbf{Reproduction Priority: }
This scheme aims to identify models that can be readily adopted and deployed. Accordingly, open-sourcing and reproducibility is assigned a relatively high weight.

\textbf{Data Resources: }
This scheme focuses on the data and resource advantages of a model, emphasizing the broader SAR adaptability enabled by richer training data. It is intended for model-selection scenarios such as SAR foundation construction. Therefore, SAR adaptability and training Data are assigned higher weights.

\textbf{Frontier Innovation: }
This scheme treats the advancement of the model architecture as the primary rating factor and is intended for researchers seeking to explore the frontiers of the field.

\textbf{Performance Priority: }
This scheme primarily evaluates models based on the performance reported in their original publications, while assigning relatively low weights to all other dimensions. Notably, setting the weight of SAR benchmark task performance to 1.00 would make this scheme consistent with conventional experimental benchmark evaluation, in which models are compared solely on the basis of their benchmark performance.

\section{Analysis of Experimental and Rating Results}
\subsection{Implementation Details}
We apply the evaluation tool to rate different types of SAR-related models, reporting the ratings along the six dimensions as well as the final ratings obtained under different predefined weight allocation schemes. More fine-grained ratings at the item level will be incorporated into the proposed tool. Among the six assessment dimensions, Influence and Open-Sourcing and Reproducibility are based on quantitative indicators and require information to be retrieved and consolidated from multiple online sources. These two dimensions are therefore evaluated through manual data collection and scoring. The remaining four dimensions are assessed using a mainstream general-purpose multi-modal large language model (MLLM), supplemented by human verification. Specifically, we first familiarize GPT-5.6-Sol with the assessment criteria defined in Tables \ref{tab:VLM-bench} and \ref{tab:VFM-bench}. We then provide it with the publications associated with each model under evaluation and instruct it to generate recommended scores for individual items, along with supporting rationales. Human reviewers subsequently examine these recommendations and make the final scoring decisions. The utilization of a fixed auxiliary MLLM combined with human review is motivated by two considerations. First, employing the same auxiliary model throughout the assessment process helps ensure consistency and maintains a relatively stable scoring scale for qualitative items. Second, this approach balances the subjectivity of human evaluators with that of a mainstream MLLM. The MLLM’s involvement reduces the time required for assessment and prevents the ratings from being dominated by the subjective judgment of any single evaluator. Human review, meanwhile, mitigates the risk of hallucinated outputs arising from multi-turn interactions with the model and enables timely corrections. The feasibility and stability of MLLM-assisted rating are discussed in Section \uppercase\expandafter{\romannumeral4}. E. In addition, when a publication introduces multiple variants of the same model, we evaluate only the variant with the largest parameter count. The data referenced in the evaluation process is up to date as of August 2026.
\subsection{VLM Track Rating Results}
We rate 40 representative SAR-related VLMs, and the results are shown in Table \ref{tab:VLM-result}. The selection of foundation models spans the four years from 2023 to 2026 and covers three types. We bold the results that obtain an L5 rating in the overall rating. It can be observed that although some RS-VLMs with SAR are not pretrained solely using the SAR modality, they still obtain an overall rating of L5 under multiple weight allocation schemes. For example, EarthDial \cite{model-EarthDial-Soni} receives L5 ratings under the “Reproduction First”, “Data Resources”, and “Frontier Innovation” schemes. TerraScope \cite{model-TerraScope-Shu} and EarthMind \cite{model-EarthMind-Shu} both obtain L4 ratings under all weight allocation schemes, reflecting their overall excellent performance across the rating dimensions. Since the proposed rating framework targets the SAR field, SAR-VLMs, as models natively designed for this field, are more advantageous than other types of models in multiple rating dimensions. Nevertheless, the overall ratings of most models still have difficulty reaching L5, indicating that development in this field remains relatively lagging. 

\begin{figure}[htbp]
\centering
\includegraphics[width=0.95\columnwidth]{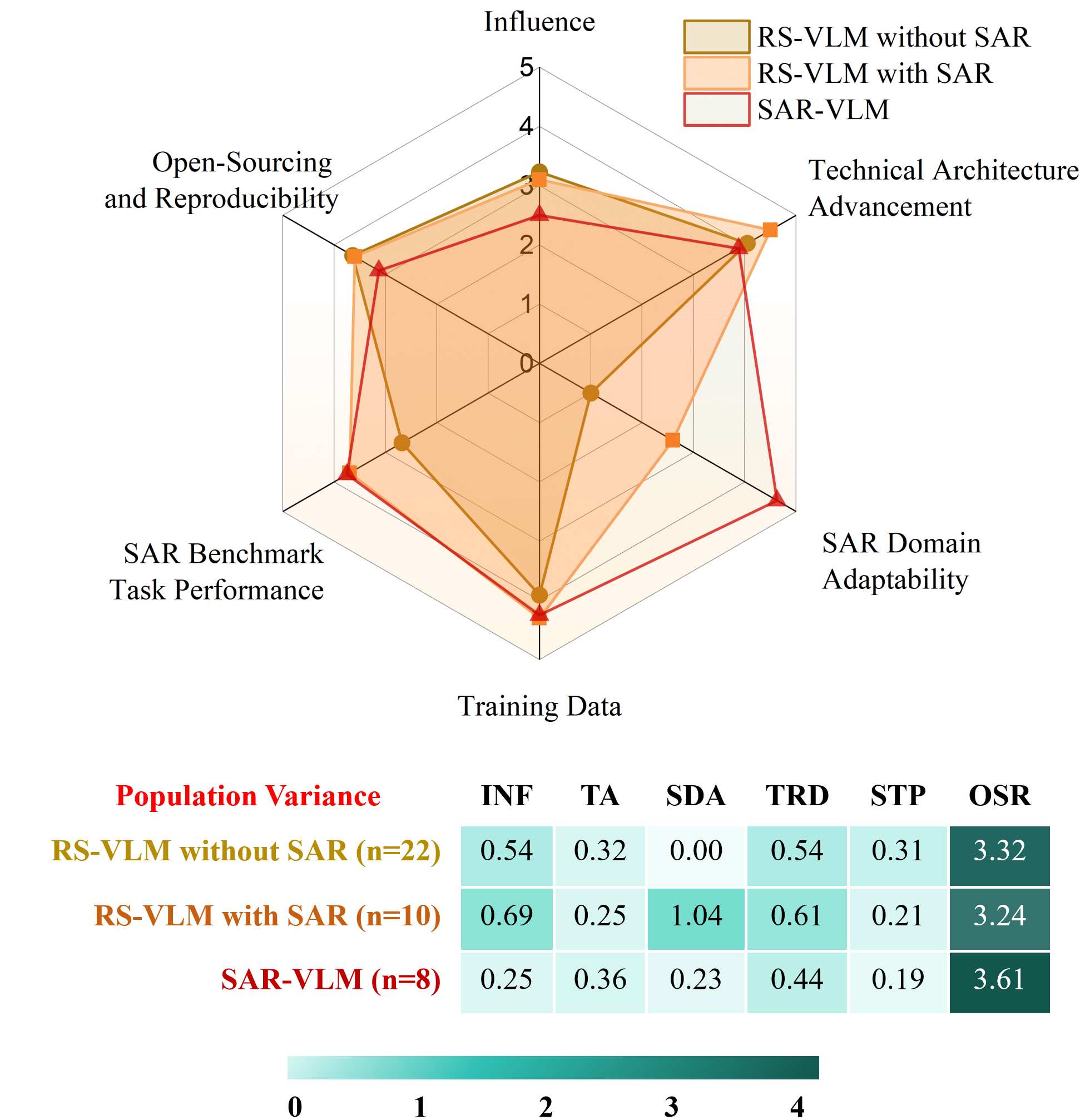}
\caption{Multi-dimensional mean rating radar chart and variance table for SAR-related VLMs.}
\label{fig5}
\end{figure}

To better analyze the overall strengths and weaknesses that each type of model exhibits across dimensions, we average the ratings of all models of each type in each individual dimension, use the six resulting values to plot a radar chart, and plot the radar charts of the three types of models in the same figure for comparison, as shown in Fig. \ref{fig5}. The resulting hexagonal profiles provide an intuitive overview of their respective capabilities. In terms of influence, SAR-VLMs rate substantially lower than RS-VLMs, which may partly reflect the narrower application scope and smaller research community of the SAR domain. In terms of technical architecture, RS-VLM with SAR achieves a higher average rating than RS-VLM without SAR. This advantage arises partly because RS-VLMs with SAR often integrate multiple modalities or observations from multiple sensors, which require more sophisticated architectures for cross-modal interaction and representation alignment. By contrast, existing SAR-VLMs generally adopt simpler architectures.

In terms of SAR-domain adaptation, SAR-VLMs achieve the highest average rating, followed by RS-VLM with SAR and RS-VLM without SAR, which aligns with objective expectations. Regarding training data, SAR-inclusive RS-VLMs outperform their SAR-free counterparts because they typically incorporate more diverse data sources and sensor modalities. However, for SAR task performance, RS-VLM with SAR achieves an average rating comparable to that of SAR-VLM. This result indicates that VLMs trained or fine-tuned exclusively on SAR data do not yet establish a clear performance advantage over broader RS foundation models on SAR tasks. Therefore, RS-VLMs with SAR remain a competitive paradigm for SAR image analysis. Moreover, these findings show that performance on SAR image analysis tasks depends not only on the proportion or modality purity of SAR training data, but also on model architecture, overall training scale, cross-modal knowledge transfer, task diversity, and training strategy.

To understand what produces this result, we examine scores at the item level rather than at the dimension level. On the items “Parameter scale” and “Training data scale”, SAR-VLMs score markedly below RS-VLMs with SAR. We read this gap as a plausible explanation for the absent downstream advantage: a SAR-native model built on a smaller backbone and a smaller pretraining corpus carries less capacity and less coverage than a broader RS model that absorbs SAR data alongside other modalities, and modality purity does not by itself compensate for that deficit. The gap has an upstream cause. SAR acquisition is subject to release restrictions and is expensive to obtain, and annotation of SAR imagery is slow and demands specialist interpretation, so publicly available, high-quality annotated SAR data remain scarce. This scarcity, rather than any deficiency in the SAR-native design, is what limits the scale at which SAR-VLMs are built. Assembling large-scale, high-quality, openly released annotated SAR corpora therefore remains a priority for the SAR-VLM community, and it is a precondition that architectural refinement alone cannot replace.

\subsection{VFM Track Rating Results}
We rate 75 representative SAR-related VFMs, with the results summarized in Table \ref{tab:VFM-result}. The selected foundation models span five years, from 2022 to 2026, and cover all three types. Models that achieve an overall L5 rating are highlighted in bold. Similar to the observations for VLMs, several RS-VFMs with SAR achieve overall L5 ratings under multiple weight allocation schemes, including SkySense++ \cite{model-SkySense++-Wu} and TESSERA \cite{model-tessera-Feng}. However, compared with RS-VLMs with SAR, a relatively smaller proportion of RS-VFMs with SAR achieve high overall ratings within their respective model categories. One possible explanation is that VLMs can benefit from rich linguistic knowledge acquired during large-scale language pretraining, which provides a strong semantic prior and a favorable initialization for subsequent training or fine-tuning, thereby reducing their dependence on the proportion of SAR imagery in the visual training data. VFMs, by comparison, do not have an equivalent text modality and rely predominantly on visual data to learn transferable representations. Consequently, the proportion, diversity, and modality composition of SAR imagery in the pretraining data exert a more direct influence on their ability to adapt to SAR-specific characteristics and downstream tasks. This may explain why the purity and prominence of the SAR modality are more critical for VFM evaluation than for VLM evaluation.

\begin{figure}[htbp]
\centering
\includegraphics[width=0.95\columnwidth]{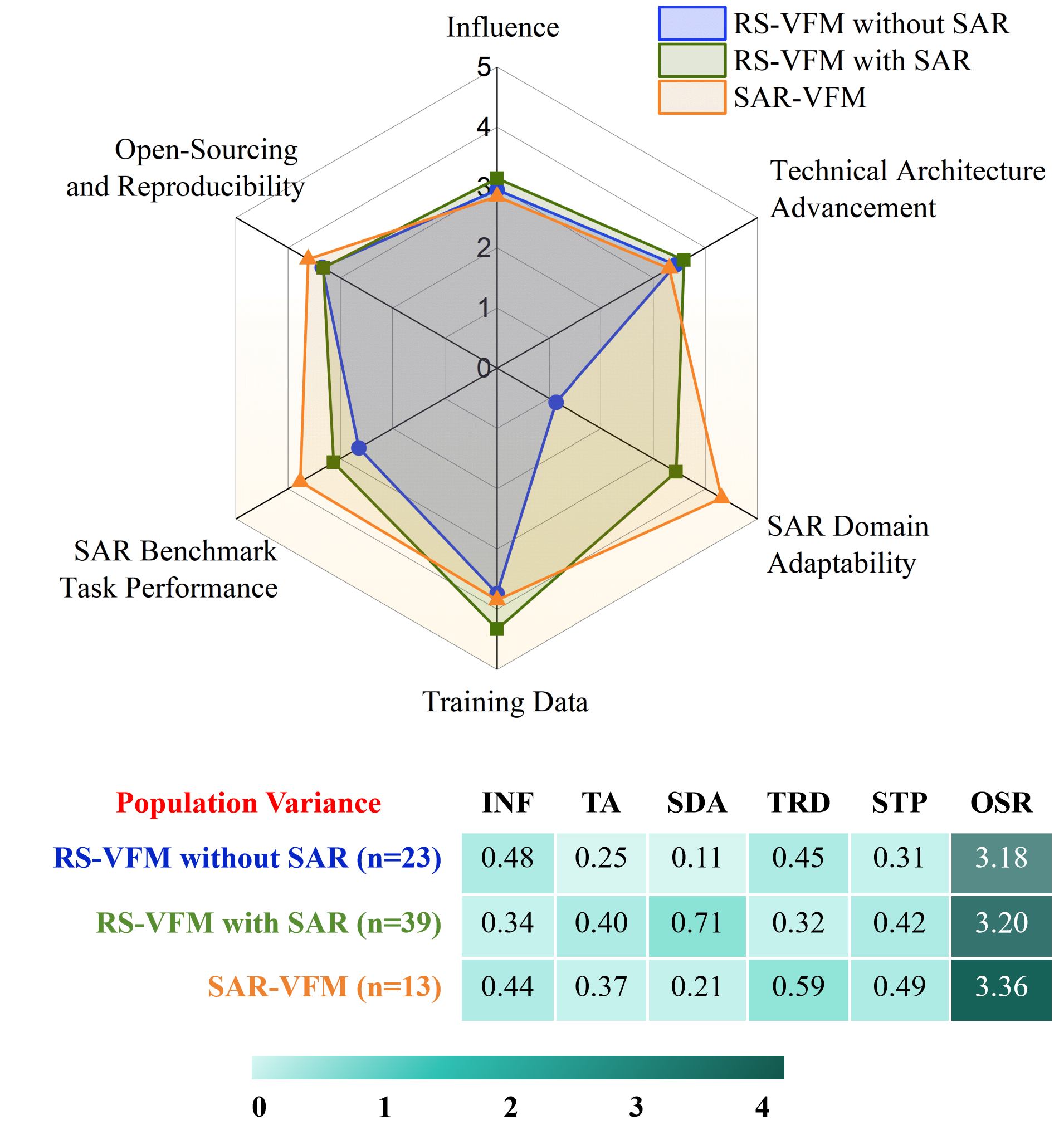}
\caption{Multi-dimensional mean rating radar chart and variance table for SAR-related VFMs.}
\label{fig6}
\end{figure}

Since RS-VFMs without SAR are trained on datasets that do not contain SAR imagery, their transferability to SAR-related applications is inherently limited. Consequently, none of these models achieves an overall L5 rating under any of the weight allocation schemes. In contrast, SAR-VFMs are developed with SAR as their primary and often exclusive modality, and are trained solely on SAR data. This modality-specific design makes them more likely to receive high ratings in the dimensions of SAR domain adaptation, training data, and SAR benchmark task performance. 

Notably, the SAR-VFMs with relatively strong overall ratings: SARATR-X-V2 \cite{model-SARATR-X-v2-Li}, SAR-MAE \cite{model-SARMAE-Liu}, SARATR-X \cite{model-SAR-ATR-X-Li}, and SUMMIT \cite{model-SUMMIT-Du}--all employ MIM. This observation suggests that MIM remains a fundamental strategy for developing SAR-VFMs. Nevertheless, the straightforward application of MIM to SAR images is unlikely to be sufficient on its own. A key direction for future research is therefore to incorporate the physical characteristics of SAR imaging more explicitly into the model-learning process. Specifically, SUMMIT \cite{model-SUMMIT-Du}, SARATR-X \cite{model-SAR-ATR-X-Li}, and SAR-MAE \cite{model-SARMAE-Liu} mitigate the adverse effects of speckle on pretraining from different perspectives, utilizing multi-task auxiliary supervision, structural gradient targets, and physical noise modeling, respectively. CrossEarth-SAR \cite{model-CrossEarth-SAR-Ye} goes a step further by explicitly modeling variations in geographic region, polarization, frequency band, and imaging platform as sources of domain shift, thereby addressing cross-domain generalization. SARATR-X-V2 \cite{model-SARATR-X-v2-Li}, in contrast, focuses on the design of pretraining targets and introduces the principles of physics-grounded stability and semantic scale compatibility. These developments indicate that SAR-VFM research is gradually moving away from the direct transfer of optical-image pretraining techniques toward domain-specific representation learning that is more deeply informed by the SAR imaging mechanism.

Similar to our analysis of VLMs, we calculate the mean rating of all models within each category for each evaluation dimension and visualized the results using radar charts, as shown in Fig. \ref{fig6}. Regarding influence dimension, SAR-VFMs receive lower ratings than RS-VFMs. Because SAR-VFMs are highly specialized, their citations are concentrated primarily within the SAR research community, whereas general-purpose RS-VFMs have broader application scenarios and consequently attract a wider range of citations. In terms of technical architecture, the relative pattern is consistent with that observed for VLMs. Moreover, a clear ordering emerges in SAR benchmark task performance dimension: SAR-VFMs achieve the highest average ratings, followed by RS-VFMs with SAR, while RS-VFMs without SAR receive the lowest ratings. This result suggests that the presence and proportion of SAR data in VFM pretraining are directly related to the models’ transferability to SAR downstream tasks. Within the scope of our evaluation, SAR-only pretraining appears to provide a stronger advantage for SAR-specific tasks than multi-modal pretraining that combines SAR with other RS modalities.

\begin{table*}[htbp]
\centering
\caption{Dimension-Wise Rating Results of SAR-related VLMs and Overall Ratings under Different Weight Allocation Schemes (Abbreviations adopted: INF=Influence, TA=Technical Architecture, SDA=SAR Domain Adaptation, TRD=Training Data, STP=SAR Benchmark Task Performance, OSR=Open-Sourcing and Reproducibility).}

\label{tab:VLM-result}

\begin{tabular}{c|c|cccccc|ccccccc}
\toprule
\multirow{2}{*}{Category}&\multirow{2}{*}{Model}&\multicolumn{6}{c|}{\textbf{Dimension Scores}}&\multicolumn{7}{c}{\textbf{Overall Rating}}\\
\cline{3-8}
\cline{9-15}
& & INF & TA & SDA & TRD & STP & OSR &No. 1&No. 2&No. 3&No. 4&No. 5&No. 6&No. 7\\

\midrule
\multirow{22}{*}{\makecell{RS-VLM\\without SAR}} &RemoteCLIP \cite{model-RemoteCLIP-Liu} & L4 & L4 & L1 & L5 & L3 & L5 & L3 &L4 &L3 &L4 &L4 &L3 &L3\\
&GeoChat \cite{model-GeoChat-Kuckreja} & L4 & L5 & L1 & L4 & L3 & L5&L4 &L4 &L3 &L4 & L3 & L4 & L3\\
&SkySense-O \cite{model-SkySense-O-Zhu} & L4 & L4 & L1 & L5 & L3 & L5  & L3 &L4 &L3 &L4 &L4 &L3 &L3\\
&ScaleEarth \cite{model-ScaleEarth-Zhang} & L2 & L5 & L1 & L4 & L3 & L1& L3 &L2&L2&L2&L3&L3&L2 \\
&SkyEyeGPT \cite{model-SkyEyeGPT-Zhan} & L4 & L4 & L1 & L4 & L3 & L4& L3 &L3&L3&L3&L3&L3&L3 \\
&SkySenseGPT \cite{model-SkySenseGPT-Luo} & L3 & L4 & L1 & L4 & L3 & L4& L3 &L3&L2&L3&L3&L3&L3 \\

&GeoRSCLIP \cite{model-GeoRSCLIP-Zhang} & L4 & L3 & L1 & L4 & L3 & L5& L3 &L4&L3&L4&L3&L3&L3 \\
&GRAFT \cite{model-graft-Mall} & L3 & L3 & L1 & L5 & L2 & L1& L2 &L2&L2&L2&L3&L2&L2 \\
&EarthMarker \cite{model-EarthMarker-Zhang} & L3 & L4 & L1 & L4 & L2 & L1& L3 &L2&L2&L2&L2&L3&L2 \\
&RingMoGPT \cite{model-RingMoGPT-Wang} & L3 & L5 & L1 & L3 & L2 & L1& L3 &L2&L2&L2&L2&L3&L2 \\
&ZoomEarth \cite{model-ZoomEarth-Liu} & L3 & L4 & L1 & L2 & L3 & L5& L3 &L3&L2&L3&L3&L3&L3 \\
&SkyMoE \cite{model-skymoe-Liu} & L3 & L4 & L1 & L4 & L2 & L1& L3 &L2&L2&L2&L2&L3&L2 \\
&ScoreRS \cite{model-ScoreRS-Muhtar} & L3 & L4 & L1 & L4 & L3 & L5& L3 &L4&L3&L4&L3&L3&L3 \\
&GeoRSMLLM \cite{model-GeoRSMLLM-Zhang} & L2 & L4 & L1 & L3 & L1 & L1& L2 &L1&L1&L1&L2&L2&L1 \\
&RS-LVLM \cite{model-RS-LVLM-Park} & L2 & L4 & L1 & L4 & L2 & L1& L2 &L2&L2&L2&L2&L3&L2 \\
&RemoteSAM \cite{model-RemoteSAM-Yao} & L3 & L3 & L1 & L4 & L3 & L5& L3 &L3&L2&L4&L3&L3&L3 \\
&FUSE-RSVLM \cite{model-FUSE-RSVLM-Dang} & L2 & L4 & L1 & L4 & L3 & L5& L3 &L3&L3&L4&L3&L3&L3 \\

&GeoPixel \cite{model-GeoPixel-Shabbir} & L4 & L5 & L1 & L3 & L3 & L5& L4 &L4&L3&L4&L3&L4&L3 \\
&GeoLLaVA \cite{model-GeoLLaVA-Wang} & L4 & L4 & L1 & L3 & L3 & L5& L3 &L4&L3&L4&L3&L3&L3 \\
&SATtxt \cite{model-SATtxt-Do} & L3 & L4 & L1 & L4 & L3 & L5& L3 &L4&L3&L4&L3&L3&L3 \\
&LHRS-Bot \cite{model-LHRS-Bot-Muhtar} & L4 & L4 & L1 & L5 & L3 & L5& L3 &L4&L3&L4&L4&L3&L3 \\
&LHRS-Bot-Nova \cite{model-LHRS-Bot-Nova-Li} & L4 & L4 & L1 & L4 & L3 & L5& L3 &L4&L3&L4&L3&L3&L3 \\
\hline
\multirow{10}{*}{\makecell{RS-VLM\\with SAR}} &EarthGPT \cite{model-EarthGPT-Zhang} &L4	&L5	&L2	&L5	&L4	&L1& L4 &L3&L3&L3&L4&L4&L3 \\
&EarthDial \cite{model-EarthDial-Soni} &L4	&L5	&L4	&L5	&L4	&L5& L4 &L4&L4&\textbf{L5}&\textbf{L5}&\textbf{L5}&L4 \\
&TerraScope \cite{model-TerraScope-Shu} &L4	&L5	&L3	&L5	&L4	&L5& L4 &L4&L4&L4&L4&L4&L4 \\
&Falcon \cite{model-falcon-Yao} &L3	&L4	&L1	&L5	&L3	&L3& L3 &L3&L2&L3&L3&L3&L3 \\

&REO-VLM \cite{model-REO-VLM-Xue} &L2&L4	&L4	&L3	&L4	&L1& L3 &L3&L4&L3&L3&L3&L3 \\
&ViTP \cite{model-ViTP-Li} &L3&L4	&L2	&L4	&L4	&L5& L4 &L4&L3&L4&L4&L4&L4 \\

&VHM \cite{model-VHM-Pang} &L4	&L4	&L1	&L4	&L3	&L5& L3 &L4&L3&L4&L3&L3&L3 \\
&DOFA-CLIP \cite{model-DOFA-CLIP-Xiong} &L2	&L4	&L3	&L4	&L3	&L5& L3 &L4&L3&L4&L4&L3&L3 \\
&EarthMind \cite{model-EarthMind-Shu} &L2	&L5	&L3	&L5	&L4	&L5& L4 &L4&L4&L4&L4&L4&L4 \\
&EarthGPT-X \cite{model-EarthGPT-X-Zhang} &L3 &L5	&L3	&L3	&L4	&L1& L4 &L3&L3&L3&L3&L4&L3 \\
\hline
\multirow{8}{*}{\makecell{SAR-VLM}}&FUSAR-KLIP \cite{model-FUSAR-KLIP-Yi} &L2	&L4	&L5	&L5	&L4	&L2& L4 &L3&L4&L3&L4&L4&L4 \\
&SARCLIP \cite{model-SARCLIP-Jiang} &L3	&L4	&L5	&L4	&L4	&L5& L4 &L4&L4&\textbf{L5}&L4&L4&L4 \\
&FUSAR-GPT \cite{model-FUSAR-GPT-Zhang} &L3	&L4	&L5	&L3	&L4	&L1& L4 &L3&L4&L3&L3&L4&L3 \\   
&SARVLM \cite{model-SARVLM-Ma} &L2	&L4	&L4	&L5	&L4	&L5& L4 &L4&L4&\textbf{L5}&L4&L4&L4 \\
&FUSAR-R1 \cite{model-FUSAR-R1-Yi} &L2	&L5	&L5	&L4	&L3	&L1& L4 &L3&L4&L3&L3&L4&L3 \\
&TACMT \cite{model-TACMT-Li} &L3	&L3	&L4	&L4	&L4	&L5& L4 &L4&L4&L4&L4&L4&L4 \\
&TVLightFormer \cite{model-TVLightFormer-Zhong} &L3	&L3	&L5	&L4	&L3	&L1& L3 &L3&L4&L3&L3&L3&L3 \\
&SAR-GPT \cite{model-SAR-TEXT-He}&L2	&L4	&L4	&L5	&L4	&L5& L4 &L4&L4&L4&L4&L4&L4\\
\bottomrule
\end{tabular}

\end{table*}

\begin{table*}[htbp]
\centering
\caption{Dimension-Wise Rating Results of SAR-related VFMs and Overall Ratings under Different Weight Allocation Schemes.}
\label{tab:VFM-result}

\begin{tabular}{c|c|cccccc|ccccccc}
\toprule
\multirow{2}{*}{Category}&\multirow{2}{*}{Model}&\multicolumn{6}{c|}{\textbf{Dimension Scores}}&\multicolumn{7}{c}{\textbf{Overall Rating}}\\
\cline{3-8}
\cline{9-15}
& & INF & TA & SDA & TRD & STP & OSR &No. 1&No. 2&No. 3&No. 4&No. 5&No. 6&No. 7\\

\midrule
\multirow{23}{*}{\makecell{RS-VFM\\without SAR}} &Alliance \cite{model-Alliance-Zhao} &L3&L3&L1&L4	&L2	&L1& L2 &L2&L2&L2&L2&L2&L2 \\
&RingMo \cite{model-Ringmo-Xian} &L3&L3&L1&L4	&L2	&L2& L2 &L2&L2&L2&L3&L2&L2 \\
&RVSA \cite{model-RVSA-Wang} &L4&L3&L1&L4	&L3	&L5& L3 &L4&L3&L4&L3&L3&L3 \\
&AgriFM \cite{model-AgriFM-Li} &L3&L3&L1&L4	&L3	&L5& L3 &L3&L2&L4&L3&L3&L3 \\
&SwiMDiff \cite{model-SwiMDiff-Tian} &L3&L3&L1&L3&L2	&L1& L2 &L2&L2&L2&L2&L2&L2 \\
&RS-DFM \cite{model-RS-DFM-Wang} &L2&L3&L1&L2	&L2	&L1& L2 &L2&L2&L2&L2&L2&L2 \\
&PIS \cite{model-PIS-An} &L3&L3&L1&L4	&L3	&L5& L3 &L3&L2&L4&L3&L3&L3 \\
&MM-VSF \cite{model-MM-VSF-Praveen} &L2&L4&L1&L4	&L2	&L3& L3 &L2&L2&L3&L3&L3&L2 \\
&GeRSP \cite{model-GeRSP-Huang} &L3&L3&L1&L4	&L3	&L5& L3 &L3&L2&L4&L3&L3&L3 \\
&RingMo-Lite \cite{model-RingMo-Lite-Wang} &L3&L3&L1&L3	&L2	&L1& L2 &L2&L2&L2&L2&L2&L2 \\
&U-BARN \cite{model-U-BARN-Dumeur} &L2&L4&L1&L4	&L3	&L2& L3 &L2&L2&L2&L3&L3&L3 \\
&SpectralGPT \cite{model-SpectralGPT-Hong} &L4&L4&L1&L4	&L3	&L5& L3 &L4&L3&L4&L3&L3&L3 \\
&SatMAE++ \cite{model-SatMAE++-Noman} &L4&L4&L1&L4	&L3	&L5& L3 &L4&L3&L4&L3&L3&L3 \\
&RingMo-Galaxy \cite{model-RingMo-Galaxy-Wang} &L3&L3&L1&L2	&L2	&L1& L2 &L2&L2&L2&L2&L2&L2 \\
&RoMA \cite{model-RoMA-Wang} &L4&L3&L1&L4	&L3	&L5& L3 &L4&L3&L4&L3&L3&L3 \\
&MOMO \cite{model-MOMO-Purohit} &L3&L4&L1&L5	&L3	&L5& L3 &L4&L3&L4&L4&L3&L3 \\
&RingMo-Aerial \cite{model-RingMo-Aerial-Diao} &L3&L4&L2&L4	&L3	&L1& L3 &L2&L3&L2&L3&L3&L3 \\
&DeepAndes \cite{model-DeepAndes-Guo} &L3&L3&L1&L3	&L2	&L5& L3 &L3&L2&L3&L3&L3&L3 \\
&HyperSIGMA \cite{model-HyperSIGMA-Wang} &L4&L4&L1&L4	&L3	&L5& L3 &L4&L3&L4&L3&L3&L3 \\
&TiMo \cite{model-TiMo-Qin} &L2&L4&L2&L4	&L3	&L2& L3 &L3&L3&L3&L3&L3&L3 \\
&SatMamba \cite{model-SatMamba-Duc} &L2&L3&L1&L4	&L2	&L2& L2 &L2&L2&L2&L2&L2&L2 \\
&DynamicVis \cite{model-DynamicVis-Chen} &L2&L4&L2&L4	&L4	&L5& L3 &L4&L3&L4&L4&L3&L4 \\
&GeoLink \cite{model-GeoLink-DBelgrave} &L3&L4&L1&L4	&L3	&L5& L3 &L4&L3&L4&L3&L3&L3 \\
\hline

\multirow{39}{*}{\makecell{RS-VFM\\with SAR}} 
&TerraFM \cite{model-TerraFM-Muhammad} &L4	&L4	&L4	&L5	&L3	&L4& L4 &L4&L4&L4&L4&L4&L4 \\
&RingMoE \cite{model-RingMoE-Bi} &L3	&L5	&L4	&L4	&L3	&L1& L4 &L3&L3&L3&L3&L4&L3 \\
&RingMo-Agent \cite{model-RingMo-Agent-Hu} &L3	&L4&L3	&L4	&L3	&L1& L3 &L2&L3&L2&L3&L3&L3 \\
&OlmoEarth \cite{model-OlmoEarth-Henry} &L4&L4&L4&L4	&L4	&L5& L4 &L4&L4&L4&L4&L4&L4 \\
&FlexiMo \cite{model-FlexiMo-Li} &L3	&L3&L3	&L4	&L3	&L2& L3 &L3&L3&L3&L3&L3&L3 \\
&MAPEX \cite{model-MAPEX-Hanna} &L3	&L4&L4	&L4	&L3	&L1& L3 &L3&L3&L3&L3&L3&L3 \\
&TESSERA \cite{model-tessera-Feng} &L4	&L4&L4	&L5	&L4	&L5& L4 &L4&L4&\textbf{L5}&L4&L4&L4 \\
&DINO-MM \cite{model-DINO-MM-Wang} &L3	&L3&L3	&L4	&L3	&L5& L3 &L4&L3&L4&L4&L3&L3 \\
&RS-VHEAT \cite{model-RS-vHeat-Hu} &L3	&L3&L3	&L5	&L3	&L5& L3 &L4&L3&L4&L4&L3&L4 \\
&DOFA \cite{model-DOFA-Xiong} &L3	&L4 &L4	&L5	&L4	&L5& L4 &L4&L4&L4&L4&L4&L4 \\
&msGFM \cite{model-msGFM-Han} &L3	&L4&L4	&L5	&L3	&L2& L3 &L3&L4&L3&L4&L4&L3 \\
&SkySense \cite{model-SkySense-Guo} &L4	&L4&L4	&L4	&L3	&L4& L4 &L4&L4&L4&L4&L4&L4 \\
&MTP \cite{model-MTP-Wang} &L3	&L3&L1	&L4	&L3	&L5& L3 &L3&L2&L4&L3&L3&L3 \\
&OFA-Net \cite{model-OFA-Net-Xiong} &L3	&L3&L3	&L4	&L2	&L1& L3 &L2&L3&L2&L3&L3&L2 \\
&OmniSat \cite{model-omnisat-Astruc} &L4	&L4&L3	&L4	&L3	&L2& L3 &L3&L3&L3&L3&L3&L3 \\
&SoftCon \cite{model-SoftCon-Wang} &L3&L4&L3	&L4	&L3	&L4& L3 &L3&L3&L4&L4&L4&L3 \\
&MURLF \cite{model-MURLF-Du} &L3&L3&L4	&L4	&L3	&L2& L3 &L3&L3&L3&L3&L3&L3 \\
&AlphaEarth \cite{model-AlphaEarth-F} &L3&L5&L5	&L5	&L4	&L1& L4 &L3&L4&L3&L4&L4&L4 \\
&Copernicus-FM \cite{model-Copernicus-FM-Wang} &L4&L3&L3	&L5	&L4	&L5& L4 &L4&L4&L4&L4&L4&L4 \\
&LESS-ViT \cite{model-LESS-ViT-Si} &L3&L3&L4	&L4	&L3	&L1& L3 &L3&L3&L2&L3&L3&L3 \\
&AllSpark \cite{model-AllSpark-Shao} &L3&L3&L2	&L4	&L2	&L2& L3 &L2&L2&L2&L3&L3&L2 \\
&RobSense \cite{model-RobSense-Do} &L3&L4&L4	&L4	&L3	&L2& L3 &L3&L3&L3&L3&L4&L3 \\
&OReole-FM \cite{model-OReole-FM-Dias} &L2&L4&L1	&L4	&L2	&L1& L2 &L2&L2&L2&L2&L3&L2 \\
&SkySense-V2 \cite{model-SkySense-V2-Zhang} &L3&L4&L4&L4	&L3	&L1& L3 &L3&L3&L3&L3&L3&L3 \\
&AnySat \cite{model-AnySat-Astruc} &L4&L3&L4	&L5	&L4	&L5& L4 &L4&L4&L4&L4&L4&L4 \\
&FoMo \cite{model-FoMo-Bountos} &L3&L3&L3&L5	&L2	&L5& L3 &L3&L3&L4&L4&L3&L3 \\
&RingMamba \cite{model-RingMamba-Wang} &L3&L4&L4&L4	&L3	&L1& L3 &L3&L3&L3&L3&L3&L3 \\
&SeaMo \cite{model-SeaMo-Li} &L3&L3&L4&L4&L3	&L5& L3 &L4&L4&L4&L4&L3&L4 \\
&MoSAiC \cite{model-MoSAiC-Gupta} &L2&L3&L3&L4	&L2	&L1& L2 &L2&L3&L2&L3&L3&L2 \\
&SkySense++ \cite{model-SkySense++-Wu} &L4&L5&L4&L5	&L4	&L5& L4 &L4&L4&\textbf{L5}&\textbf{L5}&\textbf{L5}&L4 \\
&Galileo \cite{model-Galileo-Tseng} &L4&L3&L4&L5	&L4	&L5& L4 &L4&L4&L4&L4&L4&L4 \\
&Panopticon \cite{model-Panopticon-Leonard} &L3&L3&L4&L5	&L4	&L5& L4 &L4&L4&L4&L4&L4&L4 \\
&Prithvi-EO-2.0 \cite{model-Prithvi-EO-2.0-Daniela} &L4&L4&L2&L5	&L3	&L5& L4 &L4&L3&L4&L4&L4&L3 \\
&THOR \cite{model-THOR-Theodor} &L3&L4&L4&L4	&L3	&L5& L4 &L4&L4&L4&L4&L4&L4 \\
&RAMEN \cite{model-RAMEN-Nicolas} &L3&L3&L3&L5	&L4	&L5& L4 &L4&L4&L4&L4&L3&L4 \\
&MaRS \cite{model-MaRS-Yang} &L3&L3&L4&L4	&L3	&L5& L3 &L4&L4&L4&L4&L3&L4 \\
&SpectralEarth-FM \cite{model-SpectralEarth-FM-Ait} &L2&L3&L4&L5	&L3	&L1& L3 &L3&L3&L3&L3&L3&L3 \\
&FLORO \cite{model-FLORO-Jorge} &L2&L4&L3&L3	&L4	&L5& L3 &L4&L4&L4&L3&L4&L4 \\
&RingMo-SAM \cite{model-RingMo-SAM-Yan} &L3&L3&L3&L3	&L2	&L5& L3 &L3&L3&L4&L3&L3&L3 \\
\hline
\multirow{14}{*}{\makecell{SAR-VFM}}&SARFormer \cite{model-SARFormer-Prexl} &L3	&L4	&L5	&L3	&L3	&L1& L3 &L3&L4&L3&L3&L4&L3 \\
&SUMMIT \cite{model-SUMMIT-Du} &L3	&L3	&L5	&L4	&L4	&L5& L4 &L4&L4&L4&L4&L4&L4 \\
&LSMA-SAR \cite{model-LSMA-SAR-Allen} &L3	&L3	&L4	&L3	&L2	&L1& L3 &L2&L3&L2&L3&L3&L2 \\
&SAMBA \cite{model-SAMBA-Wang} &L2	&L3	&L4	&L5	&L3	&L1& L3 &L3&L3&L3&L3&L3&L3 \\
&SAR-MAE \cite{model-SARMAE-Liu} &L3	&L4	&L4	&L4	&L4	&L5& L4 &L4&L4&L4&L4&L4&L4 \\
&PICV \cite{model-PICV-Wang} &L3	&L3	&L5	&L3	&L4	&L1& L3 &L3&L4&L3&L3&L3&L3 \\
&SAR-W-MixMAE \cite{model-SAR-W-MixMAE-Ali} &L2	&L3	&L4	&L4	&L4	&L5& L3 &L4&L4&L4&L4&L3&L4 \\
&SARATR-X \cite{model-SAR-ATR-X-Li} &L4	&L3	&L4	&L4	&L4	&L5& L4 &L4&L4&L4&L4&L4&L4 \\
&SARATR-X-V2 \cite{model-SARATR-X-v2-Li} &L2	&L4	&L4	&L5	&L5	&L5& L4 &\textbf{L5}&L4&\textbf{L5} &L4&L4&\textbf{L5} \\
&SAR-JEPA \cite{model-SAR-JEPA-Li} &L4	&L4	&L4	&L4	&L4	&L5& L4 &L4&L4&L4&L4&L4&L4 \\
&CrossEarth-SAR \cite{model-CrossEarth-SAR-Ye} &L2	&L4	&L5	&L5	&L4	&L5& L4 &L4&L4&\textbf{L5}&L4&L4&L4 \\
&Nereus-SAR-1 \cite{model-Nereus-SAR-1-Thomas} &L3	&L3	&L4	&L3	&L4	&L4& L4 &L4&L4&L4&L4&L3&L4 \\
&WV-Net \cite{model-WV-Net-Glaser} &L3	&L2	&L4	&L3	&L4	&L4& L3 &L4&L4&L4&L4&L3&L4 \\
\bottomrule
\end{tabular}

\end{table*}
\subsection{Model Rating Case Demonstration}
To provide an intuitive illustration of our evaluation process for a specific model, we take VHM \cite{model-VHM-Pang} as an example and present the complete procedure, from item-level scoring to the final rating. First, because VHM explicitly utilizes the SAR dataset MSAR \cite{dataset-MSAR-Chen,dataset-MSAR-2-Xia} during training and is designed as a VLM for multiple types of RS data, we classify it as an RS-VLM with SAR and evaluate it according to the VLM criteria listed in Table \ref{tab:VLM-bench2}. We then score VHM item by item, with the rationale behind each score presented in Table \ref{tab:vhm_evaluation}.

Next, we apply the weights specified in Table \ref{tab:VLM-bench2} to the individual evaluation items to obtain the dimension score for each dimension, as shown in Table \ref{tab:weight}. We then use the user-selected weight allocation scheme to assign weights to the scores of these six dimensions, yielding the model’s overall rating under that scheme. Finally, this rating is compared with the results of other VLMs evaluated under the same weight allocation scheme, thereby producing a model ranking tailored to the specific requirements.
\begin{table*}[t]
\centering
\caption{Evaluation Results for VHM Under the Proposed Rating Framework.}
\label{tab:vhm_evaluation}
\renewcommand{\arraystretch}{1.15}
\setlength{\tabcolsep}{5pt}
\small
\begin{tabularx}{\textwidth}{
    @{}
    >{\raggedright\arraybackslash}p{0.21\textwidth}
    >{\centering\arraybackslash}p{0.07\textwidth}
    X
    @{}
}
\toprule
\textbf{Evaluation Items} &
\textbf{Scores} &
\textbf{Justification} \\
\midrule

Publication recency &
L4 &
Published 1 year and 4 months ago. \\

Publication venue ranking &
L4 &
Published in the Proceedings of AAAI 2025 (leading conference). \\

Citation count &
L3 &
Google citations: 120 (as of our statistics cutoff). \\

Community popularity &
L3 &
GitHub repository stars: 125 (as of our statistics cutoff). \\
\hline
Architecture paradigm &
L4 &
Extends the standard CLIP-projector-Vicuna architecture with multi-level visual feature fusion, but introduces no fundamentally new backbone or interaction paradigm. \\

Parameter scale&
L4 &
Utilizes Vicuna-v1.5-7B together with CLIP-Large and projection modules. The main backbone falls within the 1-10 B range. \\


Training paradigm &
L4 &
Employs a two-stage training approach with RS adaptation and multitask instruction tuning, using captioning, dialogue, reasoning, and task-specific data, which is a common paradigm.\\

Multimodal capability &
L5 &
Supports image-level, region-level, and geometry-level interactions, as well as multi-turn dialogue and unified handling of multiple RS tasks. \\

\hline

Degree of SAR-specific design &
L1 &
Contains no SAR-specific encoder, polarization or phase branch, complex-valued module, scattering module, or dedicated SAR adapter. \\

SAR physical characteristic handling &
L1 &
Learns SAR modality labels but does not explicitly model speckle, backscatter, polarization, phase, incidence angle, layover, or radar shadow. \\

Proportion of SAR training data &
L2 &
SAR data appear only in a small mixed modality-recognition subset. The confirmed SAR sample count is below 10 K and is not dominant. \\
\hline
Training data Scale &
L4 &
Uses approximately 1.4 M RS images and additional instruction data, placing the total scale within the 1 M--10 M range. \\

Data diversity &
L4 &
Covers multiple sensors, spatial resolutions, and geographic environments, but still lacks global data coverage. \\

Data accessibility &
L5 &
All training data are publicly available. \\

\hline

SAR task coverage breadth &
L2 &
SAR-related evaluation is largely limited to modality recognition, without systematic validation on SAR interpretation tasks. However, it performs prominently in image modality recognition. \\

Rigor of comparisons \& baselines &
L4 &
Includes broad baseline comparisons, multitask evaluations, and ablation studies, but lacks SAR-specific quantitative breakdowns. \\

Experimental reproducibility &
L4 &
Reports the main training configurations and computational resources, while exact data splits, random seeds, and complete checkpoints are not fully specified. \\

Zero-shot capability on SAR tasks&
L3 &
Correctly identifies SAR image modalities, but zero-shot interpretation on unseen SAR sensor data, SAR characteristics, and traditional SAR tasks remains unevaluated. \\
\hline
Open-source level&
L5 &
Source code, pretrained weights, and training scripts are all publicly available.\\

\bottomrule
\end{tabularx}

\vspace{2pt}
\end{table*}

\begin{table*}[htbp]
\centering
\caption{Weighted Dimension Score Results for VHM.}
\label{tab:weight}
\begin{tabular}{llccc}
\hline
\textbf{Dimensions}&\textbf{Evaluation Items} & \textbf{Scores} & \textbf{Weight} & \textbf{Dimension Scores} \\
\hline
\multirow{4}{*}{Influence}&Publication recency &
L4 &0.25&\multirow{4}{*}{L4}\\

&Publication venue ranking &
L4 & 0.25&\\

&Citation count &
L3 &0.25&\\

&Community popularity &
L3 & 0.25&\\
\hline
\multirow{4}{*}{Technical Architecture}&Architecture paradigm &
L4 & 0.25&\multirow{4}{*}{L4}\\

&Parameter scale&
L4 &0.25&\\

&Training paradigm &
L4 &0.25&\\

&Multimodal capability &
L5 &0.25&\\

\hline

\multirow{3}{*}{SAR Domain Adaptation}&Degree of SAR-specific design &
L1 &0.4&\multirow{3}{*}{L1}\\

&SAR physical characteristic handling &
L1 &0.3&\\

&Proportion of SAR training data &
L2 & 0.3&\\
\hline
\multirow{3}{*}{Training Data}&Training data Scale &
L4 &0.4 &\multirow{3}{*}{L4}\\

&Data diversity &
L4 &0.3&\\

&Data accessibility &
L5 & 0.3&\\

\hline

\multirow{4}{*}{SAR Benchmark Task Performance}&SAR task coverage breadth &
L2 & 0.2&\multirow{4}{*}{L3}\\

&Rigor of comparisons \& baselines &
L4 & 0.3&\\

&Experimental reproducibility &
L4 &0.2&\\

&Zero-shot capability on SAR tasks&
L3 &0.3&\\
\hline
\multirow{1}{*}{Open-Sourcing and Reproducibility}&Open-source level&
L5 &1.0 &\multirow{1}{*}{L5}\\
\hline
\end{tabular}
\end{table*}

\subsection{Feasibility and Stability of MLLM-assisted Rating Analysis}
\begin{table*}[htbp]
\centering
\caption{Item-level Scores of SARVLM Assigned by MLLMs and Human Reviewers (SD=standard deviation).}
\label{tab:human-and-MLLM}
\resizebox{\textwidth}{!}{%
\begin{tabular}{l|>{\centering\arraybackslash}m{1.5cm}>{\centering\arraybackslash}m{1.5cm}>{\centering\arraybackslash}m{1.5cm}>{\centering\arraybackslash}m{1.5cm}>{\centering\arraybackslash}m{1.5cm}>{\centering\arraybackslash}m{1.5cm}|c|cccc|c}
\toprule
Evaluation Items & Claude-Opus-5.5 & Gemini-3.7-Flash & GPT-6-Sol & Claude-Opus-4.8& GPT-5.6-Sol & DeepSeek-V4-Pro & Mean ± SD & Subject 1 & Subject 2 & Subject 3 & Subject 4 & Mean ± SD \\
\midrule
Architecture paradigm & L3 & L3 & L3 & L2 & L3 & L3 & 2.83 $\pm$  0.41& L3 & L3 & L3 & L3 & 3.00 $\pm$ 0.00 \\
Training paradigm & L4 & L3 & L4 & L3 & L4 & L4 & 3.67 $\pm$ 0.52 & L4 & L3 & L4 & L4 & 3.75 $\pm$  0.50 \\
Multimodal capability & L3 & L4 & L3 & L3 & L4 & L3 & 3.33 $\pm$ 0.52 & L4 & L4 & L4 & L4 & 4.00 $\pm$ 0.00 \\
Degree of SAR-specific design  & L4 & L4 & L4 & L4 & L4 & L4 & 4.00 $\pm$ 0.00 & L3 & L3 & L3 & L3 & 3.00 $\pm$ 0.00 \\
SAR physical characteristic handling & L3 & L2 & L2 & L2 & L3 & L4 & 2.67 $\pm$ 0.82 & L2 & L3 & L2 & L2 & 2.25 $\pm$ 0.50 \\
Proportion of SAR training data & L5 & L4 & L5 & L5 & L5 & L5 & 4.83 $\pm$ 0.41 & L5 & L5 & L5 & L5 & 5.00 $\pm$ 0.00\\
Data diversity & L5 & L5 & L4 & L4 & L5 & L5 & 4.67 $\pm$ 0.52 & L5& L4 & L4 & L4 & 4.25 $\pm$ 0.50 \\
SAR task coverage breadth & L5 & L5 & L5 & L5 & L5 & L5 & 5.00 $\pm$ 0.00 & L4 & L5 & L4 & L5 & 4.50 $\pm$ 0.58 \\
Rigor of comparisons \& baselines & L4 & L5 & L4 & L4 & L4 & L5 & 4.33 $\pm$ 0.52 & L4 & L4 & L3 & L4 & 3.75 $\pm$  0.50 \\
Experimental reproducibility & L4 & L4 & L4 & L4 & L4 & L4 & 4.00 $\pm$ 0.00& L3 & L4 & L4 & L3 & 3.50 $\pm$ 0.58 \\
Zero-shot capability on SAR tasks & L4 & L4 & L4 & L5 & L4 & L4 & 4.17 $\pm$ 0.41 & L4 & L3 & L4 & L4 & 3.75 $\pm$ 0.50 \\
\bottomrule
\end{tabular}%
}

\end{table*}

Because the proposed rating tool includes a number of qualitative indicators, the resulting dimension-level and overall ratings are inevitably subject to a degree of subjectivity. Our rating procedure combines MLLM assistance with human review, and these two stages introduce the subjectivity of the MLLMs and of the human reviewers, respectively. To investigate how strongly such subjectivity affects the ratings, and how clearly the boundaries between adjacent levels of the proposed criteria are defined, we select the items that are affected by subjectivity and ask six recently released MLLMs and four scholars working on SAR foundation models to score these items for the same model. The evaluated model is SARVLM \cite{model-SARVLM-Ma}. The MLLMs are Claude-Opus-5.5, Gemini-3.7-Flash, GPT-6-Sol, Claude-Opus-4.8, GPT-5.6-Sol, and DeepSeek-V4-Pro. The prompt used is: “Please rate the model on the following dimensions according to the criteria below:” followed by the scoring criteria for each item. None of the participating scholars is an author of this paper. The selected qualitative items and the resulting scores are presented in Table \ref{tab:human-and-MLLM}. The remaining criteria are quantitative and can be assigned clear and consistent rating levels based on the information reported in the original papers; therefore, no additional subjective-rating analysis is conducted for these criteria. As can be seen, most items receive identical or similar ratings across the different MLLMs, with differences of no more than one level. Moreover, since a dimension score follows from weighting several items, the effect of a discrepancy in any single item on the resulting dimension score is further reduced.

Notably, SAR physical characteristic handling is the item that exhibits the largest scoring discrepancy, which stems from differences in how the MLLMs construe whether a model handles SAR characteristics in a targeted manner. MLLMs that apply strict criteria regard explicit modeling as a mandatory requirement for the item to count, whereas more permissive MLLMs also count cases in which part of the evaluated model’s design accounts for SAR characteristics or mitigates their effects (e.g., mitigating the impact of speckle noise on model performance). Overall, MLLM-assisted scoring improves rating efficiency and achieves good consistency. In future work, we consider integrating the MLLM into the proposed tool and fixing the prompt input used for rating, so as to keep the criteria consistent.

Among the human scores, the item Degree of SAR-specific design shows the most pronounced gap with the MLLMs. Because the evaluated model lacks an explicitly proposed SAR-specific module, the human raters tend to judge that it satisfies “without SAR-specific architectural improvements” at level L3. The MLLMs, by contrast, consider that the model supports joint optical–SAR modeling and training, and that it designs a training method to bridge the gap between the two domains. They therefore rate it L4 on this item. Overall, the discrepancy between human and MLLM ratings falls within an acceptable range, which indicates that the proposed criteria have good hierarchical separability.

Although human scoring and MLLM scoring each perform reliably on their own, we believe that combining the two can further improve rating accuracy. For some evaluated models, when the model defies accurate description by either of two adjacent rating criteria, the MLLM returns a recommended rating range, which a human must adjudicate on the basis of experience. In addition, for some dimensions the MLLM cannot reach a judgment from the information available in the paper, as is the case for experimental reproducibility. Some MLLMs note that the paper itself omits certain experimental details (e.g., the number of training epochs, the learning rate) and that providing them would improve the rating. However, the evaluated model may supply the missing details in its corresponding GitHub repository, which requires manual re-checking. Furthermore, compared with the MLLMs, human scoring is more prone to undercounting counting-type items (e.g., SAR task coverage breadth), whereas the MLLMs can enumerate the relevant content fairly exhaustively and thereby assist humans in judging quantities. For these reasons, all evaluation results reported in this paper come from MLLM assistance combined with human review and adjudication.

\begin{figure*}[htbp]
\centering
\includegraphics[width=0.70\textwidth]{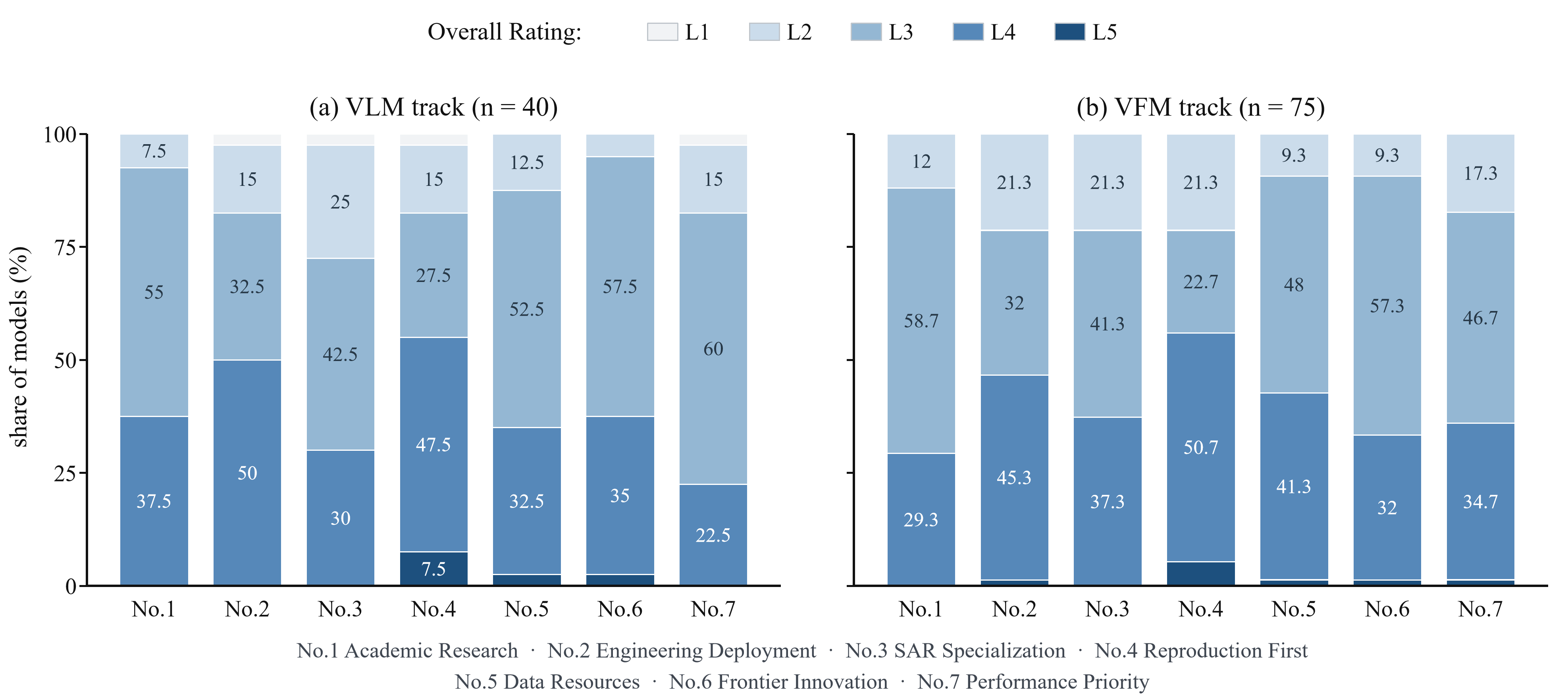}
\caption{Distribution of evaluated models across rating levels under the seven predefined weight allocation schemes.}
\label{fig7}
\end{figure*}

\subsection{Discriminative Power and Reliability of the Predefined Weight Allocation Schemes}
To examine whether the predefined weight allocation schemes are sufficiently discriminative, we visualize how the evaluated models are distributed across rating levels under each assignment, as shown in Fig. \ref{fig7}. No two schemes produce closely matching distributions. The differences between schemes are systematic rather than incidental, indicating that each scheme shifts the emphasis of the rating toward the requirements it is designed to foreground, and that the seven schemes are therefore well separated from one another. Across all schemes the distributions are sparse at both ends and fullest in the middle, with overall ratings concentrated between L2 and L4. This indicates that the proposed criteria provide a reasonably wide grading space across models, while the small number of L5 ratings leaves room at the top of the scale for more advanced foundation models to be rated later.

\section{Conclusion}

In this paper, we presented a systematic taxonomy and hierarchical evaluation framework for SAR-related foundation models. The proposed taxonomy organizes existing SAR-related models according to their modality, application domain, and model paradigm, covering RS-VFMs without SAR, RS-VFMs with SAR, RS-VLMs without SAR, RS-VLMs with SAR, SAR-VFMs, and SAR-VLMs. To provide a comprehensive assessment beyond task-specific performance, we established an L1--L5 hierarchical rating framework with six evaluation dimensions and multiple evaluation items.

Considering the fundamental differences between VLMs and VFMs in terms of their input modalities, output formats, pretraining objectives, and evaluation protocols, we designed two independent evaluation tracks for VFMs and VLMs. In addition, we introduced a dynamic weight allocation strategy based on user profiling to reflect different requirements from academic researchers and engineering practitioners. This strategy enables the final rating and ranking of models to be adapted to different application scenarios, such as academic research, performance priority, reproducibility first, and engineering deployment.

\subsection{Overall Insights}
Using the proposed framework, we rate 115 representative SAR-related foundation models, including 40 VLMs and 75 VFMs. The rating demonstrates that no single model category is uniformly superior across all dimensions. In the VLM track, RS models incorporating SAR data can remain competitive with SAR-specific models because their broader multimodal training and linguistic knowledge support semantic transfer across domains. In the VFM track, however, direct exposure to SAR imagery during pretraining plays a more decisive role, highlighting the importance of modality-specific visual representation learning.

The absence of a clear advantage for SAR-VLMs can be traced to the item level. On the items that rate parameter scale and training data scale, SAR-VLMs score markedly below RS-VLMs with SAR, a deficit that plausibly accounts for the absence of a downstream advantage. Because SAR acquisition is subject to release restrictions and annotation requires specialist interpretation, high-quality openly available annotated SAR data remain scarce, and this scarcity, rather than any weakness in SAR-native design, is what limits the scale at which SAR-VLMs can be built.

More broadly, the results indicate that SAR capability is not determined by any single factor. It emerges from the interaction among data composition, architectural design, pretraining objectives, cross-modal knowledge transfer, and the extent to which SAR imaging characteristics are explicitly considered. The increasing use of SAR-aware and physics-informed learning strategies therefore represents an important transition from the direct reuse of general visual pretraining methods toward more specialized SAR foundation models.

We also examine the rating instrument itself. Its levels prove hierarchical and discriminative, and ratings assigned by 6 different MLLMs and by 4 different practitioners agree well. The seven predefined weight assignments produce clearly separated level distributions, so that distinct requirements are reflected as corresponding shifts in emphasis rather than as noise. Under every allocation scheme the distribution is sparse at both ends and fullest between L2 and L4, which indicates a grading space wide enough to separate current models while leaving room at the top for more advanced ones.

The analysis also shows that strong task capability does not necessarily imply high reproducibility or deployment readiness. Limited access to training data, source code, pretrained weights, and complete implementation details remains a major obstacle to independent verification and practical adoption. Evaluating openness and reproducibility alongside technical performance is therefore necessary for a more realistic understanding of model maturity.

\subsection{Limitations}

The proposed framework has several limitations. First, although most criteria are based on explicit evidence, some qualitative items still require expert interpretation. The combination of MLLM-assisted scoring and human review improves efficiency and consistency, but it cannot completely eliminate evaluator subjectivity. Second, the ratings depend on publicly available information. Incomplete reporting in papers or repositories may result in conservative scores, even when additional implementation details exist but are not publicly accessible. The framework consequently evaluates not only the underlying capability of a model, but also the transparency of the evidence available to support that capability. Third, the current rating results provide a comparative meta-evaluation rather than a substitute for controlled benchmark experiments. Differences in downstream datasets, task definitions, training protocols, and reporting practices continue to restrict direct performance comparisons among some models. In addition, several operational factors, including inference latency, energy consumption, long-term stability, security, and maintenance cost, cannot yet be evaluated consistently because such information is rarely reported. Finally, the predefined weight allocation schemes represent typical user priorities but cannot cover every application scenario. Model selection for a specific operational system may require customized weights and additional constraints.

\subsection{Future works}

The rating tool will be released as a public website, and we intend to maintain and extend it on a continuing basis. The planned work proceeds along the following directions.

\textbf{Integrating the auxiliary MLLM into the tool and exposing item-level ratings:} We plan to embed the auxiliary MLLM directly into the evaluation tool through a fixed prompt, so that the criteria are applied more consistently and the workflow becomes more reproducible. In addition to the current dimension-level outputs, we also plan to release the fine-grained item-level scores on which those outputs are based.

\textbf{Monitoring model open-sourcing status continuously:} The weights of many recently released models are still in the process of being opened and may become publicly available in the future. We will therefore continue to monitor the release status of currently unopened models and dynamically update the ratings for the open-sourcing and reproducibility dimension as new resources appear.

\textbf{Complementing meta-analysis with experimental benchmarks:} The SAR benchmark task performance dimension is currently rated through meta-analysis. Although this approach is cost-efficient, it cannot fully substitute for experimental benchmarking. We plan to develop comprehensive experimental benchmarks covering both VFMs and VLMs and to integrate them into the rating tool, so that users can choose among multiple rating options for this dimension.

\textbf{Scaling the evaluated model set:} This paper evaluates and analyzes only a limited number of representative models. Because the framework is designed to be extensible, in-scope existing models as well as newly released ones can be incorporated efficiently under the same taxonomy and weight allocation schemes, allowing the community to track progress along multiple dimensions simultaneously.

We hope that the proposed taxonomy and rating framework will help different groups of users make defensible, requirement-aware model selection decisions, thereby contributing to a more transparent and sustainable development of SAR-related foundation models.


\bibliographystyle{IEEEtran}
\bibliography{reference}

@ARTICLE{survey-fifty-zhou,
  author={Zhou, Jie and Liu, Yongxiang and Liu, Li and Li, Weijie and Yang, Taoli and Peng, Bowen and Song, Yafei and Kuang, Gangyao and Li, Xiang},
  journal={IEEE Geoscience and Remote Sensing Magazine}, 
  title={Fifty Years of Synthetic Aperture Radar Automatic Target Recognition: The road forward}, 
  year={2026},
  volume={},
  number={},
  pages={2-34}}

@ARTICLE{survey-A-tutorial-Alberto,
  author={Moreira, Alberto and Prats-Iraola, Pau and Younis, Marwan and Krieger, Gerhard and Hajnsek, Irena and Papathanassiou, Konstantinos P.},
  journal={IEEE Geoscience and Remote Sensing Magazine}, 
  title={A tutorial on synthetic aperture radar}, 
  year={2013},
  volume={1},
  number={1},
  pages={6-43}}

@article{survey-Landslide-Nicola,
  author  = {Casagli, Nicola and Intrieri, Emanuele and Tofani, Veronica and Gigli, Giovanni and Raspini, Federico},
  title   = {Landslide detection, monitoring and prediction with remote-sensing techniques},
  journal = {Nature Reviews Earth \& Environment},
  volume  = {4},
  pages   = {51--64},
  year    = {2023},
  month   = jan}

@ARTICLE{survey-Urban-flood-Jie,
  author={Zhao, Jie and Li, Ming and Li, Yu and Matgen, Patrick and Chini, Marco},
  journal={IEEE Geoscience and Remote Sensing Magazine}, 
  title={Urban Flood Mapping Using Satellite Synthetic Aperture Radar Data: A review of characteristics, approaches, and datasets}, 
  year={2025},
  volume={13},
  number={1},
  pages={237-268}}

@article{survey-Earthquake-Eleanor,
  author = {Ainscoe, Eleanor A. and Swaminathan, Rohini and Way, Lin and Modugno, Sirio and Chin, Shi Tong and Panta, Niroj and Crevoisier, Thierry and Yun, Sang-Ho},
  title = {Earthquake damage mapped more comprehensively and accurately by radar satellites than optical imagery},
  journal = {Communications Earth \& Environment},
  volume = {6},
  number = {1},
  pages = {631},
  year = {2025},
  month = {August},
  day = {05},
  note = {Article number: 631}
}

@ARTICLE{survey-recent-advance-Ping,
  author={Lang, Ping and Fu, Xiongjun and Dong, Jian and Yang, Huizhang and Yin, Junjun and Yang, Jian and Martorella, Marco},
  journal={IEEE Journal of Selected Topics in Applied Earth Observations and Remote Sensing}, 
  title={Recent Advances in Deep-Learning-Based SAR Image Target Detection and Recognition}, 
  year={2025},
  volume={18},
  number={},
  pages={6884-6915}}

@article{survey-oil-spill-Brekke,
title = {Oil spill detection by satellite remote sensing},
journal = {Remote Sensing of Environment},
volume = {95},
number = {1},
pages = {1-13},
year = {2005},
issn = {0034-4257},
author = {Camilla Brekke and Anne H.S. Solberg}
}

@ARTICLE{survey-SAR4Ocan-Asiyabi,
  author={Asiyabi, Reza Mohammadi and Ghorbanian, Arsalan and Tameh, Shaahin Nazarpour and Amani, Meisam and Jin, Shuanggen and Mohammadzadeh, Ali},
  journal={IEEE Journal of Selected Topics in Applied Earth Observations and Remote Sensing}, 
  title={Synthetic Aperture Radar (SAR) for Ocean: A Review}, 
  year={2023},
  volume={16},
  number={},
  pages={9106-9138}}

@article{survey-SARbuilding-Ge,
title = {A review on synthetic aperture radar-based building damage assessment in disasters},
journal = {Remote Sensing of Environment},
volume = {240},
pages = {111693},
year = {2020},
issn = {0034-4257},
author = {Pinglan Ge and Hideomi Gokon and Kimiro Meguro}
}

@ARTICLE{survey-Deep-Learning4sar-Liao,
  author={Liao, Jinrui and Wang, Qingsong and Huang, Zhongling and Yang, Chengran and Lai, Tao and Huang, Haifeng},
  journal={IEEE Geoscience and Remote Sensing Magazine}, 
  title={Deep Learning for Synthetic Aperture Radar Automatic Target Recognition in Complex Environmental Interference: A review}, 
  year={2026},
  volume={},
  number={},
  pages={2-49}}

@ARTICLE{survey-A-Review-Yin,
  author={Yin, Junjun and Duan, Changxian and Wang, Hongbo and Yang, Jian},
  journal={IEEE Journal of Selected Topics in Applied Earth Observations and Remote Sensing}, 
  title={A Review on the Few-Shot SAR Target Recognition}, 
  year={2024},
  volume={17},
  number={},
  pages={16411-16425}}

@article{survey-UDA-Shi,
title = {Unsupervised domain adaptation for SAR target classification based on domain- and class-level alignment: From simulated to real data},
journal = {ISPRS Journal of Photogrammetry and Remote Sensing},
volume = {207},
pages = {1-13},
year = {2024},
issn = {0924-2716},
author = {Yu Shi and Lan Du and Chen Li and Yuchen Guo and Yuang Du}
}

@ARTICLE{survey-Transfer-Adaption-Yang,
  author={Yang, Xinpeng and Jiao, Lianmeng and Pan, Quan},
  journal={IEEE Journal of Selected Topics in Applied Earth Observations and Remote Sensing}, 
  title={Transfer Adaptation Learning for Target Recognition in SAR Images: A Survey}, 
  year={2024},
  volume={17},
  number={},
  pages={13577-13601}}

@article{survey-Toward-generalizing-Liao,
title = {Toward generalizing to unseen domains in remote sensing: A survey},
journal = {Neurocomputing},
volume = {697},
pages = {134238},
year = {2026},
issn = {0925-2312},
author = {Muxin Liao and Meiying Liao and Zhihong Lu and Yuting Sun and Yinglong Wang and Yingqiong Peng}
}

@misc{benchmark-SARChat-Bench-2M-Ma,
      title={SARChat-Bench-2M: A Multi-Task Vision-Language Benchmark for SAR Image Interpretation}, 
      author={Zhiming Ma and Xiayang Xiao and Sihao Dong and Peidong Wang and HaiPeng Wang and Qingyun Pan},
      year={2025},
      eprint={2502.08168},
      archivePrefix={arXiv},
      primaryClass={cs.CL}
}

@ARTICLE{model-RingMoE-Bi,
  author={Bi, Hanbo and Feng, Yingchao and Tong, Boyuan and Wang, Mengyu and Yu, Haichen and Mao, Yongqiang and Chang, Hao and Diao, Wenhui and Wang, Peijin and Yu, Yue and Peng, Hanyang and Zhang, Yehong and Fu, Kun and Sun, Xian},
  journal={IEEE Transactions on Pattern Analysis and Machine Intelligence}, 
  title={RingMoE: Mixture-of-Modality-Experts Multi-Modal Foundation Models for Universal Remote Sensing Image Interpretation}, 
  year={2026},
  volume={48},
  number={4},
  pages={4388-4405}}

@ARTICLE{model-Ringmo-Xian,
  author={Sun, Xian and Wang, Peijin and Lu, Wanxuan and Zhu, Zicong and Lu, Xiaonan and He, Qibin and Li, Junxi and Rong, Xuee and Yang, Zhujun and Chang, Hao and He, Qinglin and Yang, Guang and Wang, Ruiping and Lu, Jiwen and Fu, Kun},
  journal={IEEE Transactions on Geoscience and Remote Sensing}, 
  title={RingMo: A Remote Sensing Foundation Model With Masked Image Modeling}, 
  year={2023},
  volume={61},
  number={},
  pages={1-22}}

@ARTICLE{model-SAR-ATR-X-Li,
  author={Li, Weijie and Yang, Wei and Hou, Yuenan and Liu, Li and Liu, Yongxiang and Li, Xiang},
  journal={IEEE Transactions on Image Processing}, 
  title={SARATR-X: Toward Building a Foundation Model for SAR Target Recognition}, 
  year={2025},
  volume={34},
  number={},
  pages={869-884}}

@misc{model-FUSAR-KLIP-Yi,
      title={FUSAR-KLIP: Towards Multimodal Foundation Models for Remote Sensing}, 
      author={Yi Yang and Xiaokun Zhang and Qingchen Fang and Jing Liu and Ziqi Ye and Rui Li and Li Liu and Haipeng Wang},
      year={2026},
      eprint={2509.23927},
      archivePrefix={arXiv},
      primaryClass={cs.CV}
}

@misc{model-falcon-Yao,
      title={Falcon: A Remote Sensing Vision-Language Foundation Model (Technical Report)}, 
      author={Kelu Yao and Nuo Xu and Rong Yang and Yingying Xu and Zhuoyan Gao and Titinunt Kitrungrotsakul and Yi Ren and Pu Zhang and Jin Wang and Ning Wei and Chao Li},
      year={2025},
      eprint={2503.11070},
      archivePrefix={arXiv},
      primaryClass={cs.CV}
}

@InProceedings{model-FUSAR-GPT-Zhang,
    author    = {Zhang, Xiaokun and Yang, Yi and Ye, Ziqi and Baiyun, Baiyun and Guo, Xiaorong and Fang, Qingchen and Zhang, Ruyi and Zhou, Xinpeng and Wang, Haipeng},
    title     = {FUSAR-GPT: A Spatiotemporal Feature-Embedded and Two-Stage Decoupled Visual Language Model for SAR Imagery},
    booktitle = {Proceedings of the IEEE/CVF Conference on Computer Vision and Pattern Recognition (CVPR)},
    month     = {June},
    year      = {2026},
    pages     = {42158-42168}
}

@misc{model-SAR-TEXT-He,
      title={SAR-TEXT: A Large-Scale SAR Image-Text Dataset Built with SAR-Narrator and A Progressive Learning Strategy for Downstream Tasks}, 
      author={Yiguo He and Xinjun Cheng and Junjie Zhu and Chunping Qiu and Jun Wang and Xichuan Zhang and Qiangjuan Huang and Ke Yang},
      year={2025},
      eprint={2507.18743},
      archivePrefix={arXiv},
      primaryClass={cs.CV}
}

@article{model-SARCLIP-Jiang,
title = {SARCLIP: a multimodal foundation framework for SAR imagery via contrastive language-image pre-training},
journal = {ISPRS Journal of Photogrammetry and Remote Sensing},
volume = {231},
pages = {17-34},
year = {2026},
issn = {0924-2716},
author = {Chaowei Jiang and Chao Wang and Fan Wu and Peifeng Ma and Lichuan Zou and Tianyang Li and Jing Ning and Yixian Tang}
}

@article{model-VHM-Pang, 
title={VHM: Versatile and Honest Vision Language Model for Remote Sensing Image Analysis}, 
volume={39}, 
abstractNote={This paper develops a Versatile and Honest vision language Model (VHM) for remote sensing image analysis. VHM is built on a large-scale remote sensing image-text dataset with rich-content captions (VersaD), and an honest instruction dataset comprising both factual and deceptive questions (HnstD). Unlike prevailing remote sensing image-text datasets, in which image captions focus on a few prominent objects and their relationships, VersaD captions provide detailed information about image properties, object attributes, and the overall scene. This comprehensive captioning enables VHM to thoroughly understand remote sensing images and perform diverse remote sensing tasks. Moreover, different from existing remote sensing instruction datasets that only include factual questions, HnstD contains additional deceptive questions stemming from the non-existence of objects. This feature prevents VHM from producing affirmative answers to nonsense queries, thereby ensuring its honesty. In our experiments, VHM significantly outperforms various vision language models on common tasks of scene classification, visual question answering, and visual grounding. Additionally, VHM achieves competent performance on several unexplored tasks, such as building vectorizing, multi-label classification and honest question answering.},
number={6}, 
journal={Proceedings of the AAAI Conference on Artificial Intelligence}, 
author={Pang, Chao and Weng, Xingxing and Wu, Jiang and Li, Jiayu and Liu, Yi and Sun, Jiaxing and Li, Weijia and Wang, Shuai and Feng, Litong and Xia, Gui-Song and et al.}, 
year={2025}, 
month={Apr.}, 
pages={6381–6388} 
}

@inproceedings{benchmark-Choice-AN,
 author = {An, Xiao and Sun, Jiaxing and Gui, Zihan and He, Wei},
 booktitle = {Advances in Neural Information Processing Systems},
 editor = {D. Belgrave and C. Zhang and H. Lin and R. Pascanu and P. Koniusz and M. Ghassemi and N. Chen},
 pages = {},
 publisher = {Curran Associates, Inc.},
 title = {CHOICE: Benchmarking the Remote Sensing Capabilities of Large Vision-Language Models},
 volume = {38},
 year = {2025}
}

@ARTICLE{survey-Foundation-Models-Xiao,
  author={Xiao, Aoran and Xuan, Weihao and Wang, Junjue and Huang, Jiaxing and Tao, Dacheng and Lu, Shijian and Yokoya, Naoto},
  journal={IEEE Geoscience and Remote Sensing Magazine}, 
  title={Foundation Models for Remote Sensing and Earth Observation: A survey}, 
  year={2025},
  volume={13},
  number={4},
  pages={297-324}}

@ARTICLE{survey-Generative-Artificial-Intelligence-Huang,
  author={Huang, Zhongling and Zhang, Xidan and Tang, Zuqian and Xu, Feng and Datcu, Mihai and Han, Junwei},
  journal={IEEE Geoscience and Remote Sensing Magazine}, 
  title={Generative Artificial Intelligence Meets Synthetic Aperture Radar: A survey}, 
  year={2026},
  volume={14},
  number={1},
  pages={6-48}}

@article{benchmark-SAREval-Wang,
  title={SAREval: A Multi-Dimensional and Multi-Task Benchmark for Evaluating Visual Language Models on SAR Image Understanding},
  author={Wang, Z. and Liu, L. and Wan, G. and Lu, Y. and Zheng, F. and Sun, G. and Huang, Y. and Guo, S. and Li, X. and Yuan, L.},
  journal={Remote Sensing},
  volume={18},
  number={1},
  pages={82},
  year={2026},
  publisher={MDPI}
}

@ARTICLE{benchmark-SARLANG-1M-wei,
  author={Wei, Yimin and Xiao, Aoran and Ren, Yexian and Zhu, Yuting and Chen, Hongruixuan and Xia, Junshi and Yokoya, Naoto},
  journal={IEEE Transactions on Geoscience and Remote Sensing}, 
  title={SARLANG-1M: A Benchmark for Vision–Language Modeling in SAR Image Understanding}, 
  year={2026},
  volume={64},
  number={},
  pages={1-20}}

@ARTICLE{survey-VFM-in-RS-Lu,
  author={Lu, Siqi and Guo, Junlin and Zimmer-Dauphinee, James R. and Nieusma, Jordan M. and Wang, Xiao and VanValkenburgh, Parker and Wernke, Steven A. and Huo, Yuankai},
  journal={IEEE Geoscience and Remote Sensing Magazine}, 
  title={Vision Foundation Models in Remote Sensing: A survey}, 
  year={2025},
  volume={13},
  number={3},
  pages={190-215}}

@misc{survey-Open-Sourcing-Seger,
      title={Open-Sourcing Highly Capable Foundation Models: An evaluation of risks, benefits, and alternative methods for pursuing open-source objectives}, 
      author={Elizabeth Seger and Noemi Dreksler and Richard Moulange and Emily Dardaman and Jonas Schuett and K. Wei and Christoph Winter and Mackenzie Arnold and Sean O hEigeartaigh and Anton Korinek and Markus Anderljung and Ben Bucknall and Alan Chan and Eoghan Stafford and Leonie Koessler and Aviv Ovadya and Ben Garfinkel and Emma Bluemke and Michael Aird and Patrick Levermore and Julian Hazell and Abhishek Gupta},
      year={2023},
      eprint={2311.09227},
      archivePrefix={arXiv},
      primaryClass={cs.CY}
}

@misc{benchmark-OVRSISBenchV2-Li,
      title={Towards Realistic Open-Vocabulary Remote Sensing Segmentation: Benchmark and Baseline}, 
      author={Bingyu Li and Tao Huo and Haocheng Dong and Da Zhang and Zhiyuan Zhao and Junyu Gao and Xuelong Li},
      year={2026},
      eprint={2604.15652},
      archivePrefix={arXiv},
      primaryClass={cs.CV}
}

@ARTICLE{model-RingMo-Aerial-Diao,
  author={Diao, Wenhui and Yu, Haichen and Kang, Kaiyue and Ling, Tong and Liu, Di and Feng, Yingchao and Bi, Hanbo and Ren, Libo and Li, Xuexue and Mao, Yongqiang and Sun, Xian},
  journal={IEEE Transactions on Pattern Analysis and Machine Intelligence}, 
  title={RingMo-Aerial: An Aerial Remote Sensing Foundation Model With Affine Transformation Contrastive Learning}, 
  year={2025},
  volume={47},
  number={12},
  pages={10900-10913}}

@ARTICLE{model-RingMo-Galaxy-Wang,
  author={Wang, Zhechao and Wang, Zhirui and Cheng, Peirui and Zhao, Liangjin and Tian, Pengju and Wang, Yuchao and Chen, Mingxin and Li, Xinming and Sun, Xian},
  journal={IEEE Transactions on Geoscience and Remote Sensing}, 
  title={RingMo-Galaxy: A Remote Sensing Distributed Foundation Model for Diverse Downstream Tasks}, 
  year={2025},
  volume={63},
  number={},
  pages={1-18}}

@misc{model-Prithvi-Johannes,
      title={Foundation Models for Generalist Geospatial Artificial Intelligence}, 
      author={Johannes Jakubik and Sujit Roy and C. E. Phillips and Paolo Fraccaro and Denys Godwin and Bianca Zadrozny and Daniela Szwarcman and Carlos Gomes and Gabby Nyirjesy and Blair Edwards and Daiki Kimura and Naomi Simumba and Linsong Chu and S. Karthik Mukkavilli and Devyani Lambhate and Kamal Das and Ranjini Bangalore and Dario Oliveira and Michal Muszynski and Kumar Ankur and Muthukumaran Ramasubramanian and Iksha Gurung and Sam Khallaghi and Hanxi and Li and Michael Cecil and Maryam Ahmadi and Fatemeh Kordi and Hamed Alemohammad and Manil Maskey and Raghu Ganti and Kommy Weldemariam and Rahul Ramachandran},
      year={2023},
      eprint={2310.18660},
      archivePrefix={arXiv},
      primaryClass={cs.CV}
}

@ARTICLE{model-RingMo-sense-Yao,
  author={Yao, Fanglong and Lu, Wanxuan and Yang, Heming and Xu, Liangyu and Liu, Chenglong and Hu, Leiyi and Yu, Hongfeng and Liu, Nayu and Deng, Chubo and Tang, Deke and Chen, Changshuo and Yu, Jiaqi and Sun, Xian and Fu, Kun},
  journal={IEEE Transactions on Geoscience and Remote Sensing}, 
  title={RingMo-Sense: Remote Sensing Foundation Model for Spatiotemporal Prediction via Spatiotemporal Evolution Disentangling}, 
  year={2023},
  volume={61},
  number={},
  pages={1-21}}

@InProceedings{model-OlmoEarth-Henry,
    author    = {Herzog, Henry and Bastani, Favyen and Zhang, Yawen and Tseng, Gabriel and Redmon, Joseph and Sablon, Hadrien and Park, Ryan and Morrison, Jacob and Buraczynski, Alexandra and Farley, Karen and Hansen, Josh and Howe, Andrew and Johnson, Patrick Alan and Otterlee, Mark and Schmitt, Ted and Pitelka, Hunter and Daspit, Stephen and Ratner, Rachel and Wilhelm, Christopher and Wood, Sebastian and Jacobi, Mike and Kerner, Hannah and Shelhamer, Evan and Farhadi, Ali and Krishna, Ranjay and Beukema, Patrick},
    title     = {OlmoEarth: Stable Latent Image Modeling for Multimodal Earth Observation},
    booktitle = {Proceedings of the IEEE/CVF Conference on Computer Vision and Pattern Recognition (CVPR)},
    month     = {June},
    year      = {2026}
}

@inproceedings{model-CROMA-Aoh,
 author = {Fuller, Anthony and Millard, Koreen and Green, James},
 booktitle = {Advances in Neural Information Processing Systems},
 editor = {A. Oh and T. Naumann and A. Globerson and K. Saenko and M. Hardt and S. Levine},
 pages = {5506--5538},
 publisher = {Curran Associates, Inc.},
 title = {CROMA: Remote Sensing Representations with Contrastive Radar-Optical Masked Autoencoders},
 volume = {36},
 year = {2023}
}

@INPROCEEDINGS{model-SkySense-Guo,
  author={Guo, Xin and Lao, Jiangwei and Dang, Bo and Zhang, Yingying and Yu, Lei and Ru, Lixiang and Zhong, Liheng and Huang, Ziyuan and Wu, Kang and Hu, Dingxiang and He, Huimei and Wang, Jian and Chen, Jingdong and Yang, Ming and Zhang, Yongjun and Li, Yansheng},
  booktitle={2024 IEEE/CVF Conference on Computer Vision and Pattern Recognition (CVPR)}, 
  title={SkySense: A Multi-Modal Remote Sensing Foundation Model Towards Universal Interpretation for Earth Observation Imagery}, 
  year={2024},
  volume={},
  number={},
  pages={27662-27673}}

@inproceedings{model-TerraFM-Muhammad,
title={TerraFM: A Scalable Foundation Model for Unified Multisensor Earth Observation},
author={Muhammad Sohail Danish and Muhammad Akhtar Munir and Syed Roshaan Ali Shah and Muhammad Haris Khan and Rao Muhammad Anwer and Jorma Laaksonen and Fahad Shahbaz Khan and Salman Khan},
booktitle={The Fourteenth International Conference on Learning Representations},
year={2026}
}

@inproceedings{model-MaRS-Yang,
  title={MaRS: A Multi-Modality Very-High-Resolution Remote Sensing Foundation Model with Cross-Granularity Meta-Modality Learning},
  author={Ruoyu Yang and Yinhe Liu and Heng Yan and Yiheng Zhou and Yihan Fu and Han Luo and Yanfei Zhong},
  booktitle={AAAI Conference on Artificial Intelligence},
  year={2026}
}

@ARTICLE{model-FlexiMo-Li,
  author={Li, Xuyang and Li, Chenyu and Ghamisi, Pedram and Hong, Danfeng and Benediktsson, Jón Atli and Chanussot, Jocelyn},
  journal={IEEE Transactions on Geoscience and Remote Sensing}, 
  title={FlexiMo: A Flexible Remote Sensing Foundation Model}, 
  year={2026},
  volume={64},
  number={},
  pages={1-16}}

@ARTICLE{model-RingMo-SAM-Yan,
  author={Yan, Zhiyuan and Li, Junxi and Li, Xuexue and Zhou, Ruixue and Zhang, Wenkai and Feng, Yingchao and Diao, Wenhui and Fu, Kun and Sun, Xian},
  journal={IEEE Transactions on Geoscience and Remote Sensing}, 
  title={RingMo-SAM: A Foundation Model for Segment Anything in Multimodal Remote-Sensing Images}, 
  year={2023},
  volume={61},
  number={},
  pages={1-16}}

@ARTICLE{model-MAPEX-Hanna,
  author={Hanna, Joëlle and Scheibenreif, Linus and Borth, Damian},
  journal={IEEE Transactions on Geoscience and Remote Sensing}, 
  title={MAPEX: Modality-Aware Pruning of Experts for Remote Sensing Foundation Models}, 
  year={2026},
  volume={64},
  number={},
  pages={1-11}}

@InProceedings{model-LHRS-Bot-Muhtar,
author="Muhtar, Dilxat and Li, Zhenshi and Gu, Feng and Zhang, Xueliang and Xiao, Pengfeng",
title="LHRS-Bot: Empowering Remote Sensing with VGI-Enhanced Large Multimodal Language Model",
booktitle="Computer Vision -- ECCV 2024",
year="2025",
publisher="Springer Nature Switzerland",
address="Cham",
pages="440--457",
isbn="978-3-031-72904-1"
}

@ARTICLE{model-RemoteCLIP-Liu,
  author={Liu, Fan and Chen, Delong and Guan, Zhangqingyun and Zhou, Xiaocong and Zhu, Jiale and Ye, Qiaolin and Fu, Liyong and Zhou, Jun},
  journal={IEEE Transactions on Geoscience and Remote Sensing}, 
  title={RemoteCLIP: A Vision Language Foundation Model for Remote Sensing}, 
  year={2024},
  volume={62},
  number={},
  pages={1-16}}

@INPROCEEDINGS{model-GeoChat-Kuckreja,
  author={Kuckreja, Kartik and Danish, Muhammad Sohail and Naseer, Muzammal and Das, Abhijit and Khan, Salman and Khan, Fahad Shahbaz},
  booktitle={2024 IEEE/CVF Conference on Computer Vision and Pattern Recognition (CVPR)}, 
  title={GeoChat:Grounded Large Vision-Language Model for Remote Sensing}, 
  year={2024},
  volume={},
  number={},
  pages={27831-27840}}

@article{model-SkyEyeGPT-Zhan,
title = {SkyEyeGPT: Unifying remote sensing vision-language tasks via instruction tuning with large language model},
journal = {ISPRS Journal of Photogrammetry and Remote Sensing},
volume = {221},
pages = {64-77},
year = {2025},
issn = {0924-2716},
author = {Yang Zhan and Zhitong Xiong and Yuan Yuan}
}

@INPROCEEDINGS{model-SkySense-O-Zhu,
  author={Zhu, Qi and Lao, Jiangwei and Ji, Deyi and Luo, Junwei and Wu, Kang and Zhang, Yingying and Ru, Lixiang and Wang, Jian and Chen, Jingdong and Yang, Ming and Liu, Dong and Zhao, Feng},
  booktitle={2025 IEEE/CVF Conference on Computer Vision and Pattern Recognition (CVPR)}, 
  title={SkySense-O: Towards Open-World Remote Sensing Interpretation with Vision-Centric Visual-Language Modeling}, 
  year={2025},
  volume={},
  number={},
  pages={14733-14744}}

@inproceedings{model-GeoLLaVA-Wang,
 author = {Wang, Fengxiang and Chen, Mingshuo and Li, Yueying and Wang, Di and Wang, Haotian and Guo, Zonghao and Wang, Zefan and Boqi, Shan and Lan, Long and Wang, Yulin and Wang, Hongzhen and Yang, Wenjing and Du, Bo and Zhang, Jing},
 booktitle = {Advances in Neural Information Processing Systems},
 editor = {D. Belgrave and C. Zhang and H. Lin and R. Pascanu and P. Koniusz and M. Ghassemi and N. Chen},
 pages = {159185--159218},
 publisher = {Curran Associates, Inc.},
 title = {GeoLLaVA-8K: Scaling Remote-Sensing Multimodal Large Language Models to 8K Resolution},
 volume = {38},
 year = {2025}
}

@InProceedings{model-GeoPixel-Shabbir,
  title = 	 {GeoPixel: Pixel Grounding Large Multimodal Model in Remote Sensing},
  author =       {Shabbir, Akashah and Zumri, Mohammed and Bennamoun, Mohammed and Khan, Fahad Shahbaz and Khan, Salman},
  booktitle = 	 {Proceedings of the 42nd International Conference on Machine Learning},
  pages = 	 {54095--54111},
  year = 	 {2025},
  editor = 	 {Singh, Aarti and Fazel, Maryam and Hsu, Daniel and Lacoste-Julien, Simon and Berkenkamp, Felix and Maharaj, Tegan and Wagstaff, Kiri and Zhu, Jerry},
  volume = 	 {267},
  series = 	 {Proceedings of Machine Learning Research},
  month = 	 {13--19 Jul},
  publisher =    {PMLR}
}

@misc{model-SkySenseGPT-Luo,
      title={SkySenseGPT: A Fine-Grained Instruction Tuning Dataset and Model for Remote Sensing Vision-Language Understanding}, 
      author={Junwei Luo and Zhen Pang and Yongjun Zhang and Tingzhu Wang and Linlin Wang and Bo Dang and Jiangwei Lao and Jian Wang and Jingdong Chen and Yihua Tan and Yansheng Li},
      year={2024},
      eprint={2406.10100},
      archivePrefix={arXiv},
      primaryClass={cs.CV}
}

@ARTICLE{model-EarthGPT-Zhang,
  author={Zhang, Wei and Cai, Miaoxin and Zhang, Tong and Zhuang, Yin and Mao, Xuerui},
  journal={IEEE Transactions on Geoscience and Remote Sensing}, 
  title={EarthGPT: A Universal Multimodal Large Language Model for Multisensor Image Comprehension in Remote Sensing Domain}, 
  year={2024},
  volume={62},
  number={},
  pages={1-20}}

@InProceedings{model-TerraScope-Shu,
    author    = {Shu, Yan and Ren, Bin and Xiong, Zhitong and Zhu, Xiao Xiang and Demir, Beg\"um and Sebe, Nicu and Rota, Paolo},
    title     = {TerraScope: Pixel-Grounded Visual Reasoning for Earth Observation},
    booktitle = {Proceedings of the IEEE/CVF Conference on Computer Vision and Pattern Recognition (CVPR)},
    month     = {June},
    year      = {2026},
    pages     = {16712-16722}
}

@INPROCEEDINGS{model-EarthDial-Soni,
  author={Soni, Sagar and Dudhane, Akshay and Debary, Hiyam and Fiaz, Mustansar and Munir, Muhammad Akhtar and Danish, Muhammad Sohail and Fraccaro, Paolo and Watson, Campbell D and Klein, Levente J and Khan, Fahad Shahbaz and Khan, Salman},
  booktitle={2025 IEEE/CVF Conference on Computer Vision and Pattern Recognition (CVPR)}, 
  title={EarthDial: Turning Multi-sensory Earth Observations to Interactive Dialogues}, 
  year={2025},
  volume={},
  number={},
  pages={14303-14313}}

@misc{model-SARVLM-Ma,
      title={SARVLM: A Vision Language Foundation Model for Semantic Understanding in SAR Imagery}, 
      author={Qiwei Ma and Xukun Lu and Wang Liu and Puhong Duan and Xudong Kang and Shutao Li},
      year={2026},
      eprint={2510.22665},
      archivePrefix={arXiv},
      primaryClass={cs.CV} 
}

@misc{model-FUSAR-R1-Yi,
      title={FUSAR-R1: A Large-Scale Reasoning Model for Intelligent Interpretation of SAR Images}, 
      author={Yi Yang and Xiaokun Zhang and Yuxuan Li and Ruyi Zhang and Xinpeng Zhou and Haipeng Wang},
      year={2026},
      eprint={2607.16819},
      archivePrefix={arXiv},
      primaryClass={cs.AI}
}

@article{model-SUMMIT-Du,
title = {SUMMIT: A SAR foundation model with multiple auxiliary tasks enhanced intrinsic characteristics},
journal = {International Journal of Applied Earth Observation and Geoinformation},
volume = {141},
pages = {104624},
year = {2025},
issn = {1569-8432},
author = {Yuntao Du and Yushi Chen and Lingbo Huang and Yahu Yang and Pedram Ghamisi and Qian Du}
}

@InProceedings{model-SARMAE-Liu,
    author    = {Liu, Danxu and Wang, Di and Wang, Hebaixu and Chen, Haoyang and Jiang, Wentao and Cheng, Yilin and Guo, Haonan and Cui, Wei and Zhang, Jing},
    title     = {SARMAE: Masked Autoencoder for SAR Representation Learning},
    booktitle = {Proceedings of the IEEE/CVF Conference on Computer Vision and Pattern Recognition (CVPR)},
    month     = {June},
    year      = {2026},
    pages     = {6496-6507}
}

@INPROCEEDINGS{model-SARViT-Ma,
  author={Ma, Jian and Zhang, Hao and Wang, Zhengjue and Wang, Zhibing},
  booktitle={IET International Radar Conference (IRC 2023)}, 
  title={SARViT: vision transformer for SAR image interpretation with efficient model compression for time-real processing}, 
  year={2023},
  volume={2023},
  number={},
  pages={1028-1035}}

@ARTICLE{model-PICV-Wang,
  author={Wang, Mengyu and Bi, Hanbo and Feng, Yingchao and Xin, Linlin and Gong, Shuo and Wang, Tianqi and Yan, Zhiyuan and Wang, Peijin and Diao, Wenhui and Sun, Xian},
  journal={IEEE Transactions on Image Processing}, 
  title={A Complex-Valued SAR Foundation Model Based on Physically Inspired Representation Learning}, 
  year={2026},
  volume={35},
  number={},
  pages={2094-2109}}

@INPROCEEDINGS{model-SAR-W-MixMAE-Ali,
  author={Caglayan, Ali and Imamoglu, Nevrez and Kouyama, Toru},
  booktitle={IGARSS 2025 - 2025 IEEE International Geoscience and Remote Sensing Symposium}, 
  title={SAR-W-MixMAE: SAR Foundation Model Training Using Backscatter Power Weighting}, 
  year={2025},
  volume={},
  number={},
  pages={265-269}}

@misc{model-SAMBA-Wang,
      title={SAMBA: A Scatter-Guided Masked Bidirectional Mamba Foundation Model for SAR Target Recognition}, 
      author={Ke Wang and Xiaoyi Pan and Zhaoyu Gu and Xiaofeng Ai and Zhiming Xu and Feng Zhao and Shunping Xiao},
      year={2026},
      eprint={2606.31668},
      archivePrefix={arXiv},
      primaryClass={cs.CV}
}

@article{dataset-DAM-Net-Saleh,
title = {DAM-Net: Flood detection from SAR imagery using differential attention metric-based vision transformers},
journal = {ISPRS Journal of Photogrammetry and Remote Sensing},
volume = {212},
pages = {440-453},
year = {2024},
issn = {0924-2716},
author = {Tamer Saleh and Xingxing Weng and Shimaa Holail and Chen Hao and Gui-Song Xia}
}

@ARTICLE{CD-Patch-Similarity-Sun,
  author={Sun, Yuli and Lei, Lin and Li, Xiao and Tan, Xiang and Kuang, Gangyao},
  journal={IEEE Transactions on Geoscience and Remote Sensing}, 
  title={Patch Similarity Graph Matrix-Based Unsupervised Remote Sensing Change Detection With Homogeneous and Heterogeneous Sensors}, 
  year={2021},
  volume={59},
  number={6},
  pages={4841-4861}}

@ARTICLE{CD-Multitemporal-Spaceborne-Ban,
  author={Ban, Yifang and Yousif, Osama A.},
  journal={IEEE Journal of Selected Topics in Applied Earth Observations and Remote Sensing}, 
  title={Multitemporal Spaceborne SAR Data for Urban Change Detection in China}, 
  year={2012},
  volume={5},
  number={4},
  pages={1087-1094}}

@article{CD-Description-Jose,
title = {Description and validation of a new set of object-based temporal geostatistical features for land-use/land-cover change detection},
journal = {ISPRS Journal of Photogrammetry and Remote Sensing},
volume = {121},
pages = {77-91},
year = {2016},
issn = {0924-2716},
author = {Jose L. Gil-Yepes and Luis A. Ruiz and Jorge A. Recio and Ángel Balaguer-Beser and Txomin Hermosilla}
}

@inproceedings{benchmark-VRSBench-Li,
 author = {Li, Xiang and Ding, Jian and Elhoseiny, Mohamed},
 booktitle = {Advances in Neural Information Processing Systems},
 editor = {A. Globerson and L. Mackey and D. Belgrave and A. Fan and U. Paquet and J. Tomczak and C. Zhang},
 pages = {3229--3242},
 publisher = {Curran Associates, Inc.},
 title = {VRSBench: A Versatile Vision-Language Benchmark Dataset for Remote Sensing Image Understanding},
 volume = {37},
 year = {2024}
}

@INPROCEEDINGS{benchmark-XLRS-Bench-Wang,
  author={Wang, Fengxiang and Wang, Hongzhen and Guo, Zonghao and Wang, Di and Wang, Yulin and Chen, Mingshuo and Ma, Qiang and Lan, Long and Yang, Wenjing and Zhang, Jing and Liu, Zhiyuan and Sun, Maosong},
  booktitle={2025 IEEE/CVF Conference on Computer Vision and Pattern Recognition (CVPR)}, 
  title={XLRS-Bench: Could Your Multimodal LLMs Understand Extremely Large Ultra-High-Resolution Remote Sensing Imagery?}, 
  year={2025},
  volume={},
  number={},
  pages={14325-14336}}

@article{benchmark-OmniEarth-Bench-Wang,
  title   = {OmniEarth-Bench: Towards Holistic Evaluation of Earth's Six Spheres and Cross-Spheres Interactions with Multimodal Observational Earth Data},
  author  = {Fengxiang Wang and Mingshuo Chen and Xuming He and others},
  journal = {arXiv preprint arXiv:2505.23522},
  year    = {2025}
}

@misc{benchmark-RSMMVP-Adejumo,
      title={A Vision Centric Remote Sensing Benchmark}, 
      author={Abduljaleel Adejumo and Faegheh Yeganli and Clifford Broni-bediako and Aoran Xiao and Naoto Yokoya and Mennatullah Siam},
      year={2025},
      eprint={2503.15816},
      archivePrefix={arXiv},
      primaryClass={cs.CV}
}

@INPROCEEDINGS{benchmark-GEOBENCH-VLM-Danish,
  author={Danish, Muhammad Sohail and Munir, Muhammad Akhtar and Shah, Syed Roshaan Ali and Kuckreja, Kartik and Khan, Fahad Shahbaz and Fraccaro, Paolo and Lacoste, Alexandre and Khan, Salman},
  booktitle={2025 IEEE/CVF International Conference on Computer Vision (ICCV)}, 
  title={GEOBench-VLM: Benchmarking Vision-Language Models for Geospatial Tasks}, 
  year={2025},
  volume={},
  number={},
  pages={7132-7142}}

@INPROCEEDINGS{benchmark-VLEO-Bench-Zhang,
  author={Zhang, Chenhui and Wang, Sherrie},
  booktitle={2024 IEEE/CVF Conference on Computer Vision and Pattern Recognition Workshops (CVPRW)}, 
  title={Good at captioning, bad at counting: Benchmarking GPT-4V on Earth observation data}, 
  year={2024},
  volume={},
  number={},
  pages={7839-7849}}

@inproceedings{benchmark-GeoMMBench-Xiao,
  title={GeoMMBench and GeoMMAgent: Toward Expert-Level Multimodal Intelligence in Geoscience and Remote Sensing},
  author={Xiao, Aoran and Cheng, Shihao and Xu, Yonghao and Ren, Yexian and Chen, Hongruixuan and Yokoya, Naoto},
  booktitle={Proceedings of the IEEE/CVF Conference on Computer Vision and Pattern Recognition (CVPR)},
  year={2026}
}

@misc{benchmark-OVEarth-Bench-Li,
      title={OVEarth-Bench: Evaluating Category Breadth and Query Diversity for Open-Vocabulary Earth Observation}, 
      author={Kaiyu Li and Zepeng Xin and Zixuan Jiang and Jing Fu and Lanxuan Xue and Lingyu Zhang and Xiangyong Cao},
      year={2026},
      eprint={2607.27278},
      archivePrefix={arXiv},
      primaryClass={cs.CV}
}

@article{benchmark-Urbench-Zhou,
  title={Urbench: A comprehensive benchmark for evaluating large multimodal models in multi-view urban scenarios},
  author={Zhou, Baichuan and Yang, Haote and Chen, Dairong and Ye, Junyan and Bai, Tianyi and Yu, Jinhua and Zhang, Songyang and Lin, Dahua and He, Conghui and Li, Weijia},
  journal={AAAI},
  year={2025}
}

@ARTICLE{benchmark-ChatEarthBench,
  author={Yuan, Zhenghang and Xiong, Zhitong and Dujardin, Thomas and Li, Xiang and Mou, Lichao and Zhu, Xiao Xiang},
  journal={IEEE Geoscience and Remote Sensing Magazine}, 
  title={ChatEarthBench: Benchmarking multimodal large language models for Earth observation}, 
  year={2026},
  volume={14},
  number={2},
  pages={64-84}}

@inproceedings{benchmark-RSVLM-QA-Zi,
author = {Zi, Xing and Xiao, Jinghao and Shi, Yunxiao and Tao, Xian and Li, Jun and Braytee, Ali and Prasad, Mukesh},
title = {RSVLM-QA: A Benchmark Dataset for Remote Sensing Vision Language Model-based Question Answering},
year = {2025},
isbn = {9798400720352},
publisher = {Association for Computing Machinery},
address = {New York, NY, USA},
booktitle = {Proceedings of the 33rd ACM International Conference on Multimedia},
pages = {12905–12911},
numpages = {7},
location = {Dublin, Ireland},
series = {MM '25}
}

@article{benchmark-RSVideo-Zhou,
  title   = {RSVideo: Are Your Vision-Language Models Ready for Remote Sensing Videos?},
  author  = {Zhou, Hongjie and Wang, Shiqin and Chen, Haoyang and Guo, Haonan and Wang, Di and Liu, Juhua and Lin, Fu and Luo, Yong},
  journal = {arXiv preprint arXiv:2608.02039},
  year    = {2026}
}

@misc{benchmark-LongEarth-R1-Ding,
      title={LongEarth-R1: Benchmarking and Aligning Vision-Language Models for Long-Horizon Earth Observation Reasoning}, 
      author={Yupan Ding and Jing Xiao and Zhenyuan Zhang and Chaofeng Chen and Liang Liao and Gui-Song Xia and Mi Wang},
      year={2026},
      eprint={2608.13344},
      archivePrefix={arXiv},
      primaryClass={cs.AI}
}

@misc{benchmark-GeoReason-Li,
      title={GeoReason: Aligning Thinking And Answering In Remote Sensing Vision-Language Models Via Logical Consistency Reinforcement Learning}, 
      author={Wenshuai Li and Xiantai Xiang and Zixiao Wen and Guangyao Zhou and Ben Niu and Feng Wang and Lijia Huang and Qiantong Wang and Yuxin Hu},
      year={2026},
      eprint={2601.04118},
      archivePrefix={arXiv},
      primaryClass={cs.CV}
}

@misc{benchmark-SenseBench-Zhong,
      title={SenseBench: A Benchmark for Remote Sensing Low-Level Visual Perception and Description in Large Vision-Language Models}, 
      author={Chen Zhong and Xiao An and Jiaxing Sun and Zihan Gui and Guangyi Yang and Wei He},
      year={2026},
      eprint={2605.10576},
      archivePrefix={arXiv},
      primaryClass={cs.CV}
}

@ARTICLE{benchmark-PANGAEA-Valerio,
  author={Marsocci, Valerio and Jia, Yuru and Bellier, Georges Le and Kerekes, David and Zeng, Liang and Hafner, Sebastian and Gerard, Sebastian and Brune, Eric and Yadav, Ritu and Shibli, Ali and Fang, Heng and Ban, Yifang and Vergauwen, Maarten and Audebert, Nicolas and Nascetti, Andrea},
  journal={IEEE Geoscience and Remote Sensing Magazine}, 
  title={PANGAEA: Assessing Geospatial Foundation Models Capabilities through a Global and Inclusive Benchmark}, 
  year={2026},
  volume={14},
  number={1},
  pages={245-285}}

@inproceedings{benchmark-GeoBench-Lacoste,
 author = {Lacoste, Alexandre and Lehmann, Nils and Rodriguez, Pau and Sherwin, Evan and Kerner, Hannah and L\"{u}tjens, Bj\"{o}rn and Irvin, Jeremy and Dao, David and Alemohammad, Hamed and Drouin, Alexandre and Gunturkun, Mehmet and Huang, Gabriel and Vazquez, David and Newman, Dava and Bengio, Yoshua and Ermon, Stefano and Zhu, Xiaoxiang},
 booktitle = {Advances in Neural Information Processing Systems},
 editor = {A. Oh and T. Naumann and A. Globerson and K. Saenko and M. Hardt and S. Levine},
 pages = {51080--51093},
 publisher = {Curran Associates, Inc.},
 title = {GEO-Bench: Toward Foundation Models for Earth Monitoring},
 volume = {36},
 year = {2023}
}

@INPROCEEDINGS{benchmark-PhilEO-Bench-Casper,
  author={Fibaek, Casper and Camilleri, Luke and Luyts, Andreas and Dionelis, Nikolaos and Saux, Bertrand Le},
  booktitle={IGARSS 2024 - 2024 IEEE International Geoscience and Remote Sensing Symposium}, 
  title={PhilEO Bench: Evaluating Geo-Spatial Foundation Models}, 
  year={2024},
  volume={},
  number={},
  pages={2739-2744}}

@inproceedings{model-SatMAE-Cong,
 author = {Cong, Yezhen and Khanna, Samar and Meng, Chenlin and Liu, Patrick and Rozi, Erik and He, Yutong and Burke, Marshall and Lobell, David and Ermon, Stefano},
 booktitle = {Advances in Neural Information Processing Systems},
 editor = {S. Koyejo and S. Mohamed and A. Agarwal and D. Belgrave and K. Cho and A. Oh},
 pages = {197--211},
 publisher = {Curran Associates, Inc.},
 title = {SatMAE: Pre-training Transformers for Temporal and Multi-Spectral Satellite Imagery},
 volume = {35},
 year = {2022}
}

@inproceedings{model-skymoe-Liu,
author = {Liu, Jiaqi and Fu, Ronghao and Sun, Lang and Liu, Haoran and Yang, Xiao and Zhang, Weipeng and Na, Xu and Duan, Zhuoran and Yang, Bo},
title = {SkyMoE: a vision-language foundation model for enhancing geospatial interpretation with mixture of experts},
year = {2026},
isbn = {978-1-57735-906-7},
publisher = {AAAI Press},
booktitle = {Proceedings of the Fortieth AAAI Conference on Artificial Intelligence and Thirty-Eighth Conference on Innovative Applications of Artificial Intelligence and Sixteenth Symposium on Educational Advances in Artificial Intelligence},
articleno = {797},
numpages = {11},
series = {AAAI'26/IAAI'26/EAAI'26}
}

@misc{model-ScaleEarth-Zhang,
      title={Beyond GSD-as-Token: Continuous Scale Conditioning for Remote Sensing VLMs}, 
      author={Song Zhang and Yanlong Chen and Yilin Li and Yining Chen and Zili Yi and Xiaowei Zhang and Yawei Li},
      year={2026},
      eprint={2605.07562},
      archivePrefix={arXiv},
      primaryClass={cs.CV}
}

@article{model-TACMT-Li,
title = {TACMT: Text-aware cross-modal transformer for visual grounding on high-resolution SAR images},
journal = {ISPRS Journal of Photogrammetry and Remote Sensing},
volume = {222},
pages = {152-166},
year = {2025},
issn = {0924-2716},
author = {Tianyang Li and Chao Wang and Sirui Tian and Bo Zhang and Fan Wu and Yixian Tang and Hong Zhang}
}

@Article{model-TVLightFormer-Zhong,
AUTHOR = {Zhong, Yuqiao and Quan, Haoqi and Nie, Chenyu and Wei, Yingmei and Guo, Yanming},
TITLE = {TVLightFormer: A Lightweight Cross-Modal Transformer for Language-Guided Target Localization in SAR Imagery},
JOURNAL = {Remote Sensing},
VOLUME = {18},
YEAR = {2026},
NUMBER = {9},
ARTICLE-NUMBER = {1430},
ISSN = {2072-4292}
}

@article{model-LHRS-Bot-Nova-Li,
title = {LHRS-Bot-Nova: Improved multimodal large language model for remote sensing vision-language interpretation},
journal = {ISPRS Journal of Photogrammetry and Remote Sensing},
volume = {227},
pages = {539-550},
year = {2025},
issn = {0924-2716},
author = {Zhenshi Li and Dilxat Muhtar and Feng Gu and Yanglangxing He and Xueliang Zhang and Pengfeng Xiao and Guangjun He and Xiaoxiang Zhu}
}

@misc{model-EarthMind-Shu,
      title={EarthMind: Leveraging Cross-Sensor Data for Advanced Earth Observation Interpretation with a Unified Multimodal LLM}, 
      author={Yan Shu and Bin Ren and Zhitong Xiong and Danda Pani Paudel and Luc Van Gool and Begüm Demir and Nicu Sebe and Paolo Rota},
      year={2025},
      eprint={2506.01667},
      archivePrefix={arXiv},
      primaryClass={cs.CV}
}

@ARTICLE{model-EarthGPT-X-Zhang,
  author={Zhang, Wei and Cai, Miaoxin and Ning, Yaqian and Zhang, Tong and Zhuang, Yin and Lu, Shijian and Chen, He and Li, Jun and Mao, Xuerui},
  journal={IEEE Transactions on Geoscience and Remote Sensing}, 
  title={EarthGPT-X: A Spatial MLLM for Multilevel Multisource Remote Sensing Imagery Understanding With Visual Prompting}, 
  year={2025},
  volume={63},
  number={},
  pages={1-21}}

@misc{model-DOFA-CLIP-Xiong,
      title={DOFA-CLIP: Multimodal Vision-Language Foundation Models for Earth Observation}, 
      author={Zhitong Xiong and Yi Wang and Weikang Yu and Adam J Stewart and Jie Zhao and Nils Lehmann and Thomas Dujardin and Zhenghang Yuan and Pedram Ghamisi and Xiao Xiang Zhu},
      year={2025},
      eprint={2503.06312},
      archivePrefix={arXiv},
      primaryClass={cs.CV}
}

@InProceedings{model-SATtxt-Do,
    author    = {Do, Minh Kha and Xiang, Wei and Han, Kang and Wu, Di and Phan, Khoa and Chen, Yi-Ping Phoebe and Liu, Gaowen and Kompella, Ramana Rao},
    title     = {Spectrally Distilled Representations Aligned with Instruction-Augmented LLMs for Satellite Imagery},
    booktitle = {Proceedings of the IEEE/CVF Conference on Computer Vision and Pattern Recognition (CVPR)},
    month     = {June},
    year      = {2026},
    pages     = {6453-6463}
}

@ARTICLE{model-GeoRSCLIP-Zhang,
  author={Zhang, Zilun and Zhao, Tiancheng and Guo, Yulong and Yin, Jianwei},
  journal={IEEE Transactions on Geoscience and Remote Sensing}, 
  title={RS5M and GeoRSCLIP: A Large-Scale Vision- Language Dataset and a Large Vision-Language Model for Remote Sensing}, 
  year={2024},
  volume={62},
  number={},
  pages={1-23}}

@inproceedings{model-graft-Mall,
 title={Remote sensing vision-language foundation models without annotations via ground remote alignment},
 author={Mall, Utkarsh and Phoo, Cheng Perng and Liu, Meilin Kelsey and Vondrick, Carl and Hariharan, Bharath and Bala, Kavita},
 booktitle={ICLR},
 year={2024}
}

@ARTICLE{model-EarthMarker-Zhang,
  author={Zhang, Wei and Cai, Miaoxin and Zhang, Tong and Zhuang, Yin and Li, Jun and Mao, Xuerui},
  journal={IEEE Transactions on Geoscience and Remote Sensing}, 
  title={EarthMarker: A Visual Prompting Multimodal Large Language Model for Remote Sensing}, 
  year={2025},
  volume={63},
  number={},
  pages={1-19}}

@ARTICLE{model-RingMoGPT-Wang,
  author={Wang, Peijin and Hu, Huiyang and Tong, Boyuan and Zhang, Ziqi and Yao, Fanglong and Feng, Yingchao and Zhu, Zining and Chang, Hao and Diao, Wenhui and Ye, Qixiang and Sun, Xian},
  journal={IEEE Transactions on Geoscience and Remote Sensing}, 
  title={RingMoGPT: A Unified Remote Sensing Foundation Model for Vision, Language, and Grounded Tasks}, 
  year={2025},
  volume={63},
  number={},
  pages={1-20}}

@InProceedings{model-ZoomEarth-Liu,
    author    = {Liu, Ruixun and Fu, Bowen and Song, Jiayi and Li, Kaiyu and Li, Wanchen and Xue, Lanxuan and Qiao, Hui and Zhang, Weizhan and Meng, Deyu and Cao, Xiangyong},
    title     = {ZoomEarth: Active Perception for Ultra-High-Resolution Geospatial Vision-Language Tasks},
    booktitle = {Proceedings of the IEEE/CVF Conference on Computer Vision and Pattern Recognition (CVPR)},
    month     = {June},
    year      = {2026},
    pages     = {34877-34888}
}

@inproceedings{model-ScoreRS-Muhtar,
 author = {Muhtar, Dilxat and Zhang, Enzhuo and Li, Zhenshi and Gu, Feng and He, Yanglangxing and Xiao, Pengfeng and Zhang, Xueliang},
 booktitle = {Advances in Neural Information Processing Systems},
 editor = {D. Belgrave and C. Zhang and H. Lin and R. Pascanu and P. Koniusz and M. Ghassemi and N. Chen},
 pages = {171088--171133},
 publisher = {Curran Associates, Inc.},
 title = {Quality-Driven Curation of Remote Sensing Vision-Language Data via Learned Scoring Models},
 volume = {38, Main Conference},
 year = {2025}
}

@misc{model-GeoRSMLLM-Zhang,
      title={GeoRSMLLM: A Multimodal Large Language Model for Vision-Language Tasks in Geoscience and Remote Sensing}, 
      author={Zilun Zhang and Haozhan Shen and Tiancheng Zhao and Bin Chen and Zian Guan and Yuhao Wang and Xu Jia and Yuxiang Cai and Yongheng Shang and Jianwei Yin},
      year={2025},
      eprint={2503.12490},
      archivePrefix={arXiv},
      primaryClass={cs.CV}
}

@misc{model-RS-LVLM-Park,
      title={Remote Sensing Large Vision-Language Model: Semantic-augmented Multi-level Alignment and Semantic-aware Expert Modeling}, 
      author={Sungjune Park and Yeongyun Kim and Se Yeon Kim and Yong Man Ro},
      year={2025},
      eprint={2506.21863},
      archivePrefix={arXiv},
      primaryClass={cs.CV}
}

@inproceedings{model-RemoteSAM-Yao,
author = {Yao, Liang and Liu, Fan and Chen, Delong and Zhang, Chuanyi and Wang, Yijun and Chen, Ziyun and Xu, Wei and Di, Shimin and Zheng, Yuhui},
title = {RemoteSAM: Towards Segment Anything for Earth Observation},
year = {2025},
isbn = {9798400720352},
publisher = {Association for Computing Machinery},
address = {New York, NY, USA},
booktitle = {Proceedings of the 33rd ACM International Conference on Multimedia},
pages = {3027–3036},
numpages = {10},
location = {Dublin, Ireland},
series = {MM '25}
}

@misc{model-FUSE-RSVLM-Dang,
      title={FUSE-RSVLM: Feature Fusion Vision-Language Model for Remote Sensing}, 
      author={Yunkai Dang and Donghao Wang and Jiacheng Yang and Yifan Jiang and Meiyi Zhu and Yuekun Yang and Cong Wang and Qi Fan and Wenbin Li and Yang Gao},
      year={2025},
      eprint={2512.24022},
      archivePrefix={arXiv},
      primaryClass={cs.CV}
}

@misc{model-REO-VLM-Xue,
      title={REO-VLM: Transforming VLM to Meet Regression Challenges in Earth Observation}, 
      author={Xizhe Xue and Guoting Wei and Hao Chen and Haokui Zhang and Feng Lin and Chunhua Shen and Xiao Xiang Zhu},
      year={2024},
      eprint={2412.16583},
      archivePrefix={arXiv},
      primaryClass={cs.CV}
}

@misc{model-ViTP-Li,
      title={Visual Instruction Pretraining for Domain-Specific Foundation Models}, 
      author={Yuxuan Li and Yicheng Zhang and Wenhao Tang and Yimian Dai and Ming-Ming Cheng and Xiang Li and Jian Yang},
      year={2026},
      eprint={2509.17562},
      archivePrefix={arXiv},
      primaryClass={cs.CV}
}

@ARTICLE{model-RingMo-Agent-Hu,
  author={Hu, Huiyang and Wang, Peijin and Feng, Yingchao and Wei, Kaiwen and Yin, Wenxin and Diao, Wenhui and Wang, Mengyu and Bi, Hanbo and Kang, Kaiyue and Ling, Tong and Fu, Kun and Sun, Xian},
  journal={IEEE Transactions on Pattern Analysis and Machine Intelligence}, 
  title={RingMo-Agent: A Unified Remote Sensing Foundation Model for Multi-Platform and Multi-Modal Reasoning}, 
  year={2026},
  volume={},
  number={},
  pages={1-18}}

@InProceedings{model-tessera-Feng,
    author    = {Feng, Zhengpeng and Atzberger, Clement and Jaffer, Sadiq and Knezevic, Jovana and Sormunen, Silja and Young, Robin and Lisaius, Madeline C. and Immitzer, Markus and Jackson, Toby and Ball, James and Coomes, David A. and Madhavapeddy, Anil and Blake, Andrew and Keshav, Srinivasan},
    title     = {TESSERA: Temporal Embeddings of Surface Spectra for Earth Representation and Analysis},
    booktitle = {Proceedings of the IEEE/CVF Conference on Computer Vision and Pattern Recognition (CVPR)},
    month     = {June},
    year      = {2026},
    pages     = {34818-34831}
}

@INPROCEEDINGS{model-DINO-MM-Wang,
  author={Wang, Yi and Albrecht, Conrad M and Zhu, Xiao Xiang},
  booktitle={IGARSS 2022 - 2022 IEEE International Geoscience and Remote Sensing Symposium}, 
  title={Self-Supervised Vision Transformers for Joint SAR-Optical Representation Learning}, 
  year={2022},
  volume={},
  number={},
  pages={139-142}}

@INPROCEEDINGS{model-RS-vHeat-Hu,
  author={Hu, Huiyang and Wang, Peijin and Bi, Hanbo and Tong, Boyuan and Wang, Zhaozhi and Diao, Wenhui and Chang, Hao and Feng, Yingchao and Zhang, Ziqi and Wang, Yaowei and Ye, Qixiang and Fu, Kun and Sun, Xian},
  booktitle={2025 IEEE/CVF International Conference on Computer Vision (ICCV)}, 
  title={RS-vHeat: Heat Conduction Guided Efficient Remote Sensing Foundation Model}, 
  year={2025},
  volume={},
  number={},
  pages={9876-9887}}

@misc{model-DOFA-Xiong,
      title={Neural Plasticity-Inspired Multimodal Foundation Model for Earth Observation}, 
      author={Zhitong Xiong and Yi Wang and Fahong Zhang and Adam J. Stewart and Joëlle Hanna and Damian Borth and Ioannis Papoutsis and Bertrand Le Saux and Gustau Camps-Valls and Xiao Xiang Zhu},
      year={2025},
      eprint={2403.15356},
      archivePrefix={arXiv},
      primaryClass={cs.CV}
}

@InProceedings{model-msGFM-Han,
    author    = {Han, Boran and Zhang, Shuai and Shi, Xingjian and Reichstein, Markus},
    title     = {Bridging Remote Sensors with Multisensor Geospatial Foundation Models},
    booktitle = {Proceedings of the IEEE/CVF Conference on Computer Vision and Pattern Recognition (CVPR)},
    month     = {June},
    year      = {2024},
    pages     = {27852-27862}
}

@ARTICLE{model-MTP-Wang,
  author={Wang, Di and Zhang, Jing and Xu, Minqiang and Liu, Lin and Wang, Dongsheng and Gao, Erzhong and Han, Chengxi and Guo, Haonan and Du, Bo and Tao, Dacheng and Zhang, Liangpei},
  journal={IEEE Journal of Selected Topics in Applied Earth Observations and Remote Sensing}, 
  title={MTP: Advancing Remote Sensing Foundation Model via Multitask Pretraining}, 
  year={2024},
  volume={17},
  number={},
  pages={11632-11654}}

@INPROCEEDINGS{model-OFA-Net-Xiong,
  author={Xiong, Zhitong and Wang, Yi and Zhang, Fahong and Zhu, Xiao Xiang},
  booktitle={IGARSS 2024 - 2024 IEEE International Geoscience and Remote Sensing Symposium}, 
  title={One for All: Toward Unified Foundation Models for Earth Vision}, 
  year={2024},
  volume={},
  number={},
  pages={2734-2738}}

@article{model-omnisat-Astruc,
  title={OmniSat: Self-Supervised Modality Fusion for Earth Observation},
  author={Astruc, Guillaume and Gonthier, Nicolas and Mallet, Clement and Landrieu, Loic},
  journal={ECCV},
  year={2024}
}

@ARTICLE{model-SoftCon-Wang,
  author={Wang, Yi and Albrecht, Conrad M. and Zhu, Xiao Xiang},
  journal={IEEE Transactions on Geoscience and Remote Sensing}, 
  title={Multilabel-Guided Soft Contrastive Learning for Efficient Earth Observation Pretraining}, 
  year={2024},
  volume={62},
  number={},
  pages={1-16}}

@ARTICLE{model-MURLF-Du,
  author={Du, Dakuan and Liu, Tianzhu and Gu, Yanfeng},
  journal={IEEE Transactions on Geoscience and Remote Sensing}, 
  title={A Multimodal Unified Representation Learning Framework With Masked Image Modeling for Remote Sensing Images}, 
  year={2024},
  volume={62},
  number={},
  pages={1-16}}

@misc{model-AlphaEarth-F,
      title={AlphaEarth Foundations: An embedding field model for accurate and efficient global mapping from sparse label data}, 
      author={Christopher F. Brown and Michal R. Kazmierski and Valerie J. Pasquarella and William J. Rucklidge and Masha Samsikova and Chenhui Zhang and Evan Shelhamer and Estefania Lahera and Olivia Wiles and Simon Ilyushchenko and Noel Gorelick and Lihui Lydia Zhang and Sophia Alj and Emily Schechter and Sean Askay and Oliver Guinan and Rebecca Moore and Alexis Boukouvalas and Pushmeet Kohli},
      year={2025},
      eprint={2507.22291},
      archivePrefix={arXiv},
      primaryClass={cs.CV}
}

@INPROCEEDINGS{model-Copernicus-FM-Wang,
  author={Wang, Yi and Xiong, Zhitong and Liu, Chenying and Stewart, Adam J. and Dujardin, Thomas and Bountos, Nikolaos Ioannis and Zavras, Angelos and Gerken, Franziska and Papoutsis, Ioannis and Leal-Taixé, Laura and Zhu, Xiao Xiang},
  booktitle={2025 IEEE/CVF International Conference on Computer Vision (ICCV)}, 
  title={Towards a Unified Copernicus Foundation Model for Earth Vision}, 
  year={2025},
  volume={},
  number={},
  pages={9888-9899}}

@inproceedings{model-LESS-ViT-Si,
title={Towards Scalable Foundation Model for Multi-modal and Multi-spectral Geospatial Data},
author={Haozhe Si and Yuxuan Wan and Minh N. Do and Deepak Vasisht and Han Zhao and Hendrik Hamann},
booktitle={TerraBytes - ICML 2025 workshop},
year={2026}
}

@ARTICLE{model-AllSpark-Shao,
  author={Shao, Run and Yang, Cheng and Li, Qiujun and Xu, Linrui and Yang, Xiang and Li, Xian and Li, Mengyao and Zhu, Qing and Zhang, Yongjun and Li, Yansheng and Liu, Yu and Tang, Yong and Liu, Dapeng and Yang, Shizhong and Li, Haifeng},
  journal={IEEE Transactions on Geoscience and Remote Sensing}, 
  title={AllSpark: A Multimodal Spatiotemporal General Intelligence Model With Ten Modalities via Language as a Reference Framework}, 
  year={2025},
  volume={63},
  number={},
  pages={1-20}}

@INPROCEEDINGS{model-RobSense-Do,
  author={Do, Minh Kha and Han, Kang and Lai, Phu and Phan, Khoa T. and Xiang, Wei},
  booktitle={2025 IEEE/CVF Conference on Computer Vision and Pattern Recognition (CVPR)}, 
  title={RobSense: A Robust Multi-modal Foundation Model for Remote Sensing with Static, Temporal, and Incomplete Data Adaptability}, 
  year={2025},
  volume={},
  number={},
  pages={7427-7436}}

@inproceedings{model-OReole-FM-Dias,
author = {Dias, Philipe and Tsaris, Aristeidis and Bowman, Jordan and Potnis, Abhishek and Arndt, Jacob and Yang, H. Lexie and Lunga, Dalton},
title = {OReole-FM: successes and challenges toward billion-parameter foundation models for high-resolution satellite imagery},
year = {2024},
isbn = {9798400711077},
publisher = {Association for Computing Machinery},
address = {New York, NY, USA},
booktitle = {Proceedings of the 32nd ACM International Conference on Advances in Geographic Information Systems},
pages = {597–600},
numpages = {4},
location = {Atlanta, GA, USA},
series = {SIGSPATIAL '24}
}

@INPROCEEDINGS{model-SkySense-V2-Zhang,
  author={Zhang, Yingying and Ru, Lixiang and Wu, Kang and Yu, Lei and Liang, Lei and Li, Yansheng and Chen, Jingdong},
  booktitle={2025 IEEE/CVF International Conference on Computer Vision (ICCV)}, 
  title={SkySense V2: A Unified Foundation Model for Multi-Modal Remote Sensing}, 
  year={2025},
  volume={},
  number={},
  pages={9136-9146}}

@InProceedings{model-AnySat-Astruc,
    author    = {Astruc, Guillaume and Gonthier, Nicolas and Mallet, Cl\'ement and Landrieu, Loic},
    title     = {AnySat: One Earth Observation Model for Many Resolutions, Scales, and Modalities},
    booktitle = {Proceedings of the IEEE/CVF Conference on Computer Vision and Pattern Recognition (CVPR)},
    month     = {June},
    year      = {2025},
    pages     = {19530-19540}
}

@inproceedings{model-FoMo-Bountos,
author = {Bountos, Nikolaos Ioannis and Ouaknine, Arthur and Papoutsis, Ioannis and Rolnick, David},
title = {FoMo: multi-modal, multi-scale and multi-task remote sensing foundation models for forest monitoring},
year = {2025},
isbn = {978-1-57735-897-8},
publisher = {AAAI Press},
booktitle = {Proceedings of the Thirty-Ninth AAAI Conference on Artificial Intelligence and Thirty-Seventh Conference on Innovative Applications of Artificial Intelligence and Fifteenth Symposium on Educational Advances in Artificial Intelligence},
articleno = {3104},
numpages = {11},
series = {AAAI'25/IAAI'25/EAAI'25}
}

@ARTICLE{model-RingMamba-Wang,
  author={Wang, Peijin and Chang, Hao and Hu, Huiyang and Li, Xin and Liu, Xiaorui and Liu, Yu and Zhang, Zhaolong and Chen, Chen and Li, Yundu and Feng, Yingchao and Diao, Wenhui and Zheng, Qingfang and Wang, Yaowei and Sun, Xian},
  journal={IEEE Transactions on Geoscience and Remote Sensing}, 
  title={RingMamba: Remote Sensing Multisensor Pretraining With Visual State Space Model}, 
  year={2025},
  volume={63},
  number={},
  pages={1-16}}

@article{model-SeaMo-Li,
title = {SeaMo: A season-aware multimodal foundation model for remote sensing},
journal = {Information Fusion},
volume = {125},
pages = {103334},
year = {2026},
issn = {1566-2535},
author = {Xuyang Li and Chenyu Li and Gemine Vivone and Danfeng Hong}
}

@misc{model-MoSAiC-Gupta,
      title={MoSAiC: Multi-Modal Multi-Label Supervision-Aware Contrastive Learning for Remote Sensing}, 
      author={Debashis Gupta and Aditi Golder and Rongkhun Zhu and Kangning Cui and Wei Tang and Fan Yang and Ovidiu Csillik and Sarra Alaqahtani and V. Paul Pauca},
      year={2025},
      eprint={2507.08683},
      archivePrefix={arXiv},
      primaryClass={cs.CV}
}

@article{model-SkySense++-Wu,
  author    = {Wu, Kang and Zhang, Yingying and Ru, Lixiang and Dang, Bo and Lao, Jiangwei and Yu, Lei and Luo, Junwei and Zhu, Zifan and Sun, Yue and Zhang, Jiahao and Zhu, Qi and Wang, Jian and Yang, Ming and Chen, Jingdong and Zhang, Yongjun and Li, Yansheng},
  title     = {A semantic-enhanced multi-modal remote sensing foundation model for Earth observation},
  journal   = {Nature Machine Intelligence},
  year      = {2025},
  volume    = {7},
  number    = {8},
  pages     = {1235--1249},
  month     = aug,
  issn      = {2522-5839},
}

@InProceedings{model-Galileo-Tseng,
  title = 	 {Galileo: Learning Global \& amp; Local Features of Many Remote Sensing Modalities},
  author =       {Tseng, Gabriel and Fuller, Anthony and Reil, Marlena and Herzog, Henry and Beukema, Patrick and Bastani, Favyen and Green, James R and Shelhamer, Evan and Kerner, Hannah and Rolnick, David},
  booktitle = 	 {Proceedings of the 42nd International Conference on Machine Learning},
  pages = 	 {60280--60300},
  year = 	 {2025},
  editor = 	 {Singh, Aarti and Fazel, Maryam and Hsu, Daniel and Lacoste-Julien, Simon and Berkenkamp, Felix and Maharaj, Tegan and Wagstaff, Kiri and Zhu, Jerry},
  volume = 	 {267},
  series = 	 {Proceedings of Machine Learning Research},
  month = 	 {13--19 Jul},
  publisher =    {PMLR}
}

@INPROCEEDINGS{model-Panopticon-Leonard,
  author={Waldmann, Leonard and Shah, Ando and Wang, Yi and Lehmann, Nils and Stewart, Adam J. and Xiong, Zhitong and Zhu, Xiao Xiang and Bauer, Stefan and Chuang, John},
  booktitle={2025 IEEE/CVF Conference on Computer Vision and Pattern Recognition Workshops (CVPRW)}, 
  title={Panopticon: Advancing Any-Sensor Foundation Models for Earth Observation}, 
  year={2025},
  volume={},
  number={},
  pages={2195-2205}}

@ARTICLE{model-Prithvi-EO-2.0-Daniela,
  author={Szwarcman, Daniela and Roy, Sujit and Fraccaro, Paolo and Gislason, porsteinn Eli and Blumenstiel, Benedikt and Ghosal, Rinki and de Oliveira, Pedro Henrique and de Sousa Almeida, Joao Lucas and Sedona, Rocco and Kang, Yanghui and Chakraborty, Srija and Wang, Sizhe and Gomes, Carlos and Kumar, Ankur and Gaur, Vishal and Truong, Myscon and Godwin, Denys and Khallaghi, Sam and Lee, Hyunho and Hsu, Chia-Yu and Asanjan, Ata Akbari and Mujeci, Besart and Shidham, Disha and Balogun, Rufai Omowunmi and Kolluru, Venkatesh and Keenan, Trevor and Arevalo, Paulo and Li, Wenwen and Alemohammad, Hamed and Olofsson, Pontus and Mayer, Timothy and Hain, Christopher and Kennedy, Robert and Zadrozny, Bianca and Bell, David and Cavallaro, Gabriele and Watson, Campbell and Maskey, Manil and Ramachandran, Rahul and Moreno, Juan Bernabe},
  journal={IEEE Transactions on Geoscience and Remote Sensing}, 
  title={Prithvi-EO-2.0: A Versatile Multitemporal Foundation Model for Earth Observation Applications}, 
  year={2026},
  volume={64},
  number={},
  pages={1-20}}

@ARTICLE{model-Alliance-Zhao,
  author={Zhao, Boyu and Li, Wei and Wang, Junjie and Zhang, Yuxiang and Yang, Hong and Zhao, Haitao and Tao, Ran and Du, Qian},
  journal={IEEE Transactions on Pattern Analysis and Machine Intelligence}, 
  title={Alliance: All-in-One Spectral-Spatial-Frequency Awareness Foundation Model}, 
  year={2026},
  volume={48},
  number={3},
  pages={3586-3603}}

@InProceedings{model-THOR-Theodor,
    author    = {Forgaard, Theodor and Reksten, Jarle and Waldeland, Anders and Marsocci, Valerio and Longepe, Nicolas and Kampffmeyer, Michael and Salberg, Arnt-Borre},
    title     = {THOR: A Versatile Foundation Model for Earth Observation Climate and Society Applications},
    booktitle = {Proceedings of the IEEE/CVF Conference on Computer Vision and Pattern Recognition (CVPR) Workshops},
    month     = {June},
    year      = {2026},
    pages     = {7899-7908}
}

@article{model-AgriFM-Li,
title = {AgriFM: A multi-source temporal remote sensing foundation model for Agriculture mapping},
journal = {Remote Sensing of Environment},
volume = {334},
pages = {115234},
year = {2026},
issn = {0034-4257},
author = {Wenyuan Li and Shunlin Liang and Keyan Chen and Yongzhe Chen and Han Ma and Jianglei Xu and Yichuan Ma and Yuxiang Zhang and Shikang Guan and Husheng Fang and Zhenwei Shi}
}

@InProceedings{model-RAMEN-Nicolas,
    author    = {Houdr\'e, Nicolas and Marcos, Diego and de Turckheim, Hugo Riffaud and Ienco, Dino and Wendling, Laurent and Kurtz, Camille and Lobry, Sylvain},
    title     = {RAMEN: Resolution-Adjustable Multimodal Encoder for Earth Observation},
    booktitle = {Proceedings of the IEEE/CVF Conference on Computer Vision and Pattern Recognition (CVPR)},
    month     = {June},
    year      = {2026},
    pages     = {27838-27848}
}

@misc{model-SpectralEarth-FM-Ait,
      title={SpectralEarth-FM: Bringing Hyperspectral Imagery into Multimodal Earth Observation Pretraining}, 
      author={Nassim Ait Ali Braham and Aaron Banze and Conrad M. Albrecht and Julien Mairal and Jocelyn Chanussot and Xiao Xiang Zhu},
      year={2026},
      eprint={2605.21075},
      archivePrefix={arXiv},
      primaryClass={cs.CV}
}

@misc{model-FLORO-Jorge,
      title={FLORO: A Multimodal Geospatial Foundation Model for Ecological Remote Sensing Across Sensors and Scales}, 
      author={Jorge L. Rodriguez and Victor Angulo Morales and Areej Alwahas and Mariana Elias Lara and Fida Mohammad Thoker and Kasper Johansen and Bernard Ghanem and Fernando T. Maestre and Matthew F. McCabe},
      year={2026},
      eprint={2605.28174},
      archivePrefix={arXiv},
      primaryClass={cs.CV}
}

@InProceedings{model-SARFormer-Prexl,
    author    = {Prexl, Jonathan and Recla, Michael and Schmitt, Michael},
    title     = {SARFormer - An Acquisition Parameter Aware Vision Transformer for Synthetic Aperture Radar Data},
    booktitle = {Proceedings of the IEEE/CVF Conference on Computer Vision and Pattern Recognition (CVPR) Workshops},
    month     = {June},
    year      = {2025},
    pages     = {2250-2259}
}

@inproceedings{model-LSMA-SAR-Allen,
  title={Large Scale Masked Autoencoding for Reducing Label Requirements on SAR Data},
  author={Allen, Matthew J and Dorr, Francisco  and Gallego, Joseph A and Martínez-Ferrer, Laura and Kalaitzis, Freddie and Ramos-Pollan, Raul and Jungbluth, Anna},
  booktitle={NeurIPS 2023 Workshop on Tackling Climate Change with Machine Learning},
  year={2023}
}

@misc{model-SARATR-X-v2-Li,
      title={SARATR-X-v2: Scale-Aware Structural Pre-Training for SAR Foundation Models}, 
      author={Weijie Li and Yafei Song and Yongxiang Liu and Bowen Peng and Jie Zhou and Jingyuan Xia and Wei Yang and Tianpeng Liu and Zhen Liu and Li Liu},
      year={2026},
      eprint={2607.23238},
      archivePrefix={arXiv},
      primaryClass={cs.CV}
}

@article{model-SAR-JEPA-Li,
title = {Predicting gradient is better: Exploring self-supervised learning for SAR ATR with a joint-embedding predictive architecture},
journal = {ISPRS Journal of Photogrammetry and Remote Sensing},
volume = {218},
pages = {326-338},
year = {2024},
issn = {0924-2716},
author = {Weijie Li and Wei Yang and Tianpeng Liu and Yuenan Hou and Yuxuan Li and Zhen Liu and Yongxiang Liu and Li Liu}
}

@misc{model-CrossEarth-SAR-Ye,
      title={CrossEarth-SAR: A SAR-Centric and Billion-Scale Geospatial Foundation Model for Domain Generalizable Semantic Segmentation}, 
      author={Ziqi Ye and Ziyang Gong and Ning Liao and Xiaoxing Hu and Di Wang and Hongruixuan Chen and Chen Huang and Yiguo He and Yuru Jia and Xiaoxing Wang and Haipeng Wang and Xue Yang and Junchi Yan},
      year={2026},
      eprint={2603.12008},
      archivePrefix={arXiv},
      primaryClass={cs.CV}
}

@INPROCEEDINGS{model-Nereus-SAR-1-Thomas,
  author={Kerdreux, Thomas and Tuel, Alexandre and Febvre, Quentin and Mouche, Alexis and Chapron, Bertrand},
  booktitle={2025 IEEE/CVF Conference on Computer Vision and Pattern Recognition Workshops (CVPRW)}, 
  title={Efficient Self-Supervised Learning for Earth Observation via Dynamic Dataset Curation}, 
  year={2025},
  volume={},
  number={},
  pages={3008-3018}}

@article {model-WV-Net-Glaser,
      author = "Yannik Glaser and Justin E. Stopa and Linnea M. Wolniewicz and Ralph Foster and Doug Vandemark and Alexis Mouche and Bertrand Chapron and Peter Sadowski",
      title = "WV-Net: A Foundation Model for SAR Ocean Satellite Imagery",
      journal = "Artificial Intelligence for the Earth Systems",
      year = "2025",
      publisher = "American Meteorological Society",
      address = "Boston MA, USA",
      volume = "4",
      number = "4",
      pages=      "250003"
}

@ARTICLE{model-RVSA-Wang,
  author={Wang, Di and Zhang, Qiming and Xu, Yufei and Zhang, Jing and Du, Bo and Tao, Dacheng and Zhang, Liangpei},
  journal={IEEE Transactions on Geoscience and Remote Sensing}, 
  title={Advancing Plain Vision Transformer Toward Remote Sensing Foundation Model}, 
  year={2023},
  volume={61},
  number={},
  pages={1-15}}

@ARTICLE{model-SwiMDiff-Tian,
  author={Tian, Jiayuan and Lei, Jie and Zhang, Jiaqing and Xie, Weiying and Li, Yunsong},
  journal={IEEE Transactions on Geoscience and Remote Sensing}, 
  title={SwiMDiff: Scene-Wide Matching Contrastive Learning With Diffusion Constraint for Remote Sensing Image}, 
  year={2024},
  volume={62},
  number={},
  pages={1-13}}

@misc{model-RS-DFM-Wang,
      title={RS-DFM: A Remote Sensing Distributed Foundation Model for Diverse Downstream Tasks}, 
      author={Zhechao Wang and Peirui Cheng and Pengju Tian and Yuchao Wang and Mingxin Chen and Shujing Duan and Zhirui Wang and Xinming Li and Xian Sun},
      year={2024},
      eprint={2406.07032},
      archivePrefix={arXiv},
      primaryClass={cs.CV}
}

@ARTICLE{model-PIS-An,
  author={An, Xiao and He, Wei and Zou, Jiaqi and Yang, Guangyi and Zhang, Hongyan},
  journal={IEEE Transactions on Geoscience and Remote Sensing}, 
  title={Pretrain a Remote Sensing Foundation Model by Promoting Intra-Instance Similarity}, 
  year={2024},
  volume={62},
  number={},
  pages={1-15}}

@article{model-MM-VSF-Praveen,
author = {Ravirathinam, Praveen and Parthasarathy, Ajitesh and Khandelwal, Ankush and Ghosh, Rahul and Kumar, Vipin},
title = {Towards Knowledge Guided Pretraining Approaches for Multimodal Foundation Models: Applications in Remote Sensing},
year = {2026},
publisher = {Association for Computing Machinery},
address = {New York, NY, USA},
issn = {2374-0353},
note = {Just Accepted},
journal = {ACM Trans. Spatial Algorithms Syst.},
month = mar
}

@ARTICLE{model-CrossEarth-Gong,
  author={Gong, Ziyang and Wei, Zhixiang and Wang, Di and Hu, Xiaoxing and Ma, Xianzheng and Chen, Hongruixuan and Jia, Yuru and Deng, Yupeng and Ji, Zhenming and Zhu, Xiangwei and Yang, Xue and Yokoya, Naoto and Zhang, Jing and Du, Bo and Yan, Junchi and Zhang, Liangpei},
  journal={IEEE Transactions on Pattern Analysis and Machine Intelligence}, 
  title={CrossEarth: Geospatial Vision Foundation Model for Domain Generalizable Remote Sensing Semantic Segmentation}, 
  year={2026},
  volume={48},
  number={5},
  pages={5147-5164}}

@ARTICLE{model-GeRSP-Huang,
  author={Huang, Ziyue and Zhang, Mingming and Gong, Yuan and Liu, Qingjie and Wang, Yunhong},
  journal={IEEE Transactions on Geoscience and Remote Sensing}, 
  title={Generic Knowledge Boosted Pretraining for Remote Sensing Images}, 
  year={2024},
  volume={62},
  number={},
  pages={1-13}}

@ARTICLE{model-RingMo-Lite-Wang,
  author={Wang, Yuelei and Zhang, Ting and Zhao, Liangjin and Hu, Lin and Wang, Zhechao and Niu, Ziqing and Cheng, Peirui and Chen, Kaiqiang and Zeng, Xuan and Wang, Zhirui and Wang, Hongqi and Sun, Xian},
  journal={IEEE Transactions on Geoscience and Remote Sensing}, 
  title={RingMo-Lite: A Remote Sensing Lightweight Network With CNN-Transformer Hybrid Framework}, 
  year={2024},
  volume={62},
  number={},
  pages={1-20}}

@ARTICLE{model-U-BARN-Dumeur,
  author={Dumeur, Iris and Valero, Silvia and Inglada, Jordi},
  journal={IEEE Journal of Selected Topics in Applied Earth Observations and Remote Sensing}, 
  title={Self-Supervised Spatio-Temporal Representation Learning of Satellite Image Time Series}, 
  year={2024},
  volume={17},
  number={},
  pages={4350-4367}}

@ARTICLE{model-SpectralGPT-Hong,
  author={Hong, Danfeng and Zhang, Bing and Li, Xuyang and Li, Yuxuan and Li, Chenyu and Yao, Jing and Yokoya, Naoto and Li, Hao and Ghamisi, Pedram and Jia, Xiuping and Plaza, Antonio and Gamba, Paolo and Benediktsson, Jon Atli and Chanussot, Jocelyn},
  journal={IEEE Transactions on Pattern Analysis and Machine Intelligence}, 
  title={SpectralGPT: Spectral Remote Sensing Foundation Model}, 
  year={2024},
  volume={46},
  number={8},
  pages={5227-5244}}

@INPROCEEDINGS{model-SatMAE++-Noman,
  author={Noman, Mubashir and Naseer, Muzammal and Cholakkal, Hisham and Anwar, Rao Muhammad and Khan, Salman and Khan, Fahad Shahbaz},
  booktitle={2024 IEEE/CVF Conference on Computer Vision and Pattern Recognition (CVPR)}, 
  title={Rethinking Transformers Pre-training for Multi-Spectral Satellite Imagery}, 
  year={2024},
  volume={},
  number={},
  pages={27811-27819}}

@inproceedings{model-RoMA-Wang,
 author = {Wang, Fengxiang and Wang, Yulin and Chen, Mingshuo and Wang, Haotian and Wang, Hongzhen and Zhao, Haiyan and Sun, Yangang and Wang, Shuo and Wang, Di and Lan, Long and Yang, Wenjing and Zhang, Jing},
 booktitle = {Advances in Neural Information Processing Systems},
 editor = {D. Belgrave and C. Zhang and H. Lin and R. Pascanu and P. Koniusz and M. Ghassemi and N. Chen},
 pages = {4129--4150},
 publisher = {Curran Associates, Inc.},
 title = {RoMA: Scaling up Mamba-based Foundation Models for Remote Sensing},
 volume = {38, Main Conference},
 year = {2025}
}

@InProceedings{model-MOMO-Purohit,
    author    = {Purohit, Mirali and Gajera, Bimal and Mehta, Irish and Tokas, Bhanu and Adler, Jacob and Lu, Steven and Dickenshied, Scott and Diniega, Serina and Bue, Brian and Rebbapragada, Umaa and Kerner, Hannah},
    title     = {MOMO: Mars Orbital MOdel Foundation Model for Mars Orbital Applications},
    booktitle = {Proceedings of the IEEE/CVF Conference on Computer Vision and Pattern Recognition (CVPR)},
    month     = {June},
    year      = {2026},
    pages     = {27772-27782}
}

@ARTICLE{model-DeepAndes-Guo,
  author={Guo, Junlin and Zimmer-Dauphinee, James R. and Nieusma, Jordan M. and Lu, Siqi and Liu, Quan and Deng, Ruining and Cui, Can and Yue, Jialin and Lin, Yizhe and Yao, Tianyuan and Xiong, Juming and Zhu, Junchao and Qu, Chongyu and Yang, Yuechen and Wilkes, Mitchell and Wang, Xiao and VanValkenburgh, Parker and Wernke, Steven A. and Huo, Yuankai},
  journal={IEEE Journal of Selected Topics in Applied Earth Observations and Remote Sensing}, 
  title={DeepAndes: A Self-Supervised Vision Foundation Model for Multispectral Remote Sensing Imagery of the Andes}, 
  year={2025},
  volume={18},
  number={},
  pages={26983-26999}}

@ARTICLE{model-HyperSIGMA-Wang,
  author={Wang, Di and Hu, Meiqi and Jin, Yao and Miao, Yuchun and Yang, Jiaqi and Xu, Yichu and Qin, Xiaolei and Ma, Jiaqi and Sun, Lingyu and Li, Chenxing and Fu, Chuan and Chen, Hongruixuan and Han, Chengxi and Yokoya, Naoto and Zhang, Jing and Xu, Minqiang and Liu, Lin and Zhang, Lefei and Wu, Chen and Du, Bo and Tao, Dacheng and Zhang, Liangpei},
  journal={IEEE Transactions on Pattern Analysis and Machine Intelligence}, 
  title={HyperSIGMA: Hyperspectral Intelligence Comprehension Foundation Model}, 
  year={2025},
  volume={47},
  number={8},
  pages={6427-6444}}

@article{model-TiMo-Qin,
      title={TiMo: Spatiotemporal Foundation Model for Satellite Image Time Series}, 
      author={Xiaolei Qin and Di Wang and Jing Zhang and Fengxiang Wang and Xin Su and Bo Du and Liangpei Zhang},
      journal={arXiv preprint arXiv:2505.08723},
      year={2025}
}

@article{model-SatMamba-Duc,
year = {2026},
month = {may},
publisher = {IOP Publishing},
volume = {1633},
number = {1},
pages = {012014},
author = {Duc, Chuc Man and Fukui, Hiromichi},
title = {SatMamba: Development of Foundation Models for Remote Sensing Imagery Using State Space Models},
journal = {IOP Conference Series: Earth and Environmental Science}
}

@misc{model-DynamicVis-Chen,
      title={DynamicVis: Dynamic Visual Perception for Efficient Remote Sensing Foundation Models}, 
      author={Keyan Chen and Chenyang Liu and Bowen Chen and Wenyuan Li and Zhengxia Zou and Shijian Lu and Zhenwei Shi},
      year={2026},
      eprint={2503.16426},
      archivePrefix={arXiv},
      primaryClass={cs.CV}
}

@inproceedings{model-GeoLink-DBelgrave,
 author = {Bai, Lubin and Zhang, Xiuyuan and Zhang, Siqi and Zhang, Zepeng and Wang, Haoyu and Qin, Wei and Du, Shihong},
 booktitle = {Advances in Neural Information Processing Systems},
 editor = {D. Belgrave and C. Zhang and H. Lin and R. Pascanu and P. Koniusz and M. Ghassemi and N. Chen},
 pages = {168557--168587},
 publisher = {Curran Associates, Inc.},
 title = {GeoLink: Empowering Remote Sensing Foundation Model with OpenStreetMap Data},
 volume = {38, Main Conference},
 year = {2025}
}

@article{survey-SAR-CHINA-Hou,
  author  = {Hou, Yingyan and Wu, Haoran and Zhang, Yilong and Liu, Xiangyu and Ren, Chao and Lu, Wanxuan and Yu, Hongfeng and Huang, Yuhong and Diao, Wenhui and Wang, Yixiao and Li, Ya and Han, Bing and Sun, Xian},
  title   = {SAR foundation models: a comprehensive review of data, models, and applications},
  journal = {Science China Information Sciences},
  year    = {2026},
  volume  = {69},
  number  = {8},
  pages   = {180301},
  issn    = {1869-1919}
}

@misc{dataset-MSAR-Chen,
  author       = {Chen, Jie and Huang, Zhixiang and Xia, Runfan and Wu, Bocai and Sheng, Lei and Sun, Long and Yao, Baidong},
  title        = {Large-scale Multi-class SAR Image Target Detection Dataset-1.0},
  howpublished = {Journal of Radars},
  year         = {2022},
  url          = {https://radars.ac.cn/web/data/getData?dataType=MSAR},
  note         = {Online dataset}
}

@Article{dataset-MSAR-2-Xia,
AUTHOR = {Xia, Runfan and Chen, Jie and Huang, Zhixiang and Wan, Huiyao and Wu, Bocai and Sun, Long and Yao, Baidong and Xiang, Haibing and Xing, Mengdao},
TITLE = {CRTransSar: A Visual Transformer Based on Contextual Joint Representation Learning for SAR Ship Detection},
JOURNAL = {Remote Sensing},
VOLUME = {14},
YEAR = {2022},
NUMBER = {6},
ARTICLE-NUMBER = {1488},
ISSN = {2072-4292}
}

@article{dataset-ATRNet-STAR-Liu,
  title={ATRNet-STAR: A Large Dataset and Benchmark Towards Remote Sensing Object Recognition in the Wild},
  author={Liu, Yongxiang and Li, Weijie and Liu, Li and Zhou, Jie and Peng, Bowen and Song, Yafei and Xiong, Xuying and Yang, Wei and Liu, Tianpeng and Liu, Zhen and others},
  journal={IEEE Transactions on Pattern Analysis and Machine Intelligence},
  year={2026}
}

@article{atr-PGD-Li,
  title={Predicting gradient is better: Exploring self-supervised learning for SAR ATR with a joint-embedding predictive architecture},
  author={Li, Weijie and Yang, Wei and Liu, Tianpeng and Hou, Yuenan and Li, Yuxuan and Liu, Zhen and Liu, Yongxiang and Liu, Li},
  journal={ISPRS Journal of Photogrammetry and Remote Sensing},
  volume={218},
  pages={326--338},
  year={2024},
  publisher={Elsevier}
}

@ARTICLE{detect-Madinet-Zhou,
  author={Zhou, Jie and Liu, Yongxiang and Peng, Bowen and Liu, Li and Li, Xiang},
  journal={IEEE Transactions on Circuits and Systems for Video Technology}, 
  title={MaDiNet: Mamba Diffusion Network for SAR Target Detection}, 
  year={2025},
  volume={35},
  number={11},
  pages={10787-10800},
  doi={10.1109/TCSVT.2025.3574657}}

@article{Gene-HuiYanEarth-SAR-Liu,
  title={HuiYanEarth-SAR: A Foundation Model for High-Fidelity and Low-Cost Global Remote Sensing Imagery Generation},
  author={Liu, Yongxiang and Zhou, Jie and Song, Yafei and Liu, Tianpeng and Liu, Li},
  journal={arXiv preprint arXiv:2604.11444},
  year={2026}
}

\begin{IEEEbiography}[{\includegraphics[width=1in,height=1.25in,clip,keepaspectratio]{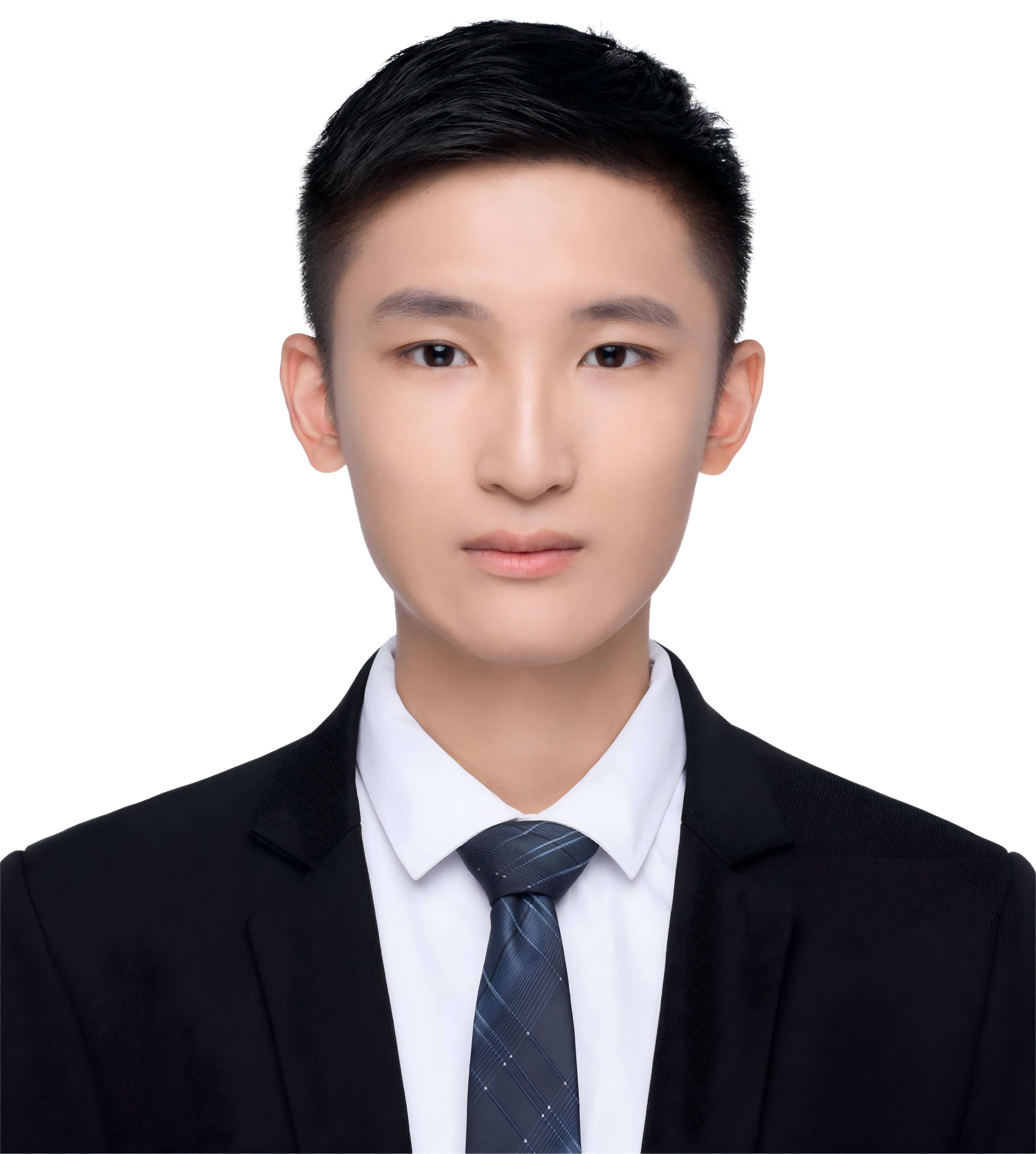}}]{Mingyao Chen}
received the B.E. degree from the College of Information and Communication, National University of Defense Technology (NUDT), China, in 2022 and the M.S. degree in information and communication engineering from National University of Defense Technology (NUDT), Changsha, China.
He is currently working toward the Ph.D. degree, majoring in information and communication engineering, with the College of Electronic Science and Technology, NUDT.

His research interests include radar automatic target recognition, pattern recognition, and remote sensing foundation model.
\end{IEEEbiography}

\vspace{-15pt} 
\begin{IEEEbiography}[{\includegraphics[width=1in,height=1.25in,clip,keepaspectratio]{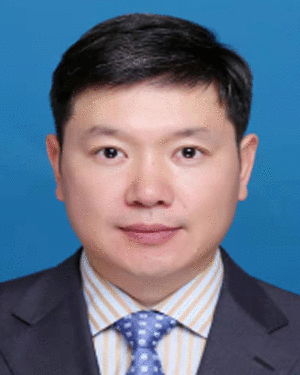}}]{Yongxiang Liu}
(Member, IEEE) received the B.S. and Ph.D. degrees in information and communication engineering from the College of Electronic Science, National University of Defense Technology, Changsha, China, in 1997 and 2004, respectively.

He is currently a Full Professor with the National University of Defense Technology. He has authored or coauthored numerous papers in respected journals, including the IEEE Transactions on Image Processing and IEEE Transactions on Geoscience and Remote Sensing. His research interests include radar imaging, SAR image interpretation, and artificial intelligence.

\end{IEEEbiography}

\vspace{-15pt} 

\begin{IEEEbiography}[{\includegraphics[width=1in,height=1.25in,clip,keepaspectratio]{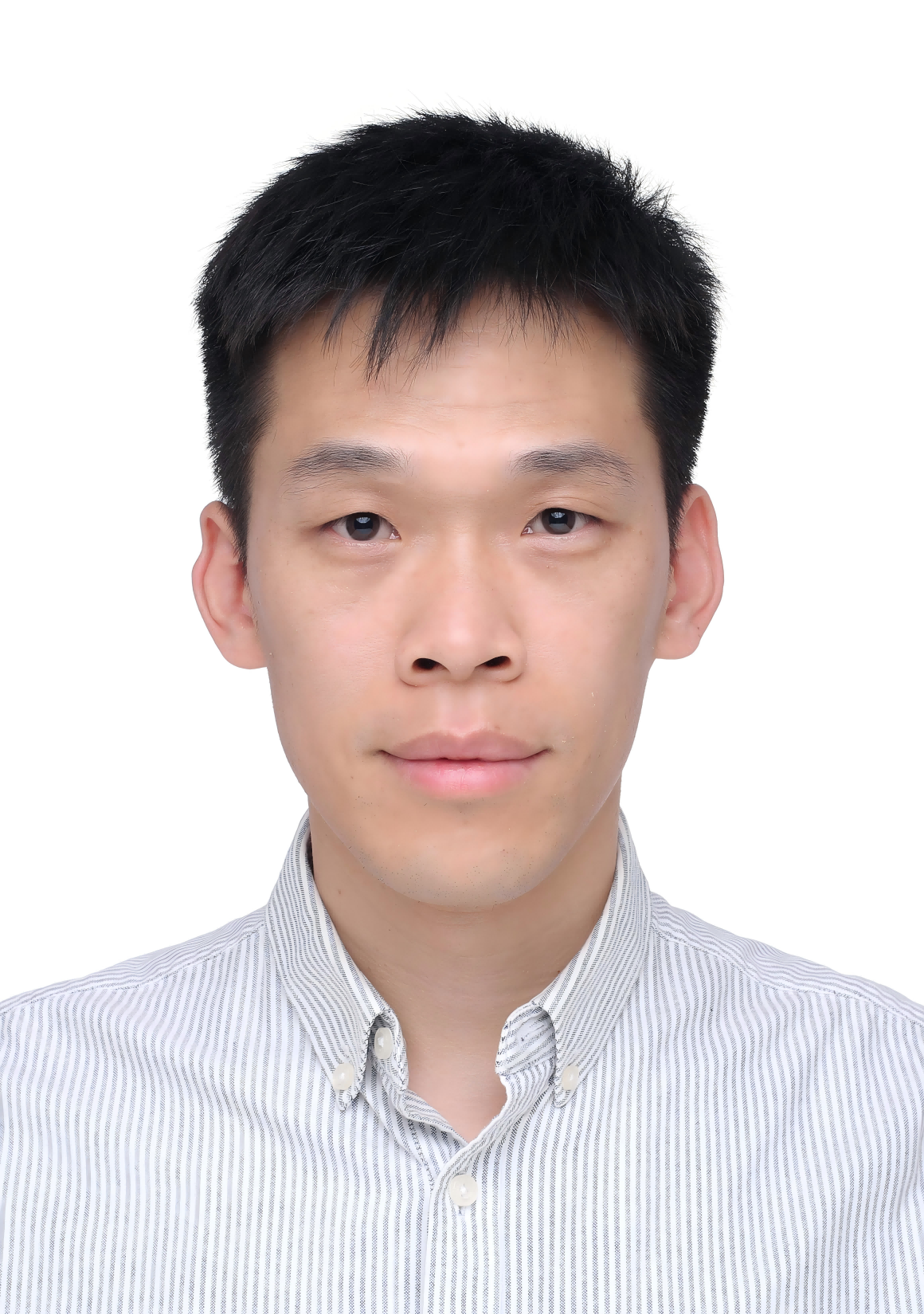}}]{Tianpeng Liu}
received the B.E., M.E., and Ph.D. degrees in information and communication engineering from the National University of Defense Technology, Changsha, China, in 2008, 2011, and 2016 respectively.

He is currently an Associate Professor with the College of Electronic Science and Technology, Changsha, China. He has authored or coauthored numerous papers in respected journals, including the IEEE Transactions on Aerospace and Electronic Systems and International Conference on Signal Processing. His research interests include radar signal processing, electronic countermeasure, and cross-eye jamming.

\end{IEEEbiography}

\vspace{-15pt} 
\begin{IEEEbiography}[{\includegraphics[width=1in,height=1.25in,clip,keepaspectratio]{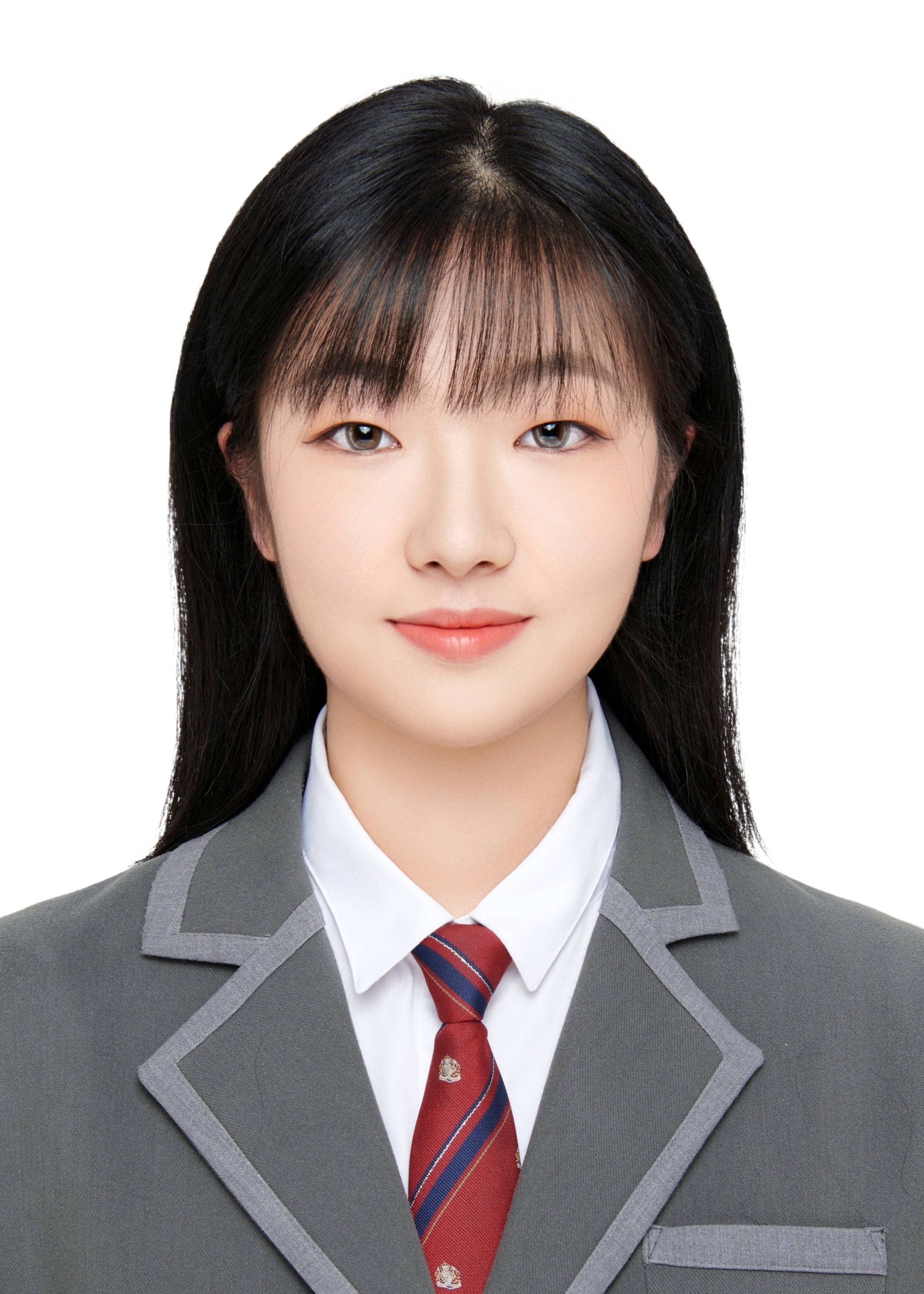}}]{Yuchen Yong}
received the M.Eng. degree from the National University of Defense Technology, Changsha, China, in 2026, where she is currently pursuing the Ph.D. degree with the College of Electronic Science and Technology. 
Her research interests include multimodal processing and visual grounding in remote sensing imagery.
\end{IEEEbiography}

\vspace{-15pt} 
\begin{IEEEbiography}[{\includegraphics[width=1in,height=1.25in,clip,keepaspectratio]{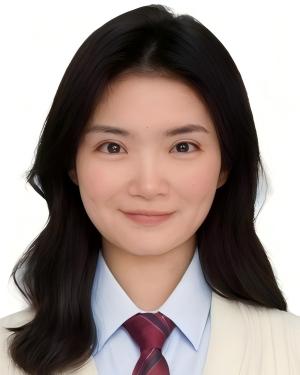}}]{Jie Zhou}
received the BS degree from the National University of Defense Technology, Changsha, China, in 2020. She is currently working toward the PhD degree with the College of Electronic Science and Technology. Her research interests include diffusion models, remote sensing image interpretation, and object detection in SAR imagery.

\end{IEEEbiography}

\vspace{-15pt} 
\begin{IEEEbiography}[{\includegraphics[width=1in,height=1.25in,clip,keepaspectratio]{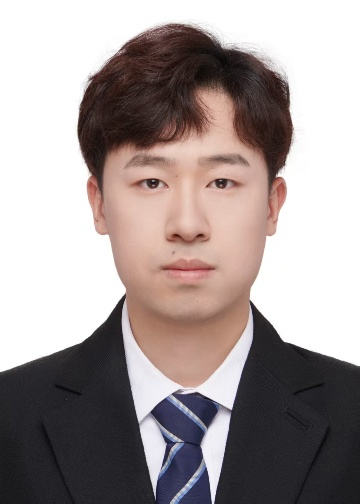}}]{Zhenhao Yang}
received the B.S. degree in Communications Engineering from Beijing Institute of Petrochemical Technology, Beijing, China, in 2023, and the M.S. degree in Communication Engineering from North China University of Technology, Beijing, China, in 2026. He is currently pursuing the Ph.D. degree with the National University of Defense Technology, Changsha, China. His research interests include synthetic aperture radar (SAR) foundation models, multi-task learning, and remote sensing image understanding.

\end{IEEEbiography}

\vspace{-15pt} 
\begin{IEEEbiography}[{\includegraphics[width=1in,height=1.25in,clip,keepaspectratio]{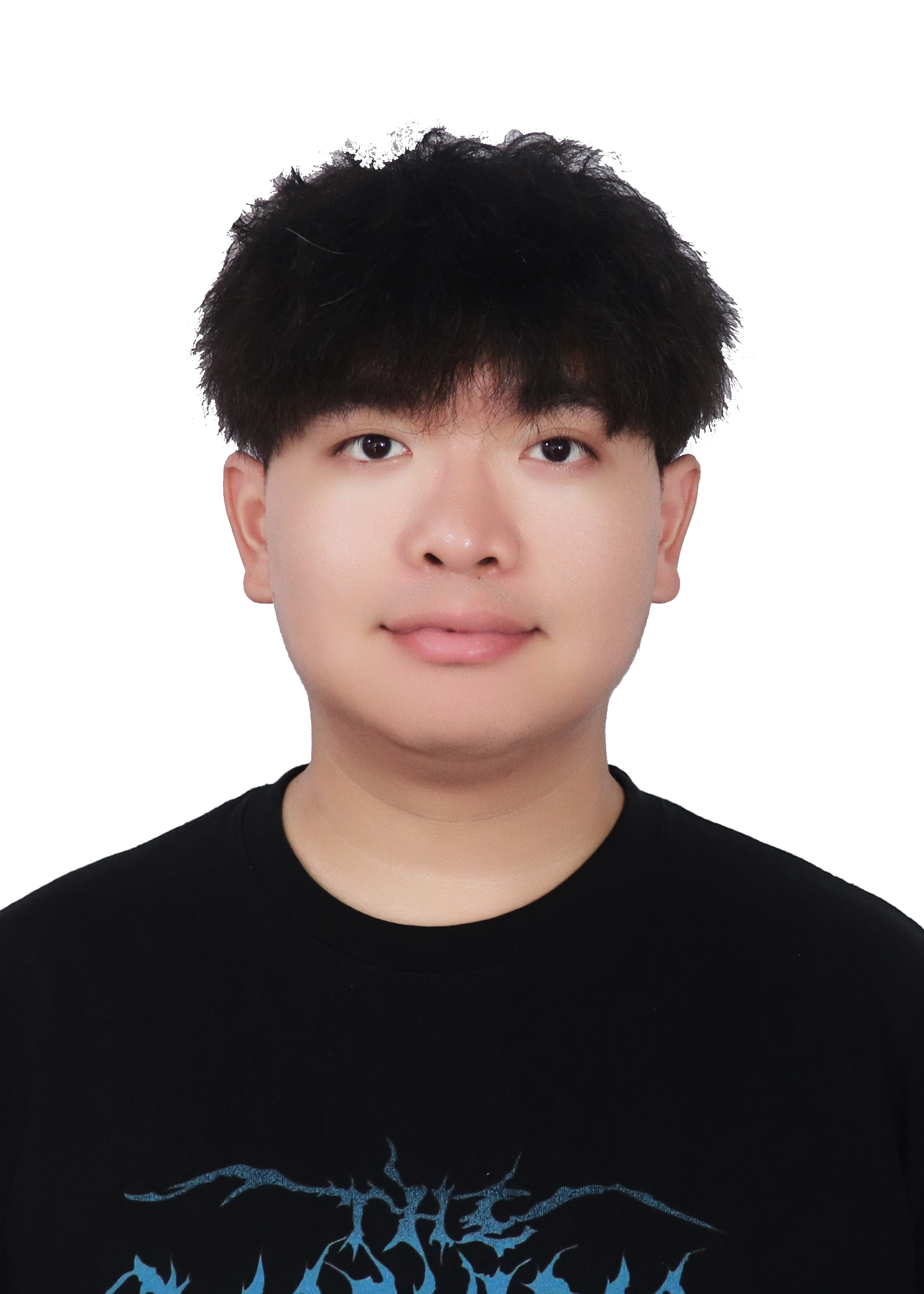}}]{Pinyi Huang}
received the M.S. degree from Central South University, Changsha, China, in 2025, and is currently pursuing a Ph.D. degree. His research interests include parameter-efficient fine-tuning (PEFT) for SAR target detection.

\end{IEEEbiography}

\vspace{-15pt} 
\begin{IEEEbiography}[{\includegraphics[width=1in,height=1.25in,clip,keepaspectratio]{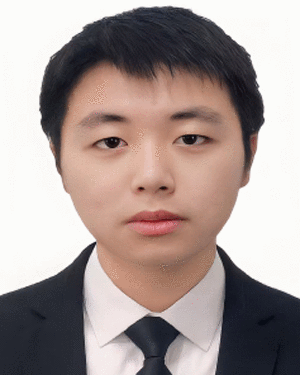}}]{Weijie Li}
received the B.E. degree in information engineering from Xi’an Jiaotong University, Xi’an, China, in 2019, and the Ph.D. degree in information and communication engineering from the National University of Defense Technology, Changsha, China, in 2026.,He is currently a Postdoctoral Researcher with the National University of Defense Technology. He has authored or coauthored papers in respected journals, including IEEE Transactions on Image Processing, ISPRS Journal P\&RS, and SCIENTIA SINICA Informationis. His research interests include radar target recognition and deep learning.

\end{IEEEbiography}

\vspace{-15pt} 
\begin{IEEEbiography}[{\includegraphics[width=1in,height=1.25in,clip,keepaspectratio]{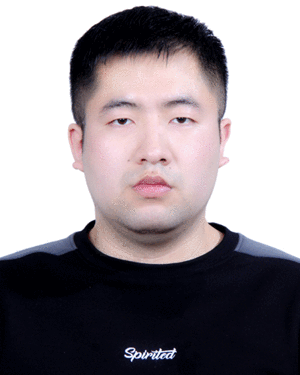}}]{Jiehua Zhang}
received the M.S. degree in aerospace science and technology from the National University of Defense Technology (NUDT), Changsha, China, in 2020, and the Ph.D. degree in computer science and engineering from the University of Oulu, Oulu, Finland, in 2025.,He is currently a Post-Doctoral Researcher with NUDT. His research interests include deep learning, computer vision, and efficient network design.

\end{IEEEbiography}

\vspace{-15pt} 
\begin{IEEEbiography}[{\includegraphics[width=1in,height=1.25in,clip,keepaspectratio]{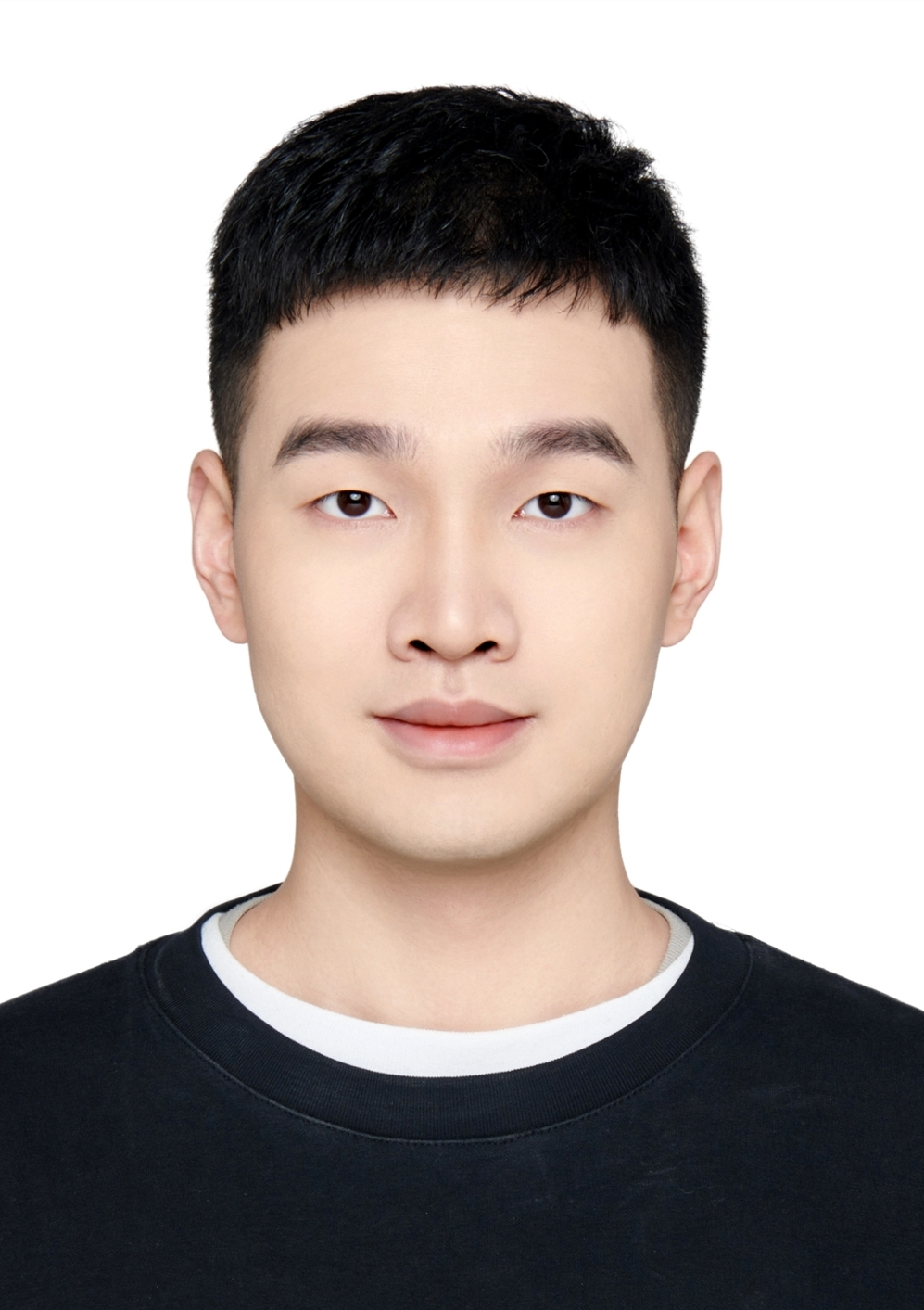}}]{Bowen Peng}
received his B.S. and M.S. degrees from NUDT, China, in 2020 and 2022, respectively, where he is pursuing the Ph.D. degree. His research interests include adversarial robustness of deep learning and trustworthy remote sensing object recognition.

\end{IEEEbiography}

\vspace{-15pt} 
\begin{IEEEbiography}[{\includegraphics[width=1in,height=1.25in,clip,keepaspectratio]{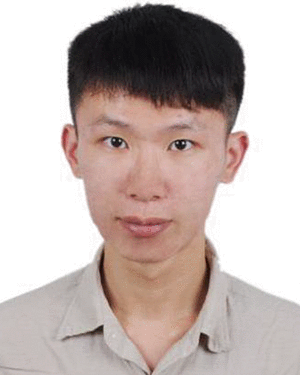}}]{Zhimao Peng}
is currently working toward the Ph.D. degree with the College of Computer Science, Nankai University, Tianjin, China, under the supervision of Prof. Ming-Ming Cheng. His research interests include deep learning, image classification, and open-world learning.

\end{IEEEbiography}

\vspace{-15pt} 
\begin{IEEEbiography}[{\includegraphics[width=1in,height=1.25in,clip,keepaspectratio]{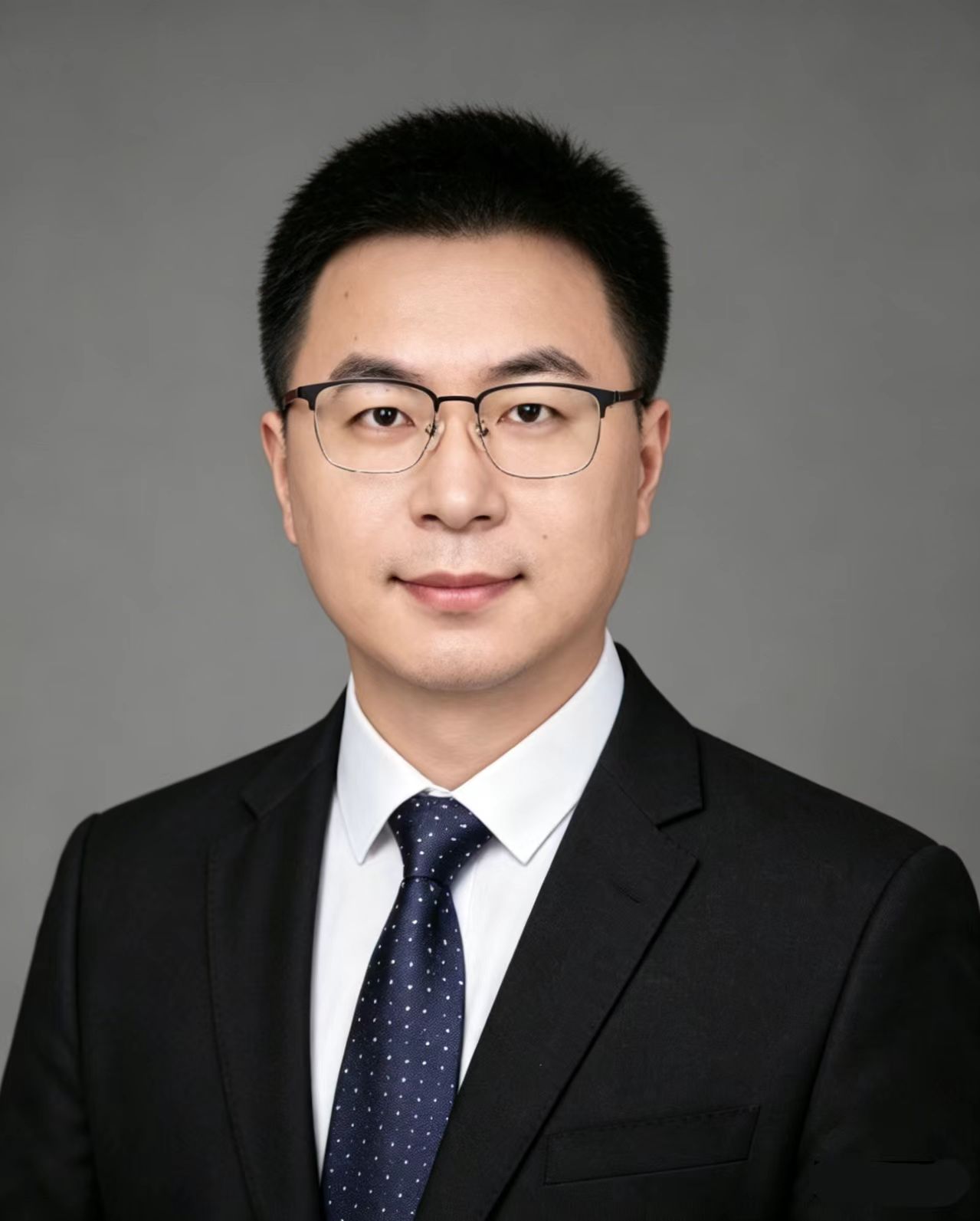}}]{Zhiyuan Zhang}
received the B.E. degree in information engineering,the M.E.degree in control science and engineering,and the Ph.D.degree in information and communication engineering from Northwestern Polytechnical University (NPU) Xi'an, China, in 2016, 2019, and 2023, respectively.His primary research interests include computer vision, point cloud,signal processing,and deep learning in artificial intelligence.

\end{IEEEbiography}

\vspace{-15pt} 
\begin{IEEEbiography}[{\includegraphics[width=1in,height=1.25in,clip,keepaspectratio]{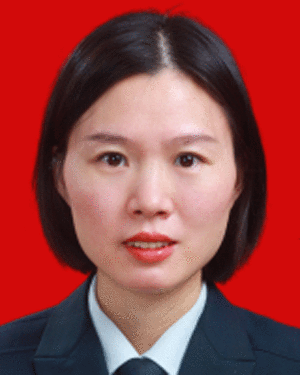}}]{Li Liu}
(Senior Member, IEEE) received the Ph.D. degree in information and communication engineering from the National University of Defense Technology (NUDT), Changsha, China, in 2012.

She is currently a Full Professor with NUDT. She has held visiting appointments with the University of Waterloo, Waterloo, ON, Canada, The Chinese University of Hong Kong, Hong Kong, and University of Oulu, Oulu, Finland, respectively. Her research interests include computer vision, pattern recognition, and machine learning.
Dr. Liu was a Co-Chair of many International Workshops along with major venues, such as CVPR and ICCV. She was the Leading Guest Editor of the special issues for IEEE Transactions on Pattern Analysis and Machine Intelligence and International Journal of Computer Vision. She is currently an Associate Editor for IEEE Transactions on Circuits and Systems for Video Technology and Pattern Recognition. Her papers currently have more than 10 000 citations, according to Google Scholar.

\end{IEEEbiography}

\end{document}